\documentstyle[cite,epsf]{crnart12}
\newcommand{\Foot}{\renewcommand{\thefootnote}{\fnsymbol{footnote}}
\footnote{\small Lectures given at the
1994 European School of High-Energy Physics (Sorrento, Italy,
29 August -- 11 September, 1994)}
\addtocounter{footnote}{-1}
\renewcommand{\thefootnote}{\arabic{footnote}}}
\newcommand{\Preprint}{\vspace*{-1.0cm} \noindent hep-ph/9505231 \hfill
  FTUV/95-19 \\ \mbox{}\hfill IFIC/95-19 \\  \mbox{}\hfill
  May 1995}
\renewcommand{\theequation}{\arabic{section}.\arabic{equation}}
\def\slashchar#1{\setbox0=\hbox{$#1$}\dimen0=\wd0%
\setbox1=\hbox{/}\dimen1=\wd1%
\ifdim\dimen0>\dimen1%
\rlap{\hbox to
\dimen0{\hfil/\hfil}}#1\else
\rlap{\hbox to \dimen1{\hfil$#1$\hfil}}/\fi}
\newcommand{\eqn}[1]{(\ref{#1})}

\newcommand{\be}{\begin{equation}}
\newcommand{\ee}{\end{equation}}
\newcommand{\ba}{\begin{array}{c}}
\newcommand{\bat}{\begin{array}{cc}}
\newcommand{\bath}{\begin{array}{ccc}}
\newcommand{\ea}{\end{array}}
\newcommand{\beqn}{\begin{eqnarray}}
\newcommand{\eeqn}{\end{eqnarray}}
\newcommand{\dis}{\displaystyle}
\newcommand{\bi}{\begin{itemize}}
\newcommand{\ei}{\end{itemize}}

\newcommand{\rms}{\rm\scriptsize}

\newcommand{\toU}{\stackrel{\mbox{\rms U(1)}}{\longrightarrow}}

\newcommand{\cL}{{\cal L}}

\newcommand{\cM}{{\cal M}}
\newcommand{\cO}{{\cal O}}
\newcommand{\cP}{{\cal P}}

\newcommand{\no}{\nonumber}
\newcommand{\bel}[1]{\be\label{#1}}
\newcommand{\e}{\mbox{\rm e}}
\begin{document}

\Preprint

\vspace*{1cm}  
\begin{flushleft}
{\bf QUANTUM CHROMODYNAMICS}\Foot\\
A. Pich\\
Departament de F\'\i sica Te\`orica
and IFIC,  Universitat de Val\`encia -- CSIC\\
Dr. Moliner 50,
E--46100 Burjassot, Val\`encia, Spain
\vspace*{1cm}  
\end{flushleft}

{\rightskip=5pc \leftskip=5pc \noindent
{\bf Abstract}\\
These lectures provide an overview of
Quantum Chromodynamics (QCD), the
$SU(3)_C$ gauge
theory of the strong interactions. After briefly reviewing the
empirical considerations which lead to the introduction of
{\it colour}, the QCD Lagrangian is discussed.
The running of the strong coupling and the associated property of
{\it Asymptotic Freedom} are analyzed.
Some selected experimental tests and the present knowledge of
$\alpha_s$ are summarized.
A short description of
the QCD flavour symmetries and the
{\it dynamical breaking of chiral symmetry} is also given.
A more detailed discussion can be found in standard textbooks
\cite{AH:89,MU:87,PT:84,YN:93}
and recent reviews \cite{HE:92,ES:94,BE:94,WE:94,SC:94}.
\vglue 2.0cm}

\section {QUARKS AND COLOUR}
\label{sec:introduction}

A fast look into the Particle Data Tables \cite{PDG:94}
reveals the richness and variety of the hadronic spectrum.
The large number of known mesonic and baryonic states clearly signals the
existence of a deeper level of elementary constituents of matter:
{\it quarks} \cite{GM:64}.
In fact, the messy hadronic world can be easily understood in terms
of a few constituent spin-$\frac{1}{2}$ quark {\it flavours}:

\begin{center}
\begin{tabular}{|c|ccc|}  \hline
$Q=+\frac{2}{3}$ & u & c & t    \\ \hline
$Q=-\frac{1}{3}$ & d & s & b    \\ \hline
\end{tabular}
\end{center}
\smallskip

\noindent
Assuming that mesons are $M\equiv q\bar q$
states, while baryons have
three quark constituents, $B\equiv qqq$, one can nicely classify
the entire hadronic spectrum:

\begin{center}
\begin{tabular}{llll}
$\pi^+ = u\bar d$,  & $K^+ = u\bar s$,    &   $K^0 = d\bar s$, &
$\pi^0 = (u\bar u - d\bar d)/\sqrt{2}$  \,\ldots
\\
$D^+ = c\bar d$,  &  $D^0 = c\bar u$,   &   $D^+_s = c\bar s$  &\ldots
\\
$B^+ = u\bar b$, & $B^0 = d\bar b$, & $B^0_s = s\bar b$, &
$B^+_c = c\bar b$  \,\ldots
\\
$p = u u d$, & $n = u d d$, &
$\Sigma^+ = u u s$, & $\Sigma^0 = u d s$  \,\ldots
\\
$\Sigma_c^+ = u d c$, & $\Sigma_c^{++} = u u c$, & $\Xi_c^+ = u s c$, &
$\Xi_c^0 = d s c$  \,\ldots
\\
$\Xi_{cc}^+ = d c c$, & $\Xi_{cc}^{++} = u c c$, &
$\Omega^+_{cc} = s c c$  &\ldots
\end{tabular}
\end{center}

There is a one--to--one correspondence between the
observed hadrons and the states predicted by this simple classification;
thus, the {\it Quark Model} appears to be a very useful
{\it Periodic Table of Hadrons}.
However, the quark picture faces a problem concerning the Fermi--Dirac
statistics of the constituents.
Since the fundamental state of a composite system is expected to have
$L=0$, the $\Delta^{++}$ baryon ($J=\frac{3}{2}$) corresponds to
$\, u^\uparrow u^\uparrow u^\uparrow \, $,
with the three quark-spins aligned into the same direction
($s_3=+\frac{1}{2}$) and all relative angular momenta equal to zero.
The wave function is symmetric and, therefore, the
$\Delta^{++}$ state obeys the wrong statistics.

The problem can be solved assuming \cite{GEL:72}
the existence of a new quantum
number, {\it colour}, such that each species of quark may have $N_C=3$
different colours: $q^\alpha$, $\alpha =1,2,3$ (red, yellow, violet).
Then one can reinterpret the $\Delta^{++}$ as
the antisymmetric state
\be\label{eq:Delta_wf}
\Delta^{++} \, = \, {1\over\sqrt{6}}\,\epsilon^{\alpha\beta\gamma}\,
|u^\uparrow_\alpha u^\uparrow_\beta u^\uparrow_\gamma\rangle
\ee
%
(notice that at least 3 colours are needed for making an antisymmetric
state).
In this picture, baryons and mesons are described by
the colour-singlet combinations
\be\label{eq:m_b_wf}
B\, =\, {1\over\sqrt{6}}\,\epsilon^{\alpha\beta\gamma}\,
|q_\alpha q_\beta q_\gamma\rangle \, ,
\qquad\qquad
M\, =\, {1\over\sqrt{3}}\,\delta^{\alpha\beta} \,
|q_\alpha \bar q_\beta \rangle \, .
\ee
In order to avoid the existence of non-observed extra states
with non-zero colour, 
one needs to further postulate that all asymptotic states
are colourless, i.e. singlets under rotations in colour space.
This assumption is known as the {\it confinement hypothesis},
because it implies the non-observability of free quarks:
since quarks carry colour they are confined within
colour-singlet bound states.

\begin{figure}[hbt]
\centerline{\epsfxsize = 6.8cm \epsfbox{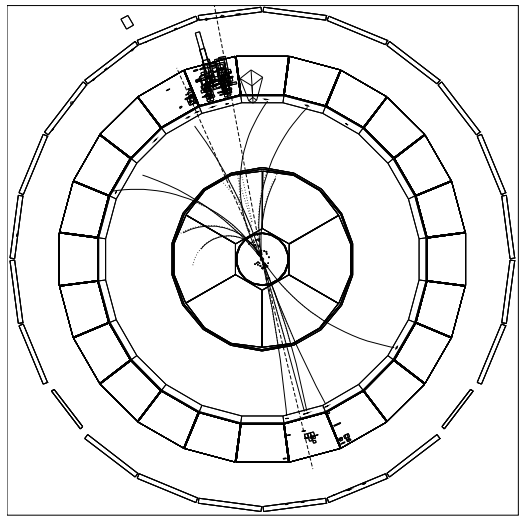}}
\caption{Two-jet event from the hadronic decay of a $Z$ boson (DELPHI).}
\label{fig:two_jet}
\end{figure}

The quark picture is not only a nice mathematical scheme to classify the
hadronic world. We have strong experimental evidence of the existence of
quarks. Fig. \ref{fig:two_jet} shows a typical
$Z\to \mbox{\rm hadrons}\, $ event,
obtained at LEP. Although there are many hadrons in the final state, they
appear to be collimated in 2 {\it jets} of particles, as expected from a
two-body decay $Z\to q\bar q$, where the $q\bar q$ pair has later
{\it hadronized}.

\subsection{Evidence of colour}

\begin{figure}[htb]
\centerline{\epsfysize = 2.2cm \epsfbox{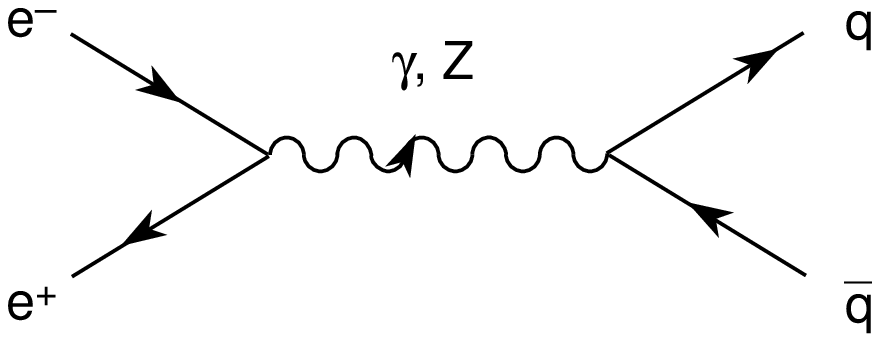}}
\caption{Feynman diagram for $e^+e^-\to \mbox{\rm hadrons}$.}
\label{fig:eediagram}
\end{figure}

A direct test of the colour quantum number can be obtained from the
ratio
\be\label{eq:R_ee}
R_{e^+e^-} \equiv
{\sigma(e^+e^-\to \mbox{\rm hadrons})\over\sigma(e^+e^-\to\mu^+\mu^-)} \, .
\ee
The hadronic production occurs through
$e^+e^-\to\gamma^*, Z^*\to q\bar q\to \mbox{\rm hadrons}$.
Since quarks are assumed
to be confined, the probability to hadronize is just one; therefore,
the sum over all possible quarks in the final state will give the total
inclusive cross-section into hadrons.
At energies well below the $Z$ peak, the cross-section is
dominated by the $\gamma$-exchange amplitude;
the ratio $R_{e^+e^-}$ is then given by the sum of the
quark electric charges squared:
\be\label{eq:R_ee_res}
R_{e^+e^-} \approx N_C \, \sum_{f=1}^{N_f} Q_f^2 \, = \,
\left\{
\begin{array}{cc}
\frac{2}{3} N_C = 2\, , \qquad & (N_f=3 \; :\; u,d,s)  \\
\frac{10}{9} N_C = \frac{10}{3}\, , \qquad & (N_f=4 \; :\; u,d,s,c)  \\
\frac{11}{9} N_C = \frac{11}{3} \, ,\qquad & (N_f=5 \; :\; u,d,s,c,b)
\ea\right. .
\ee
%

\begin{figure}[bht]
\vspace{-0.4cm}
\centerline{\epsfxsize = 12cm \epsfbox{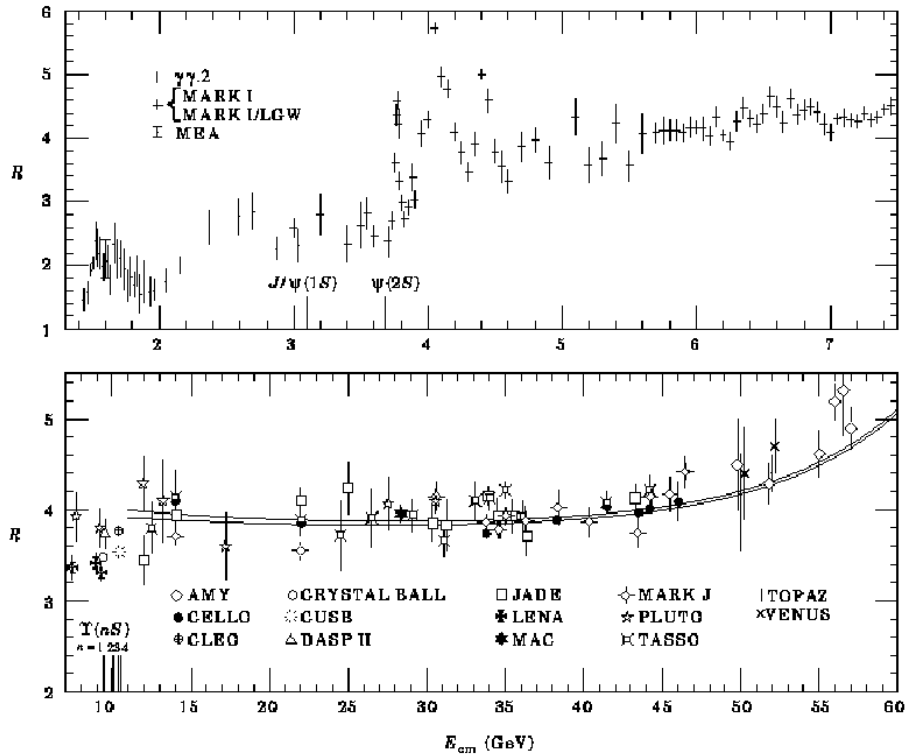}}  \vspace{-0.3cm}
\caption{Measurements of $R_{e^+e^-}$ \protect\cite{PDG:94}.
The two continuous curves are QCD fits.}
\label{fig:Ree}
\end{figure}

\begin{figure}[thb]
\centerline{\epsfysize=3cm \epsfbox{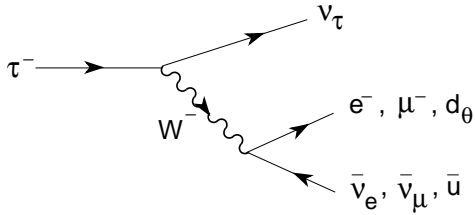}}
\caption{$\tau$-decay diagram.}
\label{fig:tau}
\end{figure}

The measured ratio is shown in Fig. \ref{fig:Ree}. Although
the simple
formula \eqn{eq:R_ee_res} cannot explain the complicated
structure around the different quark thresholds,
it gives the right average
value of the cross-section (away from the thresholds),  
provided that $N_C$ is taken to be three.
The agreement is better at larger energies.
Notice that strong
interactions have not been taken into account;
only the confinement hypothesis has been used.

The hadronic decay of the $\tau$ lepton provides additional evidence
for $N_C=3$. The decay proceeds through the $W$-emission diagram
shown in Fig.~\ref{fig:tau}. Since the $W$ coupling to the charged
current is of universal strength, there are $(2+N_C)$ equal
contributions
(if final masses and strong interactions are neglected)
to the $\tau$-decay width. Two of them correspond
to the leptonic decay modes $\tau^-\to\nu_\tau e^-\bar\nu_e$ and
$\tau^-\to\nu_\tau \mu^-\bar\nu_\mu$, while the other $N_C$
are associated with the possible colours of the quark--antiquark
pair in the $\tau^-\to\nu_\tau d_\theta u$ decay mode
($d_\theta\equiv \cos\theta_C d + \sin\theta_C s$).
Hence, the branching ratios for the different channels are expected
to be approximately:
\
\beqn\label{eq:B_l}
B_{\tau\to l}\equiv
&\!\!\!\!\!\! & \!\!\!\!\!\!
{\rm Br}(\tau^-\to\nu_\tau l^-\bar\nu_l)
\approx {1\over 2+N_C} = {1\over 5} = 20 \% \,  , \quad
\\ \label{eq:R_tau}
R_\tau &\!\!\!
\equiv &\!\!\!
{\Gamma(\tau^-\to\nu_\tau + \mbox{\rm hadrons})\over
   \Gamma(\tau^-\to\nu_\tau e^-\bar\nu_e)} \approx N_C = 3 \, ,
\eeqn
which should be compared with the experimental averages \cite{PDG:94}:
\beqn\label{eq:B_l_exp}
B_{\tau\to e} = (18.01\pm 0.18)\% \, , \qquad\qquad
B_{\tau\to \mu} = (17.65\pm 0.24)\% \, ,
\\ \label{eq:R_tau_exp}
R_\tau = (1-B_{\tau\to e}-B_{\tau\to \mu})/B_{\tau\to e}
= 3.56\pm 0.04 \, . \qquad\qquad
\eeqn
The agreement is fairly good. Taking $N_C=3$, the naive predictions
only deviate from the measured values by about 20\%.
Many other observables, such as the partial widths of the $Z$ and $W^\pm$
bosons, can be analyzed in a similar way to conclude that $N_C=3$.

A particularly strong test is obtained from the $\pi^0\to\gamma\gamma$ decay,
which occurs through the triangular quark loops in Fig.~\ref{fig:triangle}.
The crossed vertex denotes
the axial current
$A_\mu^3\equiv (\bar u \gamma_\mu\gamma_5 u - \bar d \gamma_\mu\gamma_5 d)$.
One gets:
\be\label{eq:pi_decay}
\Gamma(\pi^0\to\gamma\gamma) = \left( {N_C\over 3}\right)^2
{\alpha^2 m_\pi^3\over 64 \pi^3 f_\pi^2} = 7.73 \, {\rm eV} ,
\ee
where the $\pi^0$ coupling to $A_\mu^3$,
$f_\pi = 92.4$ MeV, is known from  the   
$\pi^-\to\mu^-\bar\nu_\mu$ decay rate (assuming isospin symmetry).
The agreement with the measured value,
$\Gamma = 7.7\pm 0.6$ eV \cite{PDG:94}, is remarkable.
With $N_C=1$, the prediction would have failed by a factor of 9.
The nice thing about this decay is that it is associated with an
{\it anomaly}: a global flavour symmetry  
which is broken
by quantum effects (the triangular loops). One can then proof that
the decay amplitude \eqn{eq:pi_decay} does not get corrected by
strong interactions \cite{AB:69}.

\begin{figure}[htb]
\centerline{\epsfysize = 3cm \epsfbox{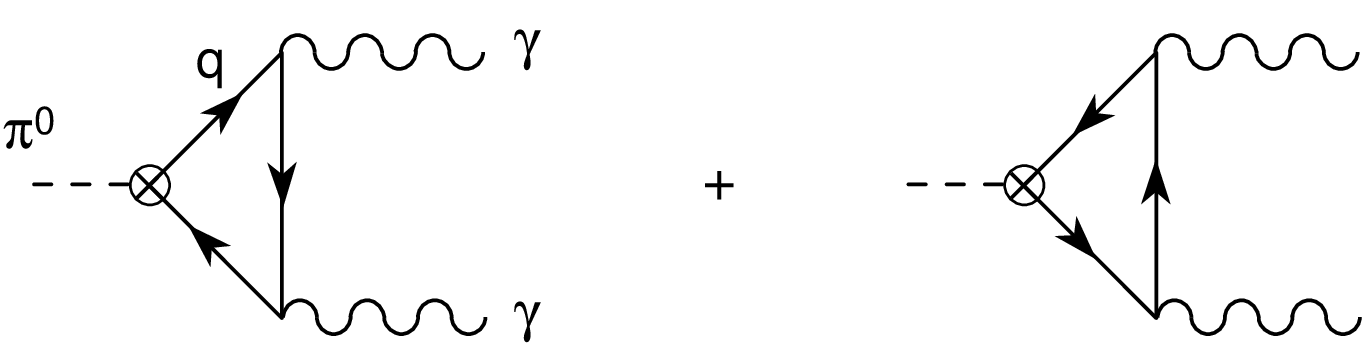}}
\caption{Triangular quark loops generating the decay $\pi^0\to\gamma\gamma$.}
\label{fig:triangle}
\end{figure}

Anomalies provide another compelling theoretical reason to
adopt $N_C=3$. The gauge symmetries of the Standard Model of
electroweak interactions have also anomalies associated with
triangular fermion loops
(diagrams of the type shown in Fig.~\ref{fig:triangle}, but with
arbitrary gauge bosons --$W^\pm,Z,\gamma$-- in the external legs and
Standard Model fermions in the internal lines).
These gauge anomalies are deathly because
they destroy the renormalizability of the theory. Fortunately,
the sum of all possible triangular loops cancels if $N_C=3$.
Thus, with three colours, anomalies are absent and the Standard
Model is well-defined.

\subsection{Asymptotic Freedom}
\label{subsec:AF}

  The structure of the proton can be probed through the scattering
$e^- p\to e^- p$. The cross-section is given by
\be\label{eq:ep_ep}
{d \sigma\over d Q^2} =
{\pi\alpha^2\cos^2\! {\theta\over 2} \over
4 E^2 \sin^4\! {\theta\over 2} E E'}\,
\left\{
{|G_E(Q^2)|^2 + {Q^2\over 4 M_p^2} |G_M(Q^2)|^2 \over 1 + {Q^2\over 4 M_p^2}}
+ {Q^2\over 2 M_p^2} |G_M(Q^2)|^2 \tan^2{\! {\theta\over 2}}
\right\} ,
\ee
where $E$ and $E'$ are the energies of the incident and scattered
electrons, respectively, in the proton rest-frame, $\theta$
the scattering angle, $M_p$ the proton mass and
\be\label{eq:Q2}
Q^2\equiv - q^2 = 4 E E'\sin^2{\! {\theta\over 2}} \, ,
\ee
with $q^\mu\equiv (k_e - k'_e)^\mu$ the momentum transfer through the
intermediate photon propagator.

$G_E$ and $G_M$ are the electric and magnetic form factors, respectively,
describing the proton electromagnetic structure; they would be equal to
one for a pointlike spin-$\frac{1}{2}$ target. Experimentally they are
known to be very well approximated by the dipole form
\be\label{eq:dipole}
G_M(Q^2)/\mu_p \approx G_E(Q^2) \approx
\left( 1 + {Q^2\over 0.7 \, \mbox{\rm GeV}^2}\right)^{-2} \, ,
\ee
where $\mu_p=2.79$ is the proton magnetic moment (in proton
Bohr magneton units). Thus, the proton is actually an extended object with a
size of the order of 1 fm.
At very low energies ($Q^2<< 1 \, \mbox{\rm GeV}^2$), the photon probe
is unable to get information on
the proton structure, $G_{M,E}(Q^2)\approx G_{M,E}(0) =1$, and
the proton behaves as a pointlike particle.
At higher energies, the photon is sensitive to shorter distances; the proton
finite size gives then rise to form factors, which suppress the elastic
cross-section at large $Q^2$, i.e. at large angles.

\begin{figure}[htb]
\centerline{\epsfysize = 4cm \epsfbox{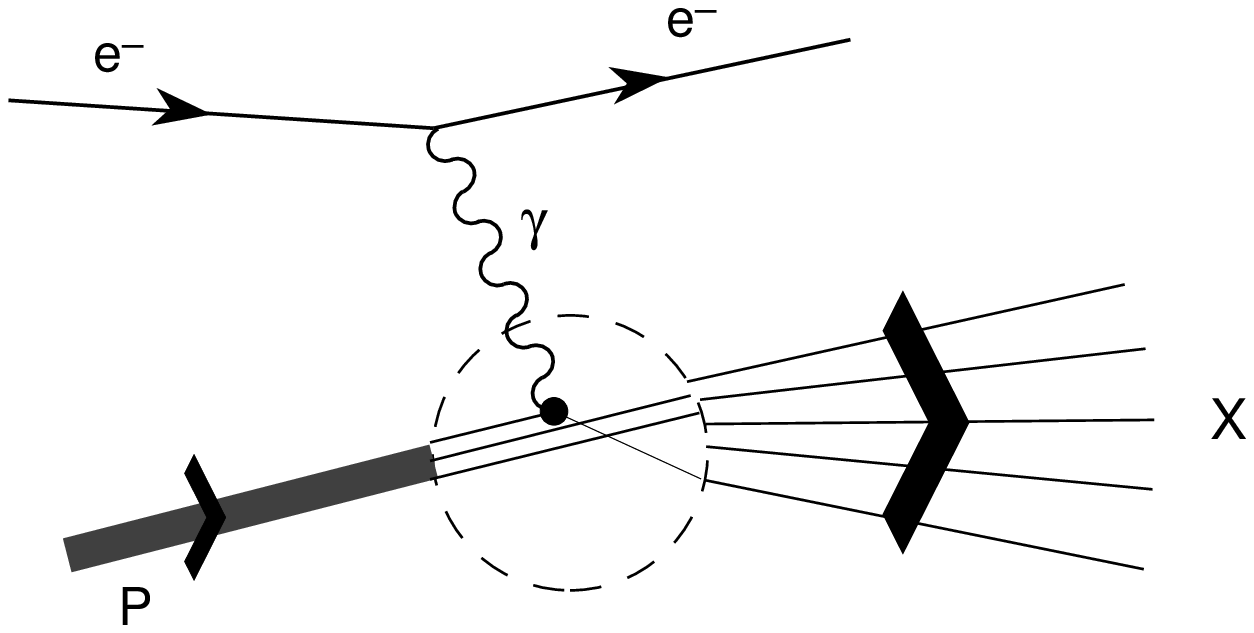}}
\caption{Inelastic $e^-p\to e^-X$ scattering.}
\label{fig:ep_eX}
\end{figure}

One can try to further resolve the proton structure, by increasing the
incident energy. The inelastic scattering
$e^- p \to e^- X$ becomes then the dominant process. Making an inclusive
sum over all hadrons produced, one has an additional kinematical variable
corresponding to the final hadronic mass, $W^2\equiv P_X^2$. The scattering
is usually described in terms of $Q^2$ and
\be\label{eq:nu}
\nu \equiv {(P\cdot q)\over M_p} = {Q^2 + W^2 - M_p^2 \over 2 M_p}
 = E - E' \, ,
\ee
where $P^\mu$ is the proton cuadrimomentum;
$\nu$ is the energy transfer in the proton rest-frame.
In the one-photon approximation, the unpolarized differential
cross-section is given by
\be\label{eq:sigma_inel}
{d \sigma\over d Q^2 \, d \nu} =
{\pi\alpha^2\cos^2\! {\theta\over 2} \over
4 E^2 \sin^4\! {\theta\over 2} E E'}\,
\left\{  W_2(Q^2,\nu) + 2\, W_1(Q^2,\nu)\, \tan^2{\! {\theta\over 2}} \right\}
{}.
\ee
The proton structure is then characterized by two measurable
{\it structure functions}.
For a pointlike proton, the elastic scattering \eqn{eq:ep_ep} corresponds to
\be\label{eq:W-elastic}
W_1(Q^2,\nu) = {Q^2\over 4 M_p^2}\,\delta\!\left(\nu - {Q^2\over 2 M_p}\right)
,
\qquad\qquad
W_2(Q^2,\nu) = \delta\!\left(\nu - {Q^2\over 2 M_p}\right) .
\ee

At low $Q^2$, the experimental data
reveals prominent resonances;
but this resonance structure quickly dies out as $Q^2$ increases.
A much softer but sizeable continuum contribution persists at large $Q^2$,
suggesting the existence of pointlike objects inside the proton.

To get an idea of the possible behaviour of the structure functions, one
can make a very rough model of the proton, assuming that it consist of
some number of pointlike spin-$\frac{1}{2}$ constituents (the so-called
{\it partons}), each one carrying a given fraction $\xi_i$ of the
proton momenta, i.e. ${p_i}^\mu = \xi_i P^\mu$. That means that we are
neglecting\footnote{\small
These approximations can be made more precise going to the
infinite momentum frame of the proton, where the transverse motion
is negligible compared with the large longitudinal boost of the partons.}
the transverse parton momenta, and $m_i = \xi M_p$.
The interaction of the photon-probe with the parton $i$ generates a
contribution to the structure functions given by:
\beqn\label{eq:W1}
W_1^{(i)}(Q^2,\nu) &=& {e_i^2 Q^2\over 4 m_i^2}\:
\delta\!\left(\nu - {Q^2\over 2 m_i}\right) =
{e_i^2\over 2 M_p}\, \delta\left(\xi_i - x\right) ,
\\ \label{eq:W2}
W_2^{(i)}(Q^2,\nu) &=& e_i^2 \delta\!\left(\nu - {Q^2\over 2 m_i}\right) =
e_i^2 {x\over\nu}\: \delta\left(\xi_i - x\right) ,
\eeqn
where $e_i$ is the parton electric charge and
\bel{eq:x_def}
x\equiv {Q^2\over 2 M_p\nu} = {Q^2\over Q^2 + W^2 - M_p^2} \, .
\ee
Thus, the parton structure functions only depend on the ratio $x$,
which, moreover, fixes the momentum fractions $\xi_i$.
We can go further, and assume that in the limit $Q^2\to\infty$,
$\nu\to\infty$, but keeping $x$ fixed, the proton structure functions
can be estimated from an incoherent sum of the parton ones
(neglecting any strong interactions among the partons). Denoting
$f_i(\xi_i)$ the probability that the parton $i$ has momentum fraction
$\xi_i$, one then has:
\beqn\label{eq:W1p}
W_1(Q^2,\nu) &\!\!\! =&\!\!\! \sum_i \int_0^1 d\xi_i\, f_i(\xi_i)\,
W_1^{(i)}(Q^2,\nu) = {1\over 2 M_p} \sum_i e_i^2 f_i(x)
\equiv {1\over M_p}\, F_1(x) \, ,\quad
\\ \label{eq:W2p}
W_2(Q^2,\nu) &\!\!\! =&\!\!\! \sum_i \int_0^1 d\xi_i\, f_i(\xi_i)\,
W_2^{(i)}(Q^2,\nu) = {x\over \nu} \sum_i e_i^2 f_i(x)
\equiv {1\over \nu}\, F_2(x) \, .
\eeqn
This simple parton description implies then the so-called Bjorken
{\it scaling} \cite{BJ:69}: the proton structure functions only depend on the
kinematical variable $x$. Moreover, one gets the Callan--Gross relation
\cite{CG:69}
\bel{eq:CG}
F_2(x) = 2 x F_1(x) \, ,
\ee
which is a consequence of our assumption of spin-$\frac{1}{2}$ partons.
It is easy to check that spin-0 partons would have lead to $F_1(x)=0$.

\begin{figure}[htb]
\vfill\centerline{
\begin{minipage}[t]{.47\linewidth}
\centerline{\mbox{\epsfxsize=7.7cm\epsfysize=7.0cm\epsffile{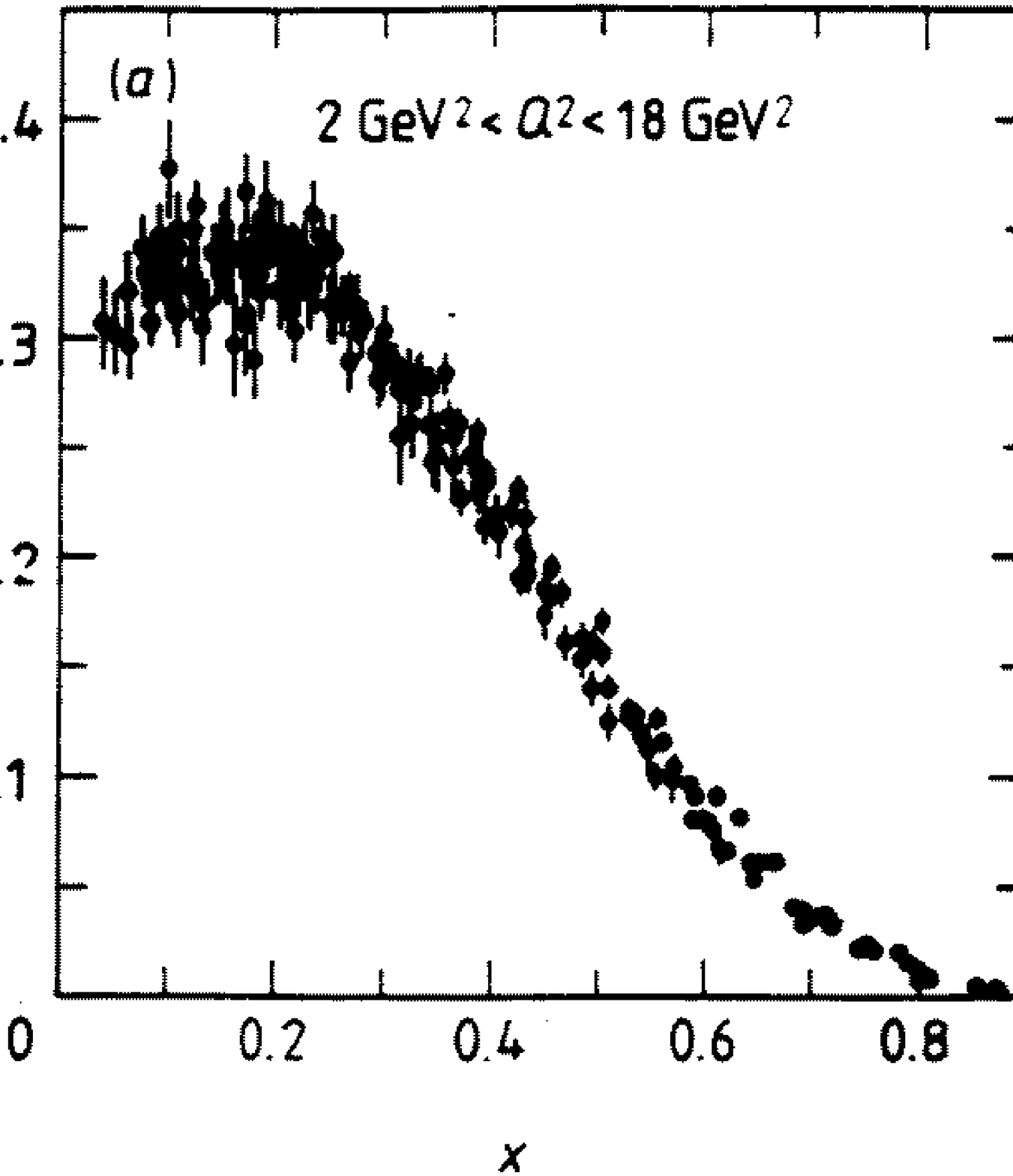}}}
\vspace{-0.3cm}
\caption{Experimental data on $\nu W_2$ as function of $x$,
  for different values of $Q^2$
  \protect\cite{AT:80} (taken from Ref.~\protect\cite{AH:89}).}
\label{fig:W2x}
\end{minipage}
\hspace{0.7cm}
\begin{minipage}[t]{.47\linewidth}
 \centerline{\mbox{\epsfysize=7.5cm\epsffile{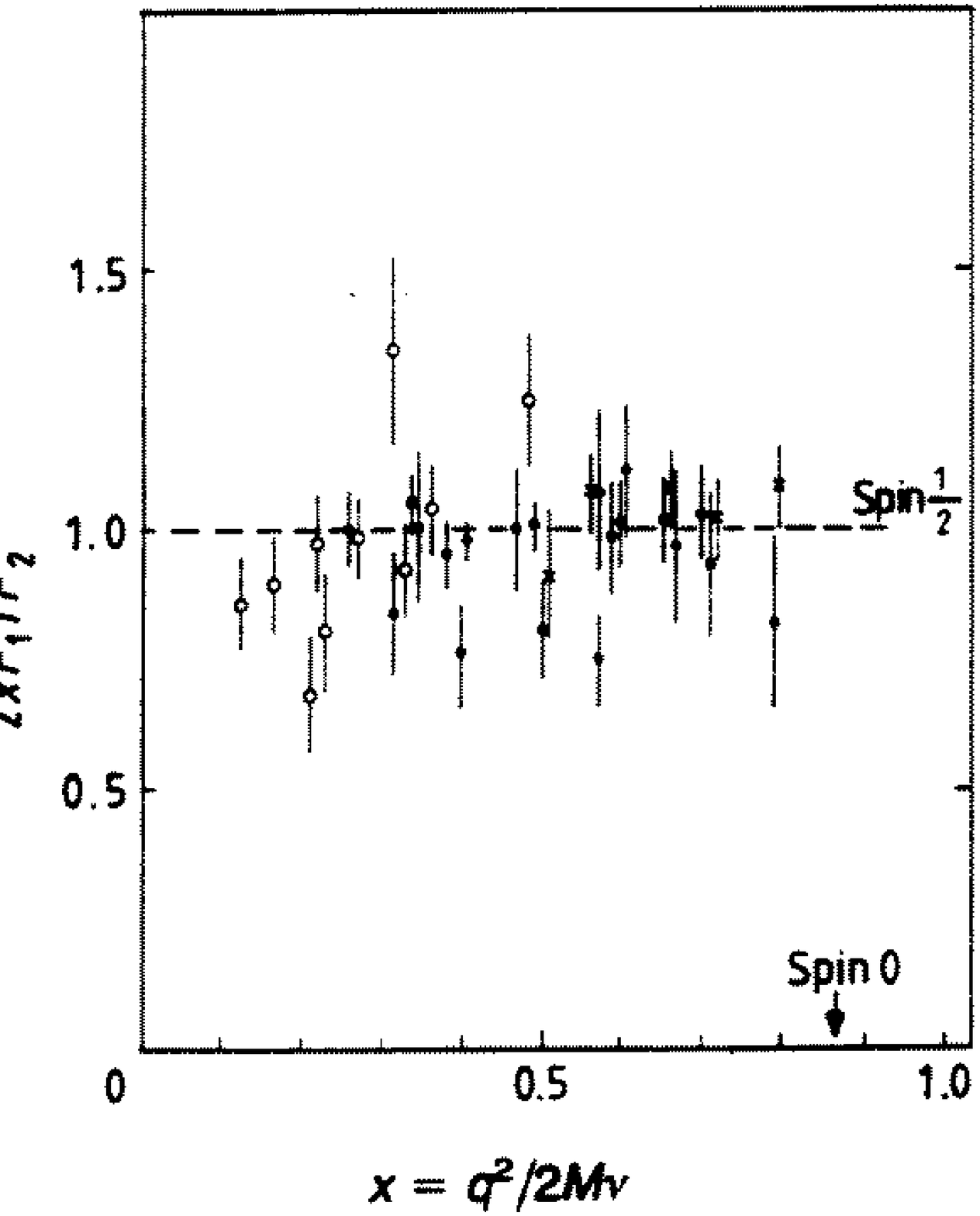}}}
\vspace{-0.3cm}
\caption{The ratio $2xF_1/F_2$ versus $x$,
for different $Q^2$ values (1.5 GeV$^2 < Q^2 < 16$ GeV$^2$)
  \protect\cite{PE:87} (taken from Ref.~\protect\cite{AH:89}).}
\label{fig:CG}
\end{minipage}
}\vfill
\end{figure}

The measured values of $\nu W_2(Q^2,\nu)$ are shown in Fig.~\ref{fig:W2x}
as function of $x$, for many different values of $Q^2$ between 2 and 18
GeV$^2$;
the concentration of data points along a curve indicates that Bjorken scaling
is correct, to a quite good approximation.
Fig.~\ref{fig:CG} shows that the Callan--Gross relation is also reasonably
well satisfied by the data, supporting the spin-$\frac{1}{2}$ assignment
for the partons.

The surprising thing of this successful predictions is that we have assumed
the existence of free independent pointlike partons inside the proton, in
spite of the fact that quarks are supossed to be confined by very strong
colour forces.
Bjorken scaling suggests that the strong interactions must have the
property of {\it asymptotic freedom}:
they should become weaker at short distances,
so that quarks behave as free particles for $Q^2\to\infty$.
This also agrees with the empirical observation in Fig.~\ref{fig:Ree},
that the free-quark description of the ratio $R_{e^+e^-}$ works
better at higher energies.

Thus, the interaction between a $q\bar q$ pair looks like some kind of
rubber band. If we try to separate the quark form the antiquark
the force joining them increases.
At some point, the energy on the elastic band is bigger than $2m_{q'}$, so
that it becomes energetically favourable to create an additional $q'\bar q'$
pair; then the band breaks down into two mesonic systems, $q\bar q'$ and
$q'\bar q$, each one with its corresponding half-band joining the quark pair.
Increasing more and more the energy, we can only produce more and more
mesons, but quarks remain always confined within colour-singlet bound states.
Conversely, if one tries to approximate two quark constituents into
a very short-distance region, the elastic band loses the energy and becomes
very soft; quarks behave then as free particles.

\subsection{Why \protect\boldmath $SU(3)$?}

Flavour-changing transitions have a much weaker strength than processes
mediated by the strong force. The quark-flavour quantum number is associated
with the electroweak interactions, while strong forces appear to be
flavour-conserving and flavour-independent.
On the other side, the carriers of the electroweak interaction
($\gamma$, $Z$, $W^\pm$) do not couple to the quark colour. Thus, it seems
natural to take colour as the charge associated with the strong forces
and try to build a quantum field theory based on it \cite{FGL:73}.
The empirical evidence described so far puts a series of requirements that
the fundamental theory of colour interactions should satisfy:
\begin{enumerate}
\item Colour is an exact symmetry $G_C$
  (hadrons do not show colour multiplicity).
\item $N_C=3$. Thus, quarks belong to the triplet representation
    $\underline{3}$ of $G_C$.
\item Quarks and antiquarks are different states. Therefore,
  $\underline{3}^*\not= \underline{3}$, i.e. the triplet representation has
  to be complex.
\item Confinement hypothesis: hadronic states are colour singlets.
\item Asymptotic freedom.
\end{enumerate}

Among all compact simple Lie groups there are only four having 3-dimensional
irreducible representations; moreover, three of them are isomorphic to
each other. Thus, we have only two choices: $SU(3)$ or
$SO(3)\simeq SU(2)\simeq Sp(1)$. Since the triplet representation of $SO(3)$
is real, only the symmetry group $SU(3)$ survives the conditions 1, 2 and 3.
The well-known $SU(3)$ decomposition of the products of $\underline{3}$
and $\underline{3}^*$ representations,
\beqn\label{eq:rep_products}
q\bar q: & & \underline{3}\otimes \underline{3}^* =
  \underline{1} \oplus \underline{8} \, ,
\no\\
qqq: && \underline{3}\otimes\underline{3}\otimes\underline{3} =
  \underline{1} \oplus \underline{8} \oplus \underline{8}
  \oplus \underline{10} \, ,
\no\\
qq: &&  \underline{3}\otimes \underline{3} =
  \underline{3}^* \oplus \underline{6} \, ,
\no\\
qqqq: &&
  \underline{3}\otimes\underline{3}\otimes\underline{3}\otimes \underline{3}=
  \underline{3}\oplus\underline{3}\oplus\underline{3}\oplus\underline{6}^*
  \oplus\underline{15}\oplus\underline{15}\oplus\underline{15}
  \oplus\underline{15'} \, ,
\eeqn
guarantees that there are colour-singlet configurations corresponding to
meson ($q\bar q$) and baryon ($qqq$) states, as required
by the confinement hypothesis. Other exotic combinations such as diquarks
($qq$, $\bar q\bar q$) or four-quark states
($qqqq$, $\bar q\bar q\bar q\bar q$) do not satisfy this requirement.

Clearly, the theory of colour interactions should be based on
the $SU(3)_C$ group.
It remains to be seen whether such a theory is able
to explain confinement and asymptotic freedom as natural dynamical
consequences of the colour forces.

\setcounter{equation}{0}
\section{GAUGE SYMMETRY: QED}
\label{sec:qed}

Let us consider the Lagrangian describing a free Dirac fermion:
\bel{eq:l_free}
\cL_0\, =\, i \,\overline{\Psi}(x)\gamma^\mu\partial_\mu\Psi(x)
\, - \, m\, \overline{\Psi}(x)\Psi(x) \, .
\ee
$\cL_0$ is invariant under {\em global}
$U(1)$ transformations
\bel{eq:global}
\Psi(x) \,\toU\, \Psi'(x)\,\equiv\,\exp{\{i Q \theta\}}\,\Psi(x) \, ,
\ee
where $Q\theta$ is an arbitrary real constant.
The phase of $\Psi(x)$ is then a pure convention-dependent
quantity without physical meaning.
However,
the free Lagrangian is no-longer invariant if one allows
the phase transformation to depend on the space-time coordinate,
i.e. under {\em local} phase redefinitions $\theta=\theta(x)$,
because
\bel{eq:local}
\partial_\mu\Psi(x) \,\toU\, \exp{\{i Q \theta\}}\,
\left(\partial_\mu + i Q \partial_\mu\theta\right)\,
\Psi(x) \, .
\ee
Thus, once an observer situated at the point $x_0$
has adopted a given phase-convention, the same convention must
be taken at all space-time points. This looks very unnatural.

The ``Gauge Principle'' is the requirement that the $U(1)$
phase invariance should hold {\em locally}.
This is only possible if one adds some additional piece to the
Lagrangian, transforming in such a way  as to cancel the
$\partial_\mu\theta$ term in Eq.~\eqn{eq:local}.
The needed modification is completely fixed by the transformation
\eqn{eq:local}: one introduces a new spin--1
(since $\partial_\mu\theta$  has a Lorentz index)
field $A_\mu(x)$, transforming as
\bel{eq:a_transf}
A_\mu(x)\,\toU\, A_\mu'(x)\,\equiv\, A_\mu(x) + {1\over e}\,
\partial_\mu\theta\, ,
\ee
and defines the covariant derivative
\bel{eq:d_covariant}
D_\mu\Psi(x)\,\equiv\,\left[\partial_\mu-ieQA_\mu(x)\right]
\,\Psi(x)\, ,
\ee
which has the required 
property of transforming like the field itself:
\bel{eq:d_transf}
D_\mu\Psi(x)\,\toU\,\left(D_\mu\Psi\right)'(x)\,\equiv\,
\exp{\{i Q \theta\}}\,D_\mu\Psi(x)\,.
\ee
The Lagrangian
\bel{eq:l_new}
\cL\,\equiv\,
i \,\overline{\Psi}(x)\gamma^\mu D_\mu\Psi(x)
\, - \, m\, \overline{\Psi}(x)\Psi(x)
\, =\, \cL_0\, +\, e Q A_\mu(x)\, \overline{\Psi}(x)\gamma^\mu\Psi(x)
\ee
is then invariant under local $U(1)$ transformations.

The gauge principle has generated an interaction
between the Dirac spinor and the gauge field $A_\mu$,
which is nothing else than the familiar QED vertex.
Note that the corresponding electromagnetic
charge $eQ$ is completely arbitrary.
If one wants $A_\mu$ to be a true propagating field, one needs to add
a gauge-invariant kinetic term
\bel{eq:l_kinetic}
\cL_{\!\!\mbox{\rms Kin}}\,\equiv\, -{1\over 4} F_{\mu\nu} F^{\mu\nu}\,,
\ee
where
$F_{\mu\nu}\,\equiv\, \partial_\mu A_\nu -\partial_\nu A_\mu$
is the usual electromagnetic field strength.
A possible mass term for the gauge field,
${1\over 2}m^2A^\mu A_\mu$, is forbidden because it would violate
gauge invariance; therefore,
the photon field is predicted to be massless.
The total Lagrangian in \eqn{eq:l_new} and \eqn{eq:l_kinetic}
gives rise to the well-known Maxwell equations.

{}From a simple gauge-symmetry requirement, we have deduced
the right QED Lagrangian, which leads to a very  successful
quantum field theory.
Remember that QED predictions have been tested
to a very high accuracy, as exemplified by the electron and
muon anomalous magnetic moments
[$a_l\equiv (g_l-2)/2$, where $\mu_l\equiv g_l \,(e \hbar/2m_l)$]
\cite{KI:90}:
\beqn\label{eq:a_e}
a_e&=&\left\{ \bat
(115 \, 965 \, 214.0\pm 2.8) \times 10^{-11} & (\mbox{\rm Theory})
\\
(115 \, 965 \, 219.3\pm 1.0) \times 10^{-11} & (\mbox{\rm Experiment})
\ea \, , \right.\\
a_\mu&=&\left\{ \bat
(1 \, 165 \, 919.2\pm 1.9)
\times 10^{-9} & (\mbox{\rm Theory})
\\
(1 \, 165 \, 923.0\pm 8.4) \times 10^{-9} & (\mbox{\rm Experiment})
\ea \, . \right.
\eeqn

\setcounter{equation}{0}
\section{THE QCD LAGRANGIAN}
\label{sec:QCDlagrangian}

Let us denote $q^\alpha_f$ a quark field of colour $\alpha$ and flavour $f$.
To simplify the equations, let us adopt a vector notation
in colour space:
   $q_f \,\equiv\, \mbox{\rm column}(q^1_f,q^2_f,q^3_f) $ .
%
The free Lagrangian
\bel{eq:L_free}
\cL_0 = \sum_f\,\bar q_f \, \left( i\gamma^\mu\partial_\mu - m_f\right) q_f
\ee
is invariant under arbitrary global $SU(3)_C$ transformations in colour
space,
\bel{eq:q_transf}
q^\alpha_f \,\longrightarrow\,
(q^\alpha_f)' = U^\alpha_{\phantom{\alpha}\beta}\, q^\beta_f \, ,
\qquad\qquad
U U^\dagger = U^\dagger U = 1 \, , \qquad\qquad
\det U = 1 \, .
\ee
The $SU(3)_C$ matrices can be written in the form
\bel{eq:U_def}
U = \exp\left\{-ig_s {\lambda^a\over 2}\theta_a\right\} \, ,
\ee
where $\lambda^a$ ($a=1,2,\ldots,8$) denote the
generators of the fundamental representation of the
$SU(3)_C$ algebra,
and $\theta_a$ are arbitrary parameters. The matrices $\lambda^a$
are traceless and
satisfy the commutation relations
\bel{eq:commutation}
\left[\lambda^a,\lambda^b\right] =
2 i f^{abc} \, \lambda^c \, ,
\ee
with $f^{abc}$ the $SU(3)_C$
structure constants, which are real and totally antisymmetric.
Some useful properties of $SU(3)$ matrices are collected in Appendix A.

As in the QED case, we can now require the Lagrangian to be also invariant
under {\it local} $SU(3)_C$ transformations, $\theta_a = \theta_a(x)$. To
satisfy this requirement, we need to change the quark derivatives by covariant
objects. Since we have now 8 independent gauge parameters, 8 different
gauge bosons $G^\mu_a(x)$, the so-called {\it gluons}, are needed:
\bel{eq:D_cov}
D^\mu q_f \,\equiv\, \left[ \partial^\mu - i g_s {\lambda^a\over 2}
G^\mu_a(x)\right]
  \, q_f \, \equiv\, \left[ \partial^\mu - i g_s G^\mu(x)\right] \, q_f \, .
\ee
Notice that we have introduced the compact matrix notation
\bel{eq:G_matrix}
  [G^\mu(x)]_{\alpha\beta}\,\equiv\,
\left({\lambda^a\over 2}\right)_{\!\alpha\beta}\, G^\mu_a(x) \, .
\ee
We want $D^\mu q_f$ to transform in exactly the same way as the colour-vector
$q_f$;
this fixes the transformation properties of the gauge fields:
\bel{eq:G_trans}
D^\mu \,\longrightarrow\, (D^\mu)'= U\, D^\mu\, U^\dagger \, ; \qquad
G^\mu \,\longrightarrow\, (G^\mu)'= U\, G^\mu\, U^\dagger
  -{i\over g_s} \, (\partial^\mu U) \, U^\dagger \, .
\ee
Under an infinitesimal $SU(3)_C$ transformation,
\beqn\label{eq:inf_transf}
q^\alpha_f &\longrightarrow & (q^\alpha_f)' =
q^\alpha_f -i g_s \left({\lambda^a\over 2}\right)_{\!\alpha\beta}
\delta\theta_a\, q^\beta_f \, ,
\no\\
G^\mu_a &\longrightarrow & (G^\mu_a)' = G^\mu_a -\partial^\mu(\delta\theta_a)
 + g_s f^{abc} \delta\theta_b \, G^\mu_c \, .
\eeqn
The gauge transformation of the gluon fields is more complicated that the one
obtained in QED for the photon.
The non-commutativity of the $SU(3)_C$ matrices gives rise to an additional
term
involving the gluon fields themselves.
For constant $\delta\theta_a$, the transformation rule for the gauge fields
is expressed in terms of the structure constants $f^{abc}$ only; thus,
the gluon fields belong to the adjoint representation of the colour group
(see Appendix~A).
Note also that there is a unique $SU(3)_C$
coupling $g_s$. In QED it was possible to assign arbitrary electromagnetic
charges to the
different fermions. Since the commutation relation \eqn{eq:commutation} is
non-linear, this freedom does not exist for $SU(3)_C$.

To build a gauge-invariant kinetic term for the gluon fields, we introduce
the corresponding field strengths:
\beqn\label{eq:G_tensor}
G^{\mu\nu}(x) & \equiv & {i\over g_s}\, [D^\mu, D^\nu] \, = \,
  \partial^\mu G^\nu - \partial^\nu G^\mu - i g_s\, [G^\mu, G^\nu] \, \equiv \,
  {\lambda^a\over 2}\, G^{\mu\nu}_a(x) \, ,
\no\\
G^{\mu\nu}_a(x) & = & \partial^\mu G^\nu_a - \partial^\nu G^\mu_a
  + g_s f^{abc} G^\mu_b G^\nu_c \, .
\eeqn
Under a gauge transformation,
\bel{eq:G_tensor_transf}
G^{\mu\nu}\,\longrightarrow\, (G^{\mu\nu})' = U\, G^{\mu\nu}\, U^\dagger \, ,
\ee
and the colour trace
Tr$(G^{\mu\nu}G_{\mu\nu}) = \frac{1}{2} G^{\mu\nu}_aG_{\mu\nu}^a$
remains invariant.

Taking the proper normalization for the gluon kinetic term, we finally have
the $SU(3)_C$ invariant QCD Lagrangian:
\bel{eq:L_QCD}
\cL_{\!\!\mbox{\rms QCD}} \,\equiv\, -{1\over 4}\, G^{\mu\nu}_aG_{\mu\nu}^a
+ \sum_f\,\bar q_f \, \left( i\gamma^\mu D_\mu - m_f\right)\, q_f \, .
\ee
It is worth while to decompose the Lagrangian into its different pieces:
\beqn\label{eq:L_QCD_pieces}
\cL_{\!\!\mbox{\rms QCD}} & = &
 -{1\over 4}\, (\partial^\mu G^\nu_a - \partial^\nu G^\mu_a)
  (\partial_\mu G_\nu^a - \partial_\nu G_\mu^a)
 + \sum_f\,\bar q^\alpha_f \, \left( i\gamma^\mu\partial_\mu - m_f\right)\,
q^\alpha_f
\qquad\no\\ && \mbox{}
 + g_s\, G^\mu_a\,\sum_f\,  \bar q^\alpha_f \gamma_\mu
 \left({\lambda^a\over 2}\right)_{\!\alpha\beta} q^\beta_f
\\ && \mbox{}
 - {g_s\over 2}\, f^{abc}\, (\partial^\mu G^\nu_a - \partial^\nu G^\mu_a) \,
  G_\mu^b G_\nu^c
 \, - \, {g_s^2\over 4} \, f^{abc} f_{ade} \, G^\mu_b G^\nu_c G_\mu^d G_\nu^e
\,
{}.
\no
\eeqn
The first line contains the correct kinetic terms for the different fields,
which
give rise to the corresponding propagators. The colour interaction between
quarks and gluons is given by the second line; it involves the $SU(3)_C$
matrices
$\lambda^a$. Finally, owing to the non-abelian character of the colour group,
the $G^{\mu\nu}_aG_{\mu\nu}^a$ term generates the cubic and quartic gluon
self-interactions shown in the last line;
the strength of these interactions is given by the same coupling $g_s$ which
appears in the fermionic piece of the Lagrangian.

In spite of the rich physics contained in it, the Lagrangian \eqn{eq:L_QCD}
looks very simple, because of its colour-symmetry properties.
All interactions are given in terms of a single universal coupling $g_s$,
which is called the {\it strong coupling constant}.
The existence of self-interactions among the gauge fields is a new feature
that was not present in the QED case; it seems then reasonable to expect that
these gauge self-interactions could explain properties
like asymptotic freedom and confinement, which do not appear in QED.

\begin{figure}[ph]
\centerline{\epsfxsize = 7cm \epsfbox{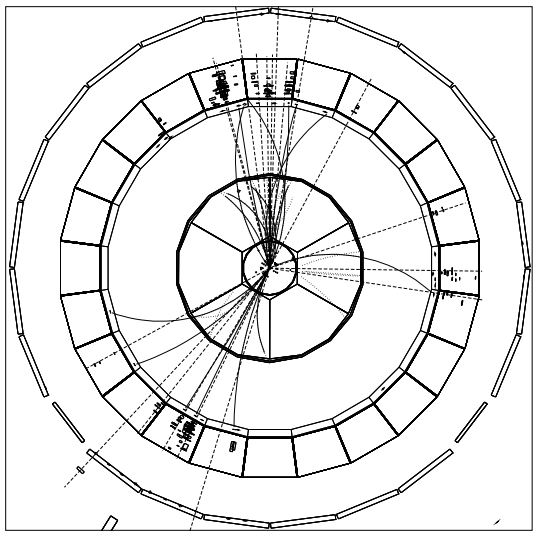}}
\caption{Three-jet event from the hadronic decay of a $Z$ boson (DELPHI).}
\label{fig:ThreeJets}
\end{figure}

Without any detailed calculation, one can already extract qualitative physical
consequences from $\cL_{\!\!\mbox{\rms QCD}}$.
Quarks can emit gluons. At lowest-order
in $g_s$, the dominant process will be the emission of a single gauge boson.
Thus, the hadronic decay of the $Z$ should result in some $Z\to q\bar q G$
events,
in addition to the dominant $Z\to q\bar q$ decays discussed in
Section~\ref{sec:introduction}
Fig.~\ref{fig:ThreeJets} clearly shows that 3-jet events, with the required
kinematics, indeed appear in the LEP data. Similar events show up in
$e^+e^-$ annihilation into hadrons, away from the $Z$ peak.

In order to properly quantize the QCD Lagrangian, one needs to add
to $\cL_{\!\!\mbox{\rms QCD}}$ the
so-called {\it Gauge-fixing} and {\it Faddeev--Popov} terms.
Since this is a rather technical issue, its discussion is relegated
to Appendix B.

\setcounter{equation}{0}
\section{QUANTUM LOOPS}
\label{sec:loops}

The QCD Lagrangian is rather economic in the sense that it involves a single
coupling $g_s$. Thus, all strong-interacting phenomena should be described in
terms of just one parameter. At lowest order in $g_s$ (tree-level),
it is straightforward
to compute all kind of scattering amplitudes involving quarks and gluons:
$q\bar q\to GG$, $qq\to qq$, $G q\to G q$, \ldots
Unfortunately, this exercise by itself does not help very much to understand
the
physical hadronic world.
First, we see hadrons instead of quarks and gluons. Second, we have learnt from
experiment that the strength of the strong forces changes with the
energy scale: the interaction is very strong (confining) at low energies, but
quarks behave as nearly free particles at high energies.
Obviously, we cannot understand both energy regimes with a single constant
$g_s$,
which is the same everywhere. Moreover, if we neglect the quark masses,
the QCD Lagrangian does not contain any energy scale;
thus, there is no way to decide when the energy
of a given process is large or small, because we do not have any
reference scale to compare with.

If QCD is the right theory of the strong interactions, it should provide
some dynamical scale through quantum effects.

\subsection{Regularization of loop integrals}
\label{sub:regularization}

The computation of perturbative corrections to the tree-level results
involves divergent loop integrals. It is then necessary to find a way of
getting finite results with physical meaning from a priori meaningless
divergent quantities.

\begin{figure}[htb]
\centerline{\epsfysize=3cm \epsfbox{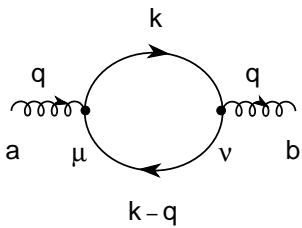}}
\caption{Gluon self-energy diagram.}
\label{fig:gluonSE}
\end{figure}

Let us consider the self-energy gluon loop in Fig.~\ref{fig:gluonSE}.
The corresponding contribution in momentum space can be easily obtained,
using standard Feynman rules techniques:
\bel{eq:gluon_se}
i \Pi^{\mu\nu}_{ab}(q) \, = \, - g_s^2 \delta_{ab} T_F \int
{d^4k\over (2\pi)^4}
{\mbox{\rm Tr}[\gamma^\mu\slashchar{k}\gamma^\nu (\slashchar{k}-\slashchar{q})]
\over k^2 (k-q)^2} \, .
\ee
The result is proportional to $g_s^2$, because there are two $q\bar q G$
vertices, and there is a trivial $SU(3)_C$ factor, $T_F = {1\over 2}$,
coming from the colour trace
${1\over 4}\mbox{\rm Tr}(\lambda^a\lambda^b) = \delta^{ab} T_F$.

The problem appears in the momentum integration, which is clearly divergent
[$\sim \int d^4k (1/k^2) = \infty$].
We can define the momentum integral in many different (and arbitrary) ways.
For instance, we could introduce a {\it cut-off} $M$, such that only
momentum scales smaller than $M$ are integrated; obviously, the resulting
integral would be an increasing function of $M$.
Instead, it has become conventional to define the loop integrals through
{\it dimensional regularization}: the calculation is performed in
$D=4+2\epsilon$ dimensions. For $\epsilon\not=0$ the resulting integral is
well-defined:
\bel{eq:dimensional}
\int {d^Dk\over (2\pi)^D}\, {k^\alpha (k-q)^\beta \over k^2 (k-q)^2} \, = \,
{-i\over 6 (4\pi)^2} \left({-q^2\over 4\pi}\right)^\epsilon
\Gamma(-\epsilon) \, \left(1-{5\over 3}\epsilon\right)
\left\{ {q^2 g^{\alpha\beta}\over 2 (1+\epsilon)} + q^\alpha q^\beta\right\} \,
{}.
\ee
The ultraviolet divergence of the loop appears at $\epsilon=0$,
through the pole of the Gamma function,
\bel{eq:gamma}
\Gamma(-\epsilon) \, = \, -{1\over\epsilon} - \gamma_E + \cO(\epsilon^2) \, ,
\ee
where $\gamma_E = 0.577215\ldots $ is the Euler constant.

Although the integral \eqn{eq:dimensional} looks somewhat funny, dimensional
regularization has many advantages because does not spoil the gauge symmetry
of QCD and, therefore, simplifies a lot the calculations.
One could argue that a cut-off
procedure would
be more {\it physical}, since the parameter $M$ could be related to some
unknown additional physics at very short distances. However, within the
QCD framework, both prescriptions are equally meaningless. One just introduces
a regularizing parameter, such that the integral is well-defined and the
divergence is recovered in some limit ($M\to\infty$ or $\epsilon\to 0$).

Since the momentum-transfer $q^2$ has dimensions, it turns out to be convenient
to introduce and arbitrary energy scale $\mu$ and write
\bel{eq:mu_scale}
\left({-q^2\over 4\pi}\right)^\epsilon\Gamma(-\epsilon) =
\mu^{2\epsilon} \left({-q^2\over 4\pi\mu^2}\right)^\epsilon\Gamma(-\epsilon) =
-\mu^{2\epsilon}\left\{{1\over\epsilon} +\gamma_E -\ln{4\pi}
+\ln{(-q^2/\mu^2)} + \cO(\epsilon)\right\} \, .
\ee
Obviously, this expression does not depend on $\mu$; but written in this form
one has a dimensionless quantity ($-q^2/\mu^2$) inside the logarithm.

The contribution of the loop diagram in Fig.~\ref{fig:gluonSE}
can finally be written as
\beqn\label{eq:g_se}
\Pi^{\mu\nu}_{ab} &\!\! = &\!\!
\delta_{ab}\,\left( -q^2 g^{\mu\nu} + q^\mu q^\nu\right)
\, \Pi(q^2) \, ,
\no\\
\Pi(q^2) &\!\! = &\!\!
 -{4\over 3} T_F \,\left({g_s\mu^\epsilon\over 4\pi}\right)^2
\left\{{1\over\epsilon} +\gamma_E -\ln{4\pi}
+\ln{(-q^2/\mu^2)} - {5\over 3} + \cO(\epsilon)\right\} \, .
\eeqn

Owing to the ultraviolet divergence, Eq.~\eqn{eq:g_se} does not determine the
wanted self-energy contribution. Nevertheless, it does show how this effect
changes with the energy scale.
If one could fix the value of
$\Pi(q^2)$ at some reference momentum transfer $q_0^2$, the result
would be known at any other scale:
\bel{eq:pi_1_2}
\Pi(q^2) = \Pi(q_0^2) - {4\over 3} T_F \, \left({g_s\over 4\pi}\right)^2
\,\ln{(q^2/q_0^2)} \, .
\ee

 We can split the self-energy contribution into a meaningless divergent piece
and a finite term, which includes the $q^2$ dependence,
\bel{eq:Pi_splitting}
\Pi(q^2) \,\equiv\, \Delta\Pi_\epsilon(\mu^2) + \Pi_R(q^2/\mu^2)\, .
\ee
This separation is of course ambiguous, because the finite $q^2$-independent
contributions can be splitted in many different ways. A given choice defines
a {\it scheme}:
\be\label{eq:Pi_epsilon}
\Delta\Pi_\epsilon(\mu^2) \, =\, \left\{
 \begin{array}{ll}
  -{T_F\over 3\pi} {g_s^2\over 4\pi}  \mu^{2\epsilon} \left[ {1\over\epsilon}
        + \gamma_E -\ln(4\pi) -{5\over 3}\right]  \quad\qquad\qquad &
     \;\;   (\mu\mbox{\rm -scheme}) , \\
  -{T_F\over 3\pi} {g_s^2\over 4\pi}  \mu^{2\epsilon} {1\over\epsilon} \qquad &
	       (\mbox{\rm MS}\mbox{\rm -scheme}) , \; \\
  -{T_F\over 3\pi} {g_s^2\over 4\pi}  \mu^{2\epsilon} \left[ {1\over\epsilon}
        + \gamma_E -\ln(4\pi)\right]  \qquad &
       (\overline{\mbox{\rm MS}}\mbox{\rm -scheme}) ,
  \ea\right.
\ee
\be\label{eq:Pi_R}
  \Pi_R(q^2/\mu^2) \, =\, \left\{
   \begin{array}{ll}
     -{T_F\over 3\pi} {g_s^2\over 4\pi} \ln{(-q^2/\mu^2)} \qquad &
    \;\;   (\mu\mbox{\rm -scheme}) , \\
     -{T_F\over 3\pi} {g_s^2\over 4\pi} \left[ \ln{(-q^2/\mu^2)}
        + \gamma_E -\ln(4\pi) -{5\over 3}\right]  \qquad &
        (\mbox{\rm MS}\mbox{\rm -scheme}) , \quad\\
     -{T_F\over 3\pi} {g_s^2\over 4\pi} \left[ \ln{(-q^2/\mu^2)}
        -{5\over 3}\right]  \qquad & (\overline{\mbox{\rm MS}}\mbox{\rm
-scheme}) .
   \ea\right.
\ee
In the $\mu$-scheme, one uses the value of $\Pi(-\mu^2)$ to define the
divergent part.
MS and $\overline{\mbox{\rm MS}}$ stand for minimal subtraction
\cite{THO:73} and
modified minimal subtraction schemes \cite{BBDM:78};
in the MS case, one subtracts only
the divergent $1/\epsilon$ term, while the $\overline{\mbox{\rm MS}}$
scheme puts also the annoying $\gamma_E-\ln(4\pi)$ factor into the
divergent part.
Notice that the logarithmic $q^2$-dependence is always the same.

\begin{figure}[htb]
\centerline{\epsfysize=4cm \epsfbox{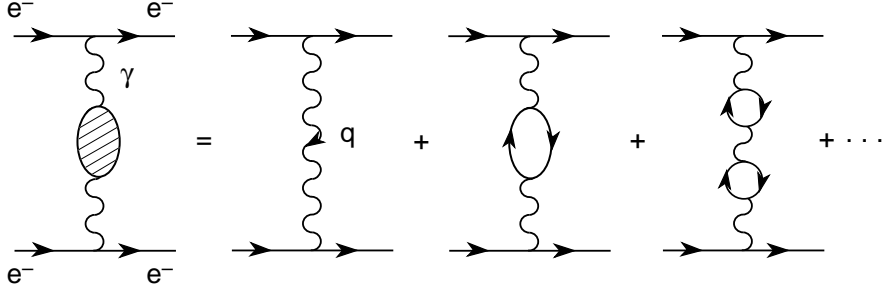}}
\caption{Photon self-energy contribution to $e^-e^-$ scattering.}
\label{fig:EEint}
\end{figure}

\subsection{Renormalization: QED}
\label{subsec:renormalization}

A Quantum Field Theory is called {\it renormalizable} if all ultraviolet
divergences can be reabsorbed through a redefinition of the original fields
and couplings.

Let us consider the electromagnetic interaction between two electrons.
At one loop,
the QED photon self-energy contribution is just given by
Eq.~\ref{eq:g_se}, with the changes $T_F\to 1$ and $g_s\to e$. The
corresponding scattering amplitude takes the form
\bel{eq:T_ee}
T(q^2) \,\sim\, -J^\mu J_\mu \, {e^2\over q^2} \, \left\{ 1 -\Pi(q^2) + \ldots
\right\} \, ,
\ee
where $J^\mu$ denotes the electromagnetic fermion current.

At lowest order,
$T(q^2)\sim \alpha/q^2$ with $\alpha = e^2/(4\pi)$.
The divergent correction generated by quantum loops can be reabsorbed into a
redefinition of the coupling:
\bel{eq:alpha_R}
{\alpha_0\over q^2}\, \left\{ 1 -\Delta\Pi_\epsilon(\mu^2)
- \Pi_R(q^2/\mu^2)\right\} \,\equiv\,
{\alpha_R(\mu^2)\over q^2}\, \left\{ 1
- \Pi_R(q^2/\mu^2)\right\} \, ,
\ee
\bel{eq:alpha_Rb}
\alpha_R(\mu^2) \, = \, \alpha_0 \,
\left\{ 1 + {\alpha_0 \over 3\pi} \mu^{2\epsilon}
\left[{1\over\epsilon} + C_{\!\!\mbox{\rms scheme}}\right] + \ldots
\right\} \, ,
\qquad\qquad
\alpha_0\,\equiv\, {e_0^2\over 4\pi}\ , \quad
\ee
%
where $e_0$ denotes the {\it bare}
coupling appearing in the QED Lagrangian;
this bare quantity is, however, not directly observable. Making the
redefinition
\eqn{eq:alpha_R}, the scattering amplitude is finite and gives rise
to a definite prediction for the cross-section, which can be compared with
experiment; thus, one actually measures the {\it renormalized} coupling
$\alpha_R$.

The redefinition \eqn{eq:alpha_R} is meaningful, provided that it can be done
in a self-consistent way: all ultraviolet divergent contributions to all
possible scattering processes should be eliminated through the same
redefinition
of the coupling (and the fields). The nice thing of gauge theories, such as
QED or QCD, is that the underlying gauge symmetry guarantees the
renormalizability of the quantum field theory.

The renormalized coupling $\alpha_R(\mu^2)$ depends
on the arbitrary scale $\mu$ and on the chosen {\it renormalization scheme}
[the constant $C_{\!\!\mbox{\rms scheme}}$ denotes the different finite terms
in
Eq.~\eqn{eq:Pi_epsilon}]. Quantum loops have introduced a scale dependence in a
quite subtle way. Both $\alpha_R(\mu^2)$ and the renormalized self-energy
correction $\Pi_R(q^2/\mu^2)$ depend on $\mu$, but the physical scattering
amplitude $T(q^2)$ is of course $\mu$-independent: ($Q^2\equiv -q^2$)
\beqn\label{eq:mu_dep}
T(q^2) &\sim & -4\pi\, J^\mu J_\mu\,
{\alpha_R(\mu^2)\over q^2}\, \left\{ 1 +
{\alpha_R(\mu^2)\over 3\pi} \left[ \ln{\left(-q^2\over\mu^2\right)} +
C'_{\!\!\mbox{\rms scheme}}\right] + \ldots  \right\}\no\\
& = & 4\pi\, J^\mu J_\mu\,
{\alpha_R(Q^2)\over Q^2} \, \left\{ 1 +
{\alpha_R(Q^2)\over 3\pi} C'_{\!\!\mbox{\rms scheme}} + \cdots \right\}\, .
\eeqn

The quantity $\alpha(Q^2)\equiv\alpha_R(Q^2)$ is called the QED
{\it running coupling}.
The ordinary fine structure constant $\alpha=1/137$ is defined through the
classical Thomson formula; therefore, it corresponds to a very low scale
$Q^2= -m_e^2$. Clearly, the value of $\alpha$ relevant for LEP experiments
is not the same
[$\alpha(M_Z^2)_{\overline{\!\!\mbox{\rms MS}}} = 1/129$].
The scale dependence of $\alpha(Q^2)$ is regulated by
the so-called $\beta$-function:
\bel{eq:beta}
\mu {d\alpha\over d\mu} \,\equiv\, \alpha \,\beta(\alpha) \, ;
\qquad\qquad \beta(\alpha)\, =\, \beta_1 {\alpha\over \pi} +
\beta_2 \left({\alpha\over\pi}\right)^2 + \cdots
\ee
At the one-loop level, the $\beta$-function reduces to the first coefficient,
which is fixed by Eq.~\eqn{eq:alpha_Rb}:
\bel{eq:beta_1}
\beta_1^{\!\!\mbox{\rms QED}} \, = \, {2\over 3}\, .
\ee
The first-order differential equation \eqn{eq:beta} can then be easily solved,
with the result:
\bel{eq:alpha_run}
\alpha(Q^2) \, = \, {\alpha(Q_0^2)\over 1 - {\beta_1 \alpha(Q_0^2)\over 2\pi}
\ln{(Q^2/Q_0^2)}} \, .
\ee
Since $\beta_1>0$, the QED running coupling increases with the energy scale:
$\alpha(Q^2)>\alpha(Q_0^2)$ if $Q^2>Q_0^2$;
i.e. the electromagnetic charge decreases at large distances. This can be
intuitively understood as the screening effect due to the virtual $e^+e^-$
pairs
generated, through quantum effects, around the electron charge.
The physical QED vacuum behaves as a polarized dielectric medium.

\begin{figure}[htb]
\centerline{\epsfysize=5cm \epsfbox{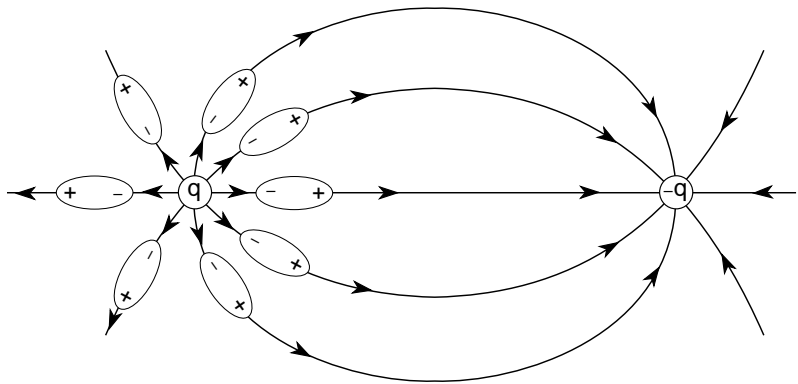}}
\caption{Electromagnetic charge screening in a dipolar medium.}
\label{fig:screening}
\end{figure}

Notice that taking $\mu^2=Q^2$ in Eq.~\eqn{eq:mu_dep} we have eliminated
all dependences on $\ln{(Q^2/\mu^2)}$ to all orders in $\alpha$.
The running coupling \eqn{eq:alpha_run} makes a resummation of all
leading logarithmic corrections, i.e
\bel{eq:alpha_logs}
\alpha(Q^2) \, = \, \alpha(\mu^2)\,\sum_{n=0}^\infty
\left[ {\beta_1 \alpha(\mu^2)\over 2\pi}
\ln{(Q^2/\mu^2)}\right]^n \, .
\ee
This higher-order logarithms correspond to the contributions from an
arbitrary number of one-loop self-energy insertions along the intermediate
photon propagator in Fig.~\ref{fig:EEint}
$[1 - \Pi_R(q^2/\mu^2) + \left(\Pi_R(q^2/\mu^2)\right)^2 + \cdots]$.

\subsection{The QCD running coupling}
\label{subsec:AlphaRunning}

\begin{figure}[htb]
\centerline{\epsfxsize=14cm \epsfbox{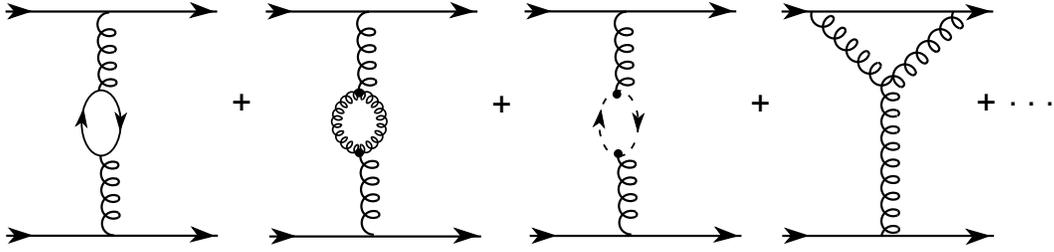}}
\caption{Feynman diagrams contributing to the renormalization of the
strong coupling.
The dashed loop indicates the {\it ghost} correction discussed in Appendix B.}
\label{fig:g_si}
\end{figure}

The renormalization of the QCD coupling proceeds in a similar
way. Owing to the non-abelian character of $SU(3)_C$, there are additional
contributions involving gluon self-interactions. From the
calculation of the relevant one-loop diagrams, shown in Fig.~\ref{fig:g_si},
one gets the value of the first $\beta$-function
coefficient \cite{GW:73,PO:73}:
\bel{eq:QCD_beta}
\beta_1 \, = \, {2\over 3} T_F N_f  -  {11\over 6} C_A
\, = \, {2 N_f - 11 N_C \over 6} \, .
\ee
The positive contribution proportional to $N_f$ is generated by the
$q$-$\bar q$ loops and corresponds to the QED result (except for the $T_F$
factor).
The gluonic self-interactions introduce the additional {\it negative}
contribution proportional to $N_C$. This second term is responsible for the
completely different behaviour of QCD: $\beta_1 < 0$ if $N_f \leq 16$.
The corresponding QCD running coupling,
\bel{eq:QCD_run}
\alpha_s(Q^2) \, = \, {\alpha_s(Q_0^2)\over 1 - {\beta_1 \alpha_s(Q_0^2)\over
2\pi}
\ln{\left({Q^2/ Q_0^2}\right)}} \, ,
\ee
decreases at short distances, i.e.
\bel{eq:af_limit}
\lim_{Q^2\to\infty} \, \alpha_s(Q^2) \, = \, 0 \, .
\ee
Thus, for $N_f\leq 16$, QCD has indeed the required property of asymptotic
freedom.
The gauge self-interactions of the gluons {\it spread out} the QCD charge,
generating an {\it antiscreening} effect. This could not happen in QED, because
photons do not carry electric charge. Only non-abelian gauge theories,
where the intermediate gauge bosons are self-interacting particles, have this
antiscreening property \cite{CG:73}.

Although quantum effects have introduced a dependence with the energy, we
still need a reference scale to decide when a given $Q^2$ can be
considered large or small. An obvious possibility is to choose the scale
at which $\alpha_s$ enters into a strong-coupling regime (i.e.
$\alpha_s\sim 1$), where perturbation theory is no longer valid.
A more precise definition can be obtained from the solution of the
$\beta$-function differential equation \eqn{eq:beta}. At one loop, one gets
\bel{eq:Lambda_def}
\ln{\mu} + {\pi\over\beta_1\alpha_s(\mu^2)} \, = \,
\ln{\Lambda} \, ,
\ee
where $\ln{\Lambda}$ is just an integration constant. Thus,
\bel{eq:alpha_Lambda}
\alpha_s(\mu^2) \, = \,
{2\pi\over -\beta_1 \ln{\left({\mu^2/\Lambda^2}\right)}} \, .
\ee
In this way, we have traded the dimensionless parameter $g_s$ by the
dimensionful scale $\Lambda$.
The number of QCD free parameters is the same (1 for massless quarks), but
quantum effects have generated an energy scale.
Although, Eq.~\eqn{eq:QCD_run} gives the impression that the
scale-dependence of
$\alpha_s(\mu^2)$ involves two parameters, $\mu_0^2$ and
$\alpha_s(\mu_0^2)$, only the combination \eqn{eq:Lambda_def}
is actually relevant, as explicitly shown in
\eqn{eq:alpha_Lambda}.

When $\mu>>\Lambda$, $\alpha_s(\mu^2)\to 0$, so that we recover asymptotic
freedom. At lower energies the running coupling gets bigger; for
$\mu\to\Lambda$, $\alpha_s(\mu^2)\to \infty$ and perturbation theory breaks
down. The scale $\Lambda$ indicates when  the strong coupling blows up.
Eq.~\eqn{eq:alpha_Lambda} suggests that confinement at low energies
is quite plausible
in QCD; however, it does not provide a proof because
perturbation theory is no longer valid when $\mu\to\Lambda$.

\subsection{Higher orders}

Higher orders in perturbation theory are much more important in QCD than in
QED, because the coupling is much bigger (at ordinary energies). Unfortunately,
the calculations are also technically more involved. Nevertheless, many
quantities have been already computed at $\cO(\alpha_s^2)$ or even
$\cO(\alpha_s^3)$. The $\beta$-function is known to three loops; in the
$\overline{\mbox{\rm MS}}$ scheme, the computed higher-order coefficients
take the values \cite{CA:74}:
\bel{eq:beta_hl}
\beta_2 = -{51\over 4} + {19\over 12} N_f \, ; \qquad\qquad
\beta_3 = {1\over 64}\left[ -2857 + {5033\over 9} N_f
- {325\over 27} N_f^2 \right] \, .
\ee
If $N_f\leq 8$, $\beta_2 <0$ \ ($\beta_3 < 0$ for $N_f\leq 5$)
which further reinforces the asymptotic freedom behaviour.

The scale dependence of the running coupling at higher-orders is given by:
\bel{eq:rsc_a} \!\!
\alpha_s(\mu^2) \, =\, \alpha_s(\mu_0^2) \,\left\{ 1 -
{\beta_1\over 2}{\alpha_s(\mu_0^2)\over\pi}
\ln{\left({\mu^2/\mu_0^2}\right)}
-{\beta_2\over 2}\left({\alpha_s(\mu_0^2)\over\pi}\right)^2
\ln{\left({\mu^2/\mu_0^2}\right)} + \cdots \right\}^{-1} \!\! ,
\ee
or, in terms of $\Lambda$,
\bel{eq:rsc_b}
\alpha_s(\mu^2) \, =\,
{2\pi\over (-\beta_1) \ln{\left({\mu^2/\Lambda^2}\right)}}
\left\{ 1 - {\beta_2 \over \beta_1}
{2 \over (-\beta_1) \ln{\left({\mu^2/\Lambda^2}\right)}}
\ln{\left[{1\over 2}\ln{\left({\mu^2/\Lambda^2}\right)}\right]}
+ \cdots \right\} \, .
\ee

When comparing different QCD fits to the data, it is worth while to have in
mind
that any given value of $\alpha_s$ refers to a particular selection of
scale and renormalization scheme. Moreover, the resulting numerical values can
be different if one works at leading (LO), next-to-leading (NLO)
or next-to-next-to-leading (NNLO) order.
Although the parameter $\Lambda$ does not depend on the scale, it is
a scheme-dependent quantity. For instance:
\bel{eq:L_scheme}
\Lambda^2_{\!\!\mbox{\rms MS}} \, = \, {\mbox{\rm e}^{\gamma_E}\over 4\pi}\,
\Lambda^2_{\overline{\!\!\mbox{\rms MS}}} \, .
\ee
Moreover, $\Lambda_{\!\!\mbox{\rms LO}} \not= \Lambda_{\!\!\mbox{\rms NLO}}
\not= \Lambda_{\!\!\mbox{\rms NNLO}}$.
In fact, slightly different definitions of
$\Lambda$ can be given at NLO, depending on the way the integration constant
is chosen when solving the $\beta$-function differential equation.
Moreover, since in the MS and $\overline{\mbox{\rm MS}}$ schemes the
$\beta$-function coefficients depend on $N_f$, $\Lambda$ takes different
values when the number of flavours is changed. At NLO,
the relation between the $\Lambda$ scales for 3 and 4 flavours
is given by:
\bel{eq:Lf}
\Lambda_4\,\approx\,\Lambda_3 \left({\Lambda_3\over m_c}\right)^{2/25}
\,\left[\ln{\left({m_c^2/\Lambda_3^2}\right)}\right]^{-107/1875} \, .
\ee

\setcounter{equation}{0}
\section{PERTURBATIVE QCD PHENOMENOLOGY}
\label{sec:QCDpert}

\subsection{\protect\boldmath $e^+e$\protect\unboldmath\mbox{}$^-$
\protect\boldmath $\to\; $
\protect\unboldmath hadrons}

\label{subsec:Rhadrons}

\begin{figure}[htb]
\centerline{\epsfxsize=14cm \epsfbox{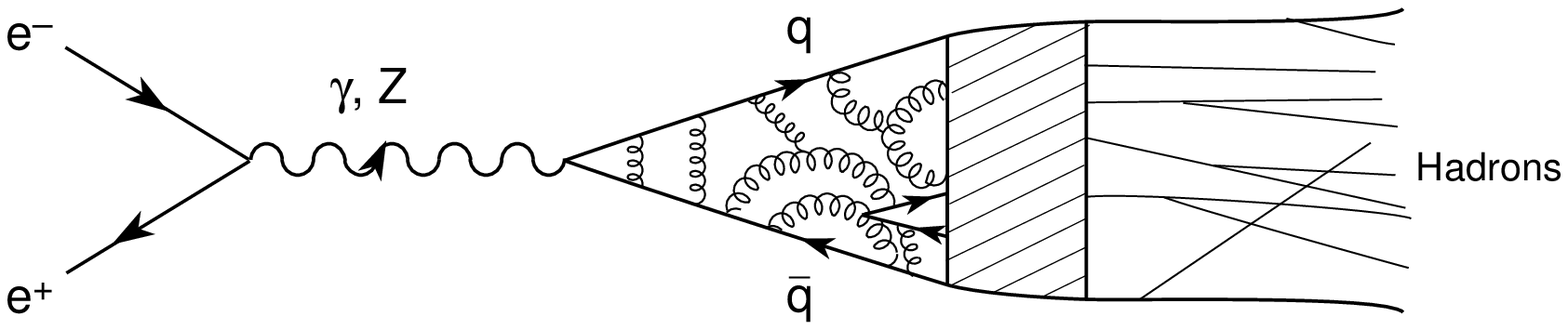}}
\caption{$e^+e^-\to\gamma^*,Z^*\to\, $ hadrons.}
\label{fig:eeqqhad}
\end{figure}

 The inclusive production of hadrons in $e^+e^-$ annihilation is a good
process for testing perturbative QCD predictions.
The hadronic production occurs through the basic mechanism
$e^+e^-\to\gamma^*,Z^*\to q\bar q$,   
where the final $q$-$\bar q$ pair interacts through the QCD forces;
thus, the quarks exchange and emit gluons (and $q'$-$\bar q'$ pairs) in all
possible ways.

At high energies, where $\alpha_s$ is small, we can use perturbative
techniques to predict the different subprocesses:
$e^+e^-\to q\bar q,q\bar q G, q\bar q GG, \ldots$ However, we still do not
have a good understanding of the way quarks and gluons hadronize.
Qualitatively, quarks and gluons are created by the
$q$-$\bar q$ current at very short distances, $x\sim 1/\sqrt{s}$. Afterwards,
they continue radiating additional soft gluons with smaller energies.
At larger
distances, $x\sim 1/\Lambda$, the interaction becomes very strong and the
hadronization process occurs. Since we are lacking a rigorous description  of
the confinement mechanism, we are unable to provide precise predictions of the
different exclusive processes, such as $e^+e^-\to 16\pi$. However, we can make
a quite accurate prediction for the total inclusive production of hadrons:
\bel{eq:sigma_total}
\sigma(e^+e^-\to\mbox{\rm hadrons}) = \sigma(e^+e^-\to
q\bar q + q\bar q G + q\bar q GG + \ldots) \, .
\ee
The details of the final hadronization are irrelevant for the inclusive sum,
because the probability to hadronize is just one owing to our confinement
assumption.

\begin{figure}[bh] 
\centerline{\hbox{\Large $\sigma\quad\;\sim\quad\; $}
\hbox{\epsfxsize=12.5cm \epsfbox[95 400 524 491]{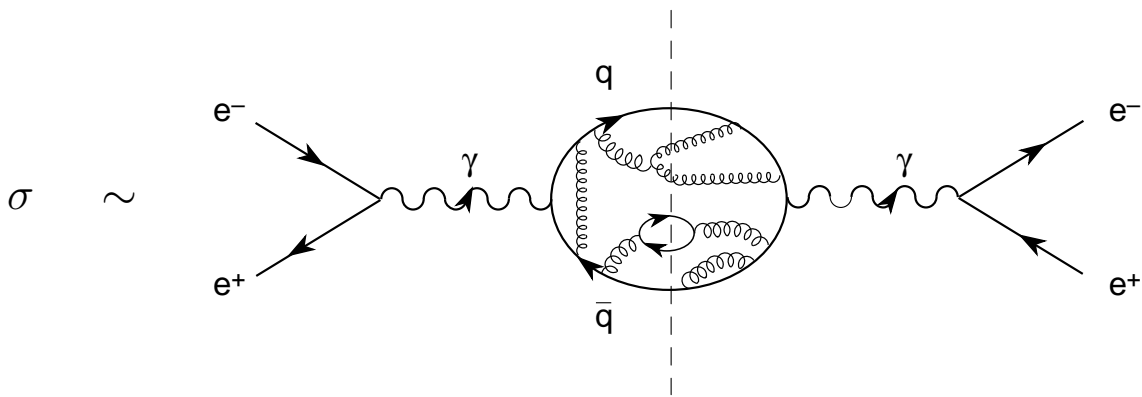}}}
\vspace{2.5cm}
\caption{Diagrammatic relation between the total hadronic-production
cross-section
and the two-point function $\Pi^{\mu\nu}(q)$.
The $q\bar q$ blob contains all possible QCD corrections. The dashed vertical
line
indicates that the blob is cut in all possible ways, so that the left and
right sides correspond to the production amplitude $T$ and its
complex-conjugate
$T^\dagger$, respectively, for a given intermediate state.}
\label{fig:PiBlob}
\end{figure}

Well below the $Z$ peak, the hadronic production is dominated by the
$\gamma$-exchange contribution.
Thus, we can compute the cross-sections of all subprocesses
$e^+e^-\to\gamma^*\to q\bar q, q\bar q G,\ldots $
(at a given order in $\alpha_s$), and make the sum. Technically,
it is much easier
to compute the QCD T-product of two electromagnetic currents
[$J^\mu_{\!\!\mbox{\rms em}} = \sum_f Q_f q_f\gamma^\mu q_f$]:
\bel{eq:Pi_def}
\Pi^{\mu\nu}(q)\equiv i \int d^4x \; \e^{iqx}\;
\langle 0| T\left( J^\mu_{\!\!\mbox{\rms em}}(x)
J^\nu_{\!\!\mbox{\rms em}}(0)^\dagger\right)|0\rangle
= \left(-g^{\mu\nu}q^2 + q^\mu q^\nu\right) \Pi_{\!\!\mbox{\rms em}}(q^2) \, .
\ee
As shown in Fig.~\ref{fig:PiBlob}, the absorptive part of this object
(i.e. the imaginary part, which results from cutting --putting {\it on shell}--
the propagators of the intermediate exchanged quarks and gluons in all
possible ways) just corresponds to the sum of the squared moduli of the
different production amplitudes. The exact relation with the total
cross-section is:
\bel{eq:R_ee_QCD}
R_{e^+e^-} \equiv
{\sigma(e^+e^-\to \mbox{\rm hadrons})\over\sigma(e^+e^-\to\mu^+\mu^-)}
= 12 \pi \,\mbox{\rm Im}\Pi_{\!\!\mbox{\rms em}}(s)
\, .
\ee

Neglecting the small (away from thresholds)
corrections generated by the non-zero quark masses, the ratio
$R_{e^+e^-}$ is given by a perturbative series in powers of $\alpha_s(s)$:
\beqn\label{eq:R_series}
R_{e^+e^-} &\!\!\! =\!\!\! & \left(\sum_{f=1}^{N_f} Q_f^2\right)
\, N_C \, \left\{ 1 +
\sum_{n\geq 1} F_n \left({\alpha_s(s)\over\pi}\right)^{\! n} \right\} \no\\
&\!\!\! =\!\!\! & \left(\sum_{f=1}^{N_f} Q_f^2\right)
 \, N_C \, \Biggl\{ 1 +
F_1 {\alpha_s(\mu^2)\over\pi} +
\left[ F_2 + F_1 {\beta_1\over 2} \ln\left({s\over\mu^2}\right)\right]
\left({\alpha_s(\mu^2)\over\pi}\right)^{\! 2}
\Biggr. \\
&\!\!\! +\!\!\! & \Biggl.
\left[ F_3 + F_2 \beta_1 \ln\left({s\over\mu^2}\right) + F_1
\left( {\beta_2\over 2} \ln\left({s\over\mu^2}\right)
 + {\beta_1^2\over 4} \ln^2\left({s\over\mu^2}\right)\right)\right]
\left({\alpha_s(\mu^2)\over\pi}\right)^{\! 3} +
\cO(\alpha_s^4) \Biggr\} .\no
\eeqn
The second expression, shows explicitly how the running coupling
$\alpha_s(s)$ sums an infinite number of
higher-order logarithmic terms.

So far, the calculation has been performed to order $\alpha_s^3$, with
the result (in the $\overline{\rm MS}$ scheme) \cite{CKT:79,GKL:91}:
\beqn\label{eq:F_results}
F_1 &\!\! = &\!\! 1 \, , \no\\
F_2 &\!\! = &\!\! 1.986 - 0.115 N_f \, , \\
F_3 &\!\! = &\!\! -6.637 - 1.200 N_f - 0.005 N_f^2 - 1.240
{\left(\sum_f Q_f \right)^2\over 3 \sum_f Q_f^2 } \, . \no
\eeqn
Note the different charge-dependence on the last term, which is due to the
contribution from three intermediate gluons (with a separate quark trace
attached to each electromagnetic current in Fig.~\ref{fig:PiBlob}).

For 5 flavours, one has:
\bel{eq:R_five}
R_{e^+e^-}(s) \, = \, {11\over 3} \left\{ 1 + {\alpha_s(s)\over\pi} +
1.411 \left({\alpha_s(s)\over\pi}\right)^2
-12.80 \left({\alpha_s(s)\over\pi}\right)^3 + \cO(\alpha_s^4) \right\} \, .
\ee

The perturbative uncertainty of this prediction is of order
$\alpha_s^4$, since the coefficient $F_4$ is unknown. This uncertainty also
includes the ambiguities related to the choice of renormalization scale and
scheme. Although, the total sum of the perturbative series is of course
independent of our renormalization conventions, different choices of scale
and/or scheme lead to slightly different numerical predictions for the
truncated series. For instance,
the perturbative series truncated at a finite order $N$,
$R_{e^+e^-}^{(N)}(s)\equiv\left(\sum_f Q_f^2 \right)\, N_C \, \left\{ 1 +
\sum_{n= 1}^N F_n \left({\alpha_s(s)\over\pi}\right)^n \right\}$,
has an explicit scale dependence of order $\alpha_s^{N+1}$:
\bel{eq:R_mu_dep}
{dR_{e^+e^-}^{(N)}\over d\mu^2}\sim\left({\alpha_s(\mu^2)\over\pi}\right)^{N+1}
{}.
\ee

The numerical values of $\alpha_s$ and the $F_n$ ($n\geq 2$) coefficients
depend on our choice of scheme
(also $\beta_n$ for $n\geq 3$).
For instance, at second order\footnote{\small
Actually, at second order a scheme is completely specified by a single
parameter. Thus, scale and scheme dependence is just the same at this order.
The relations in Eqs.~\protect\eqn{eq:scheme_dep_a} and
\protect\eqn{eq:scheme_dep_F} are equivalent to a change of scale:
$\mu^2_{\!\!\mbox{\rms MS}} = (4\pi/{\rm e}^{\gamma_E})\,
\mu^2_{\overline{\!\!\mbox{\rms MS}}}$.},
the relation
between the MS and $\overline{\mbox{\rm MS}}$ schemes is:
\beqn\label{eq:scheme_dep_a}
\alpha_s^{\!\!\mbox{\rms MS}} &\!\!\! = &\!\!\!
\alpha_s^{\overline{\!\!\mbox{\rms MS}}} \left\{
1 + {\beta_1\over 2} \left[\ln{(4\pi)}-\gamma_E\right]
{\alpha_s^{\overline{\!\!\mbox{\rms MS}}} \over\pi} + \cdots \right\} \, ,
\\ \label{eq:scheme_dep_F}
F_2^{\!\!\mbox{\rms MS}} &\!\!\! = &\!\!\!
F_2^{\overline{\!\!\mbox{\rms MS}}} -
F_1 {\beta_1\over 2} \left[\ln{(4\pi)}-\gamma_E\right]
= 7.359 - 0.441 N_f \, .
\eeqn
The difference between both schemes is obviously a higher-order effect.
With $N_f=5$, the MS scheme leads to a second-order coefficient
$F_2^{\!\!\mbox{\rms MS}} = 5.156$, which is a factor 3.6 bigger than
$F_2^{\overline{\!\!\mbox{\rms MS}}}$. Thus, the perturbative series
looks more convergent with the $\overline{\mbox{\rm MS}}$ choice.

The theoretical prediction for $R_{e^+e^-}(s)$ above the $b$-$\bar b$
threshold is compared \cite{PDG:94}
in Fig.~\ref{fig:Ree} with the measured data,
taking into account mass-corrections and electroweak ($Z$-exchange)
contributions. The two curves correspond to
$\Lambda^{(N_f=5)}_{\overline{\!\!\mbox{\rms MS}}} = 60$ MeV (lower curve)
and 250 MeV (upper curve). The rising at large energies is due to the
tail of the $Z$ peak.
A global fit to all data between 20 and 65 GeV yields \cite{HA:93}
\bel{eq:alpha_ee}
\alpha_s(34\:\mbox{\rm GeV}) \, = \, 0.146\pm0.030\, .
\ee

The hadronic width of the $Z$ boson can be analyzed in the same way:
\bel{eq:Z_hadrons}
R_Z\equiv{\Gamma(Z\to\mbox{\rm hadrons})\over \Gamma(Z\to e^+e^-)} =
R_Z^{\!\!\mbox{\rms EW}} N_C \left\{ 1 +
\sum_{n\geq 1} \tilde{F}_n \left({\alpha_s(M_Z^2)\over\pi}\right)^n
+ \cO\left({m_f^2\over M_Z^2}\right) \right\} \, .
\ee
The global factor
\bel{eq:R_Z_ew}
R_Z^{\!\!\mbox{\rms EW}}\, =\, {\sum_f (v_f^2 + a_f^2)\over v_e^2 + a_e^2}\;
(1 + \delta_{\!\!\mbox{\rms EW}} )
\ee
contains the underlying electroweak $Z\to \sum_f q_f \bar q_f$ decay amplitude.
Since both vector and axial-vector couplings are present, the QCD-correction
coefficients $\tilde{F}_n$ are slightly different from  $F_n$ for $n\geq 2$.
For instance, the $Z$ axial coupling generates the two-loop contribution
$Z\to t\bar t\to GG \to q\bar q$ (through triangular quark diagrams),
which is absent in the vector case; this leads to an additional
$\cO\left(\alpha_s^2 m_t^2/M_Z^2\right)$ correction.

In order to determine $\alpha_s$ from $R_Z$, one needs to perform a global
analysis of the LEP/SLC data, taking properly into account
the higher-order electroweak corrections \cite{BA:95,PI:94b}.
The latest $\alpha_s$ value reported by the LEP Electroweak Working Group
\cite{LEPEWG:94} is
\bel{eq:alpha_Z_fit}
\alpha_s(M_Z^2) \, = \, 0.125 \pm 0.005 \pm 0.002 \, .
\ee

\subsection{\protect\boldmath $\tau$\protect\unboldmath\mbox{}$^-$
\protect\boldmath $\to\nu$\protect\unboldmath\mbox{}$_\tau$ 
  \protect\unboldmath  +  hadrons}

\label{subsec:TauHadrons}

The calculation of QCD corrections to the inclusive decay of the $\tau$
lepton \cite{BR:88,NP:88,BNP:92,LDP:92a,PI:94}
looks quite similar from a diagrammatic point of view. One just puts
all possible gluon (and $q\bar q$) corrections to the basic decay diagram
in Fig.~\ref{fig:tau}, and computes the sum
\bel{eq:Tau_sum}
\Gamma(\tau^-\to\nu_\tau + \mbox{\rm hadrons}) =
\Gamma(\tau^-\to\nu_\tau + q\bar q) + \Gamma(\tau^-\to\nu_\tau + q\bar q G)
+ \Gamma(\tau^-\to\nu_\tau + q\bar q GG) + \cdots
\ee
As in the $e^+e^-$ case, the calculation is more efficiently performed through
the two-point-function
\bel{eq:PiTau}
\Pi_L^{\mu\nu}(q)\equiv i \int d^4x \, \e^{iqx}\;
\langle 0| T\left( L^\mu(x) L^\nu(0)^\dagger\right)|0\rangle
= \left(-g^{\mu\nu}q^2 + q^\mu q^\nu\right) \Pi_L^{(1)}(q^2)
+ q^\mu q^\nu \,\Pi_L^{(0)}(q^2)\, ,\;
\ee
which involves the T-ordered product of two left-handed currents,
$L^\mu = \bar u \gamma^\mu (1-\gamma^5) d_\theta$.
This object can be easily visualized through a diagram analogous to
Fig.~\ref{fig:PiBlob}, where the photon is replaced by a $W^-$ line and
one has a $\tau\nu_\tau$ pair in the external fermionic lines instead of
the $e^+e^-$ pair. The precise relation with the ratio $R_\tau$ is:
\beqn\label{eq:R_tau_int}
R_\tau  &\!\!\! \equiv &\!\!\!
{\Gamma(\tau^-\to\nu_\tau + \mbox{\rm hadrons})\over
   \Gamma(\tau^-\to\nu_\tau e^-\bar\nu_e)} \no\\
&\!\!\! = &\!\!\!
12 \pi \int^{m_\tau^2}_0 {ds \over m_\tau^2 } \,
 \left(1-{s \over m_\tau^2}\right)^2
\left\{ \left(1 + 2 {s \over m_\tau^2}\right)
 \mbox{\rm Im} \Pi^{(1)}_L(s)
 + \mbox{\rm Im} \Pi^{(0)}_L(s) \right\} \,  . \quad
\eeqn

The three-body character of the basic decay mechanism,
$\tau^-\to\nu_\tau u d_\theta$, shows here a crucial difference with
$e^+e^-$ annihilation. One needs to integrate over all possible neutrino
energies or, equivalently, over all possible values of the total hadronic
invariant-mass $s$. The spectral functions
$\mbox{\rm Im} \Pi^{(0,1)}_L(s)$ contain the dynamical information on the
invariant-mass distribution of the final hadrons.
The lower integration limit corresponds to the threshold for hadronic
production, i.e. $m_\pi$ (equal to zero for massless quarks). Clearly, this
lies deep into the non-perturbative region where confinement is crucial.
Thus, it is very difficult to make a reliable prediction for
the integrand in \eqn{eq:R_tau_int}.

\begin{figure}[bht]
\centerline{\epsfysize =7cm \epsfbox{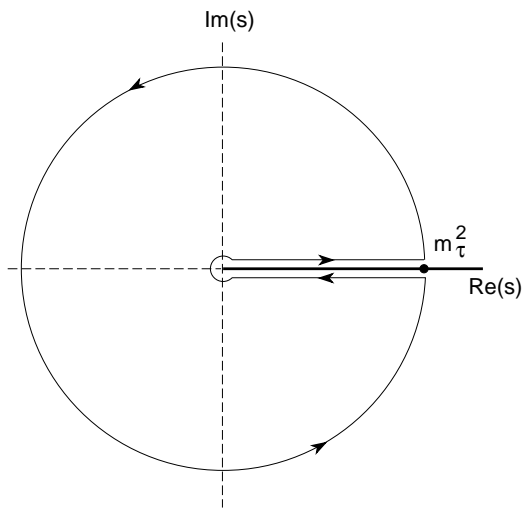}}
\vspace{-0.5cm}
\caption{Integration contour in the complex $s$-plane used to obtain
Eq.~\protect\eqn{eq:circle}.}
\label{fig:circle}
\end{figure}

Fortunately, we have precious exact (i.e. non-perturbative) information on
the dynamical functions $\Pi^{(0,1)}_L(s)$, which allows us to accurately
predict the total integral \eqn{eq:R_tau_int}:
$\Pi^{(0,1)}_L(s)$ are analytic functions in the complex $s$-plane except for a
cut in the positive real axis. The physics we are interested in lies of course
in the singular region, where hadrons are produced.
We need to know the integral along the physical cut of
$\,\mbox{\rm Im} \Pi^{(0,1)}_L(s) = -{i\over 2}
[\Pi^{(0,1)}_L(s+i\epsilon)-\Pi^{(0,1)}_L(s-i\epsilon)]$.
However, we can use Cauchy's theorem (close integrals of analytic functions are
zero if there are no singularities within the integration contour), to express
$R_\tau$ as a contour integral
in the complex $s$-plane running
counter-clockwise around the circle $|s|=m_\tau^2$
\cite{BR:88,NP:88,BNP:92}:
\bel{eq:circle}
 R_\tau =
6 \pi i \oint_{|s|=m_\tau^2} {ds \over m_\tau^2} \,
 \left(1 - {s \over m_\tau^2}\right)^2
 \left\{ \left(1 + 2 {s \over m_\tau^2}\right) \Pi^{(0+1)}_L(s)
         - 2 {s \over m_\tau^2} \Pi^{(0)}_L(s) \right\} .
\ee
The advantage of this expression
is that it requires dynamical information only for
complex $s$ of order $m_\tau^2$, which is significantly larger than the scale
associated with non-perturbative effects in QCD.  A perturbative
calculation of $R_\tau$ is then possible.

Using the so-called {\it Operator Product Expansion} techniques
it is possible to show
\cite{NP:88,BNP:92,PI:94}
that non-perturbative contributions are
very suppressed [$\sim (\Lambda/m_\tau)^6$].
Thus, $R_\tau$ is a perfect observable for determining the strong coupling.
 In fact, $\tau$ decay is probably the lowest energy process from which the
 running coupling constant can be extracted cleanly, without hopeless
 complications from non-perturbative effects.
The $\tau$ mass, $m_\tau = 1.7771{\,}^{+0.0004}_{-0.0005}$  GeV \cite{PDG:94},
lies fortuitously
 in a {\it compromise} region  where the coupling constant
 $\alpha_s$ is large enough that $R_\tau$ is very sensitive to its
 value, yet still small enough that the perturbative expansion
 still converges well.

The explicit calculation gives \cite{BNP:92,PI:94}:
\bel{eq:r_tau_total}
R_{\tau}  =
 3 \left( |V_{ud}|^2 + |V_{us}|^2 \right)
S_{\!\!\mbox{\rms EW}} \left\{ 1 + \delta_{\!\!\mbox{\rms EW}}' +
\delta^{(0)} + \delta_{\!\!\mbox{\rms NP}}\right\}
\, ,
\ee
where  $S_{\!\!\mbox{\rms EW}} = 1.0194$ and
$\delta_{\!\!\mbox{\rms EW}}' = 0.0010$
are the leading and next-to-leading electroweak corrections, and
$\delta^{(0)}$ contains the dominant perturbative-QCD contribution:
\beqn\label{eq:delta_0}
\delta^{(0)} &\!\!\! =&\!\!\! {\alpha_s(m_\tau^2)\over\pi} +
\left[ F_2 - {19\over 24}\beta_1\right]
\left( {\alpha_s(m_\tau^2)\over\pi} \right)^2
\no\\ &\!\!\!  &\!\!\! \mbox{} +
\left[ F_3' - {19\over 12} F_2\beta_1 - {19\over 24}\beta_2 +
{265\over 288} \beta_1^2\right]
\left( {\alpha_s(m_\tau^2)\over\pi} \right)^3 + \cO(\alpha_s^4) \\
 &\!\!\! = &\!\!\! {\alpha_s(m_\tau^2)\over\pi} +
5.2023 \left( {\alpha_s(m_\tau^2)\over\pi} \right)^2
+ 26.366 \left( {\alpha_s(m_\tau^2)\over\pi} \right)^3
+ \cO(\alpha_s^4) \, .
\eeqn
The remaining factor $\delta_{\!\!\mbox{\rms NP}}\approx -0.016\pm 0.005$
includes the estimated \cite{BNP:92,PI:94}
small mass-corrections and non-perturbative contributions.

Owing to its high sensitivity to $\alpha_s$ \cite{NP:88,BNP:92}
the ratio $R_\tau$ has been a subject of intensive study in recent years.
Many different sources of possible perturbative and non-perturbative
contributions have been analyzed in detail.
Higher-order logarithmic corrections have been resummed \cite{LDP:92a},
leading to very small renormalization-scheme dependences.
The size of the non-perturbative contributions has been
experimentally analyzed, through a study of the invariant-mass
distribution of the final hadrons \cite{LDP:92b}; the present data
implies \cite{ALEPH:93}
$\delta_{\!\!\mbox{\rms NP}}= (0.3\pm 0.5)\% $ confirming
the predicted \cite{BNP:92} suppression of non-perturbative
corrections.
An exhaustive summary of the $R_\tau$ analysis can be found
in Ref.~ \cite{PI:94}.

Using the Particle Data Group values for the $\tau$ lifetime and
leptonic branching ratios \cite{PDG:94}, the theoretical analysis
of $R_\tau$ results in a fitted value of $\alpha_s$ \cite{PI:94},
\bel{alpha_s_tau}
\alpha_s(m_\tau^2)\, = \, 0.33\pm 0.03 \, ,
\ee
which is significantly larger than \eqn{eq:alpha_Z_fit}.
After evolution up to the scale $M_Z$, the strong coupling constant in
Eq.~\eqn{alpha_s_tau} decreases to
$\alpha_s(M_Z^2) = 0.120^{+0.003}_{-0.004}$, in excellent agreement
with the $Z$-width determination and with a smaller error bar.
This comparison provides a beautiful test of the predicted running of
$\alpha_s$.

\subsection{\protect\boldmath $e^+e$\protect\unboldmath\mbox{}$^-$
\protect\boldmath $\to\; $
\protect\unboldmath jets}

\label{subsec:jets}

\begin{figure}[htb]
\centerline{\epsfxsize = 14.5cm \epsfbox{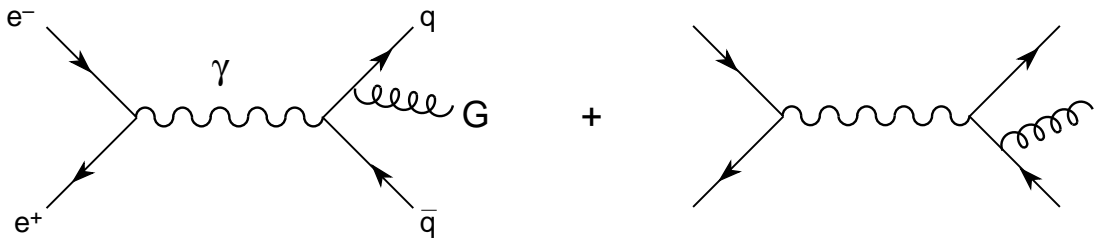}}
\caption{Gluon bremsstrahlung corrections to $e^+e^-\to q\bar q$.}
\label{fig:ee3J}
\end{figure}

At lowest-order in the strong coupling, the hadronic production in
$e^+e^-$ collisions proceeds through $e^+e^-\to q\bar q$. Thus,
at high-energies, the final hadronic states are predicted to
have mainly a two-jet structure, which agrees with the
empirical observations.
At $\cO(\alpha_s)$, the emission of a hard gluon from a quark leg
generates the $e^+e^-\to q\bar q G$ transition, leading to
3-jet configurations. For massless quarks, the differential
distribution of the 3-body final state is given by:
\bel{eq:3J_dis}
{1\over\sigma_0}\, {d^2\sigma\over dx_1 dx_2} \, = \,
{2\alpha_s\over 3\pi}\, {x_1^2 + x_2^2 \over (1-x_1) (1-x_2)} \ ,
\ee
where
\bel{eq:sigma0}
\sigma_0 \,\equiv\, {4\pi\alpha^2\over 3 s} \, N_C \,
\sum_{f=1}^{N_f} Q_f^2
\ee
is the lowest-order $e^+e^-\to\gamma^*\to q\bar q$ cross-section.
The kinematics is defined through the invariants
$s\equiv q^2$ and
$s_{ij}\equiv (p_i + p_j)^2 = (q-p_k)^2 \equiv s (1-x_k)$
($i,j,k = 1,2,3$), where $p_1$, $p_2$ and $p_3$ are the quark,
antiquark and gluon momenta, respectively,
and $q$ is the total $e^+e^-$ momentum.
For given $s$, there are only
two independent kinematical variables since
\bel{eq:x_rel}
x_1 + x_2 + x_3 = 2 \, .
\ee
In the centre-of-mass system [$q^\mu = (\sqrt{s}, \vec{0}\, )$],
$x_i = E_i/E_e = 2 E_i/\sqrt{s}$.

Eq.~\eqn{eq:3J_dis} diverges as $x_1$ or $x_2$ tend to 1.
This is a very different infinity from the ultraviolet ones
encountered before in the loop integrals.
In the present case, the tree amplitude itself is becoming singular in
the phase-space boundary. The problem originates in the infrared
behaviour of the intermediate quark propagators:
\bel{eq:infrared}
\ba
x_1\to 1 \qquad\Longleftrightarrow\qquad
(p_2+p_3)^2 = 2 \,(p_2\cdot p_3) \to 0 \ ;
\\
x_2\to 1 \qquad\Longleftrightarrow\qquad
(p_1+p_3)^2 = 2\, (p_1\cdot p_3) \to 0 \ .
\ea\ee
There are two distinct kinematical configurations leading to
{\it infrared divergences}:
\begin{enumerate}
\item {\bf Collinear gluon}: The 4-momentum of the gluon is
parallel to that of either the quark or the antiquark.
This is also called a {\it mass singularity}, since the
divergence would be absent if either the gluon or the
quark had a mass
($p_3\| p_2$ implies $s_{23}=0$ if $p_2^2=p_3^3=0$).
\item {\bf Soft gluon}: $p_3\to 0$.
\end{enumerate}

In either case, the observed final hadrons
will be detected as a 2-jet configuration, because the $qG$ or
$\bar qG$ system cannot be resolved.
Owing to the finite resolution of any detector,
it is not possible (not even in principle) to separate those
2-jet events generated by the basic $e^+e^-\to q\bar q$ process,
from $e^+e^-\to q\bar q G$ events with a collinear or soft gluon.
In order to resolve a 3-jet event, the gluon should have an energy
and opening angle (with respect to the quark or antiquark)
bigger than the detector resolution.
The observable 3-jet cross-section will never include the
problematic region $x_{1,2}\to 1$; thus, it will be finite,
although its value will depend on the detector resolution and/or
the precise definition of {\it jet}
(i.e. $\sigma$ depends on the chosen integration
limits).

On the other side, the 2-jet configurations will include both
$e^+e^-\to q\bar q\; $ and $e^+e^-\to q\bar q G$ with an unobserved
gluon. The important question is then the infrared behaviour of the
sum of both amplitudes.
The exchange of virtual gluons among the quarks
generate an $\cO(\alpha_s)$ correction to the $e^+e^-\to q\bar q$
amplitude:
\bel{eq:T_eeqq}
T[e^+e^-\to q\bar q] = T_0 + T_1 + \cdots
\ee
where $T_0$ is the lowest-order (tree-level) contribution,
$T_1$ the $\cO(\alpha_s)$ correction, and so on.
The interference of $T_0$ and $T_1$ gives rise to an
$\cO(\alpha_s)$ contribution to the $e^+e^-\to q\bar q$
cross-section.

\begin{figure}[htb]
\centerline{\epsfxsize = 14.5cm \epsfbox{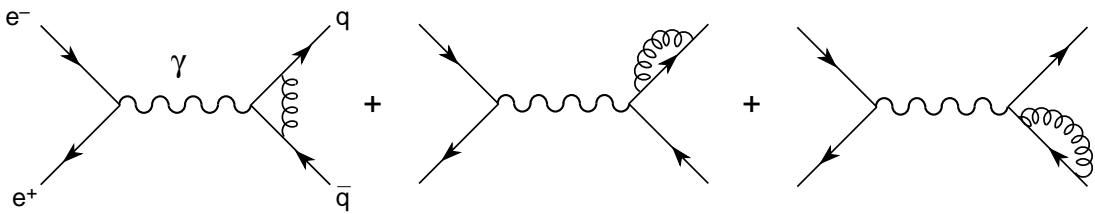}}
\caption{1-loop gluonic corrections to $e^+e^-\to q\bar q$.}
\label{fig:eeqq(g)}
\end{figure}

We know already that loop diagrams have ultraviolet divergences
which must be renormalized. In addition, they also have infrared
divergences associated with collinear and soft configurations
of the virtual gluon.
One can explicitly check that the $\cO(\alpha_s)$ infrared divergence
of $\sigma(e^+e^-\to q\bar q)$ exactly cancels the one in
$\sigma(e^+e^-\to q\bar q G)$, so that the sum is well-defined:
\bel{eq:infrared_sum}
\sigma(e^+e^-\to q\bar q) + \sigma(e^+e^-\to q\bar q G) + \cdots
= \sigma_0\,\left( 1 + {\alpha_s\over\pi} + \cdots\right) \ .
\ee
This is precisely the inclusive result discussed
in Sect.~\ref{subsec:Rhadrons}.
This remarkable cancellation of infrared divergences is
actually a general result (Bloch-Nordsieck \cite{BN:37}
and Kinoshita--Lee--Nauenberg \cite{KLN:62} theorems):
for inclusive enough cross-sections both the
soft and collinear infrared divergences cancel.

\begin{figure}[tbh]
\centerline{\epsfysize = 4.5cm \epsfbox{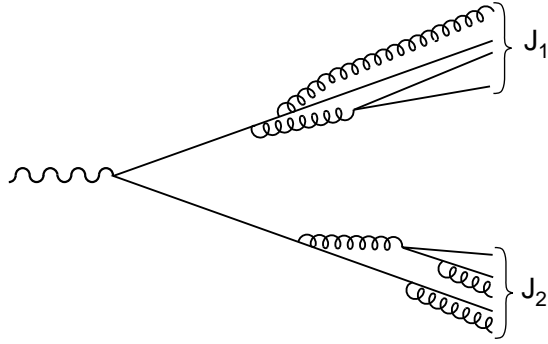}}
\caption{2-jet configuration.}
\label{fig:jetdef}
\end{figure}

While the total hadronic cross-section is unambiguously defined,
we need a precise definition of jet in order to classify a given
event as a 2-, 3-, \ldots, or n-jet configuration.
Such a definition should be free of infrared singularities, and
insensitive to the details of the non-perturbative fragmentation
into hadrons.
A popular example of jet definition is the so-called
JADE algorithm \cite{JADE:86}, which makes use of an invariant-mass cut $y$:
\bel{eq:jade}
\mbox{\rm 3 jet} \qquad\Longleftrightarrow\qquad
s_{ij} \equiv (p_i + p_j)^2 > y s \qquad (\forall i,j=1,2,3) \ .
\ee

Clearly, both the theoretical predictions and the experimental
measurements depend on the adopted jet definition.
With the JADE algorithm, the fraction of 3-jet events
is predicted to be:
\be\label{eq:R3}
R_3  =  {2\alpha_s\over 3\pi} \,\biggl\{
(3-6y) \ln{\!\left({y\over 1-2y}\right)}
+ 2 \ln^2{\!\left({y\over 1-y}\right)} + {5\over 2}
-6y -\frac{9}{2} y^2
+ 4 \mbox{\rm Li}_2\!\left({y\over 1-y}\right) -{\pi^2\over 3}
\biggr\}  ,
\ee
where
\bel{eq:Li}
\mbox{\rm Li}_2(z)\,\equiv\, -\int_0^z\,{d\xi\over 1-\xi}\,
\ln{\xi} \ .
\ee
The corresponding fraction of 2-jet events is given by
$R_2 = 1 - R_3$.
The fraction of 2- or 3-jet events obviously depends on the
chosen cut $y$.
The infrared singularities are manifest in the divergent behaviour
of $R_3$ for $y\to 0$.

At higher-orders in $\alpha_s$ one needs to define the
different multi-jet fractions. For instance, one can
easily generalize the JADE algorithm an classify a
$\{p_1,p_2,\ldots,p_n\}$ event as a n-jet configuration
provided that $s_{ij}>y s$ for all $i,j=1,\ldots,n$.
If a pair of momenta does not satisfy this constraint, they are
combined into a single momentum and the event is considered
as a $(n-1)$ jet configuration (if
the constraint is satisfied by all other combinations
of momenta).
The general expression for the fraction of n-jet events takes the
form:
\bel{eq:R_n_frac}
R_n(s,y) \, = \, \left({\alpha_s(s)\over\pi}\right)^{n-2}
\sum_{j=0} C_j^{(n)}(y)\,\left({\alpha_s(s)\over\pi}\right)^j \ ,
\ee
with $\sum_n R_n = 1$.

A few remarks are in order here:
\bi
\item The jet fractions have a
high sensitivity to $\alpha_s$ [$R_n\sim \alpha_s^{n-2}$].
 Although the sensitivity
increases with $n$, the number of events decreases
with the jet multiplicity.
\item Higher-order $\alpha_s(\mu^2)^j\ln^k(s/\mu^2)$ terms have
been summed into $\alpha_s(s)$. However, the coefficients
$C_j^{(n)}(y)$ still contain $\ln^k(y)$ terms.
At low values of $y$, the infrared divergence ($y\to 0$) reappears
and the perturbative series becomes unreliable.
For large $y$, the jet fractions $R_n$ with $n\geq 3$ are small.
\item Experiments measure hadrons rather than partons. Therefore,
since these observables are not fully inclusive, there is an
unavoidable dependence on the non-perturbative fragmentation into
hadrons. This is usually modelled through Monte Carlo analyses,
and introduces theoretical uncertainties which need to be estimated.
\ei

Many different jet algorithms and jet variables
(jet rates, event shapes, energy correlations, \ldots) have
been introduced to optimize the perturbative analysis.
In some cases, a resummation of
$\alpha_s(s)^n\ln^m\! (y)$ contributions with $m>n$ has been performed
to improve the predictions at low $y$ values \cite{CA:91}.

\begin{figure}[bth]
\centerline{\epsfysize=8.5cm \epsfbox{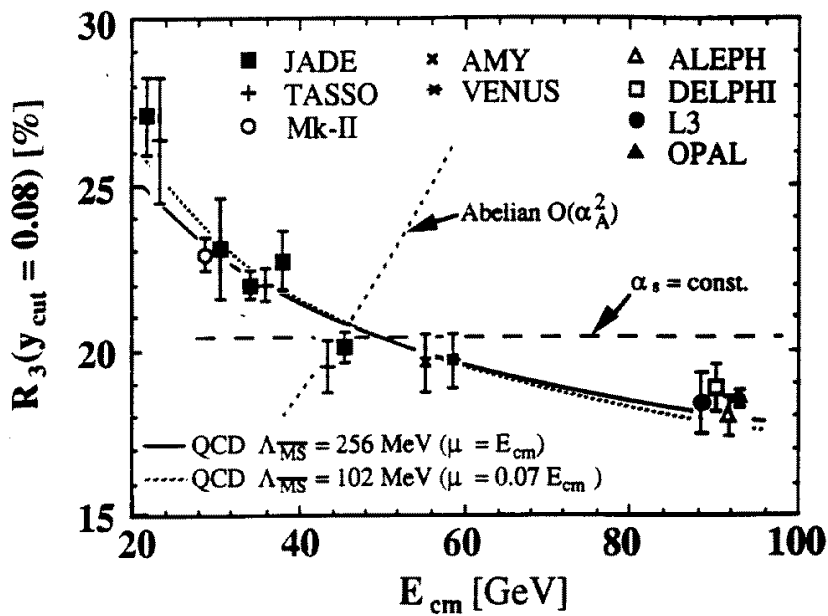}}
\vspace{-0.5cm}
\caption{Energy dependence of 3-jet event production rates
$R_3(y=0.8)$, compared with predictions of analytic $\cO(\alpha_s^2)$
QCD calculations, with the hypothesis of an energy independent $\alpha_s$
and with the abelian vector theory in $\cO(\alpha_A^2)$
(taken from Ref.~\protect\cite{BE:93}).}
\label{fig:r3_e}
\end{figure}

Fig.~\ref{fig:r3_e} \cite{BE:93} shows the energy dependence of the
measured 3-jet
production fraction $R_3$ ($y=0.08$), compared with QCD
predictions. The data is in good agreement with QCD and
fits very well the predicted energy-dependence of the
running coupling. A constant value of $\alpha_s$ cannot
describe the observed production rates. The figure
shows also the predictions obtained with an abelian vector
theory at $\cO(\alpha_A^2)$, which are clearly excluded.


Several measurements of $\alpha_s$, using different jet variables, have
been performed.
All measurements are in good agreement, providing a
good consistency test of the QCD predictions.
Combining the results from all experiments at LEP and SLC, one gets
the average value \cite{BE:94}:
\bel{eq:alpha_s_shapes}
\alpha_s(M_Z^2)\, = \,\left\{
\bat 0.119\pm 0.006 \qquad & \left(\cO(\alpha_s)^2\right) \\
0.123\pm 0.006 \qquad & \left(\mbox{\rm resummed calculations}\right)
\ea\right. \, .
\ee
The two numbers correspond to different theoretical approximations
used in the fits to extract $\alpha_s$.

3-jet events can also be used to test the gluon spin.
For a spin-0 gluon, the differential distribution is still
given by Eq.~\eqn{eq:3J_dis}, but changing the $x_1^2+x_2^2$ factor
in the numerator to $x_3^2/4$.
In general, one cannot readily be sure which hadronic jet emerges
via fragmentation from a quark (or antiquark), and which from a gluon.
Therefore, one adopts instead a jet ordering,
$x_1>x_2>x_3$, where $x_1$ refers to the most energetic jet, 2 to the next
and 3 to the least energetic one, which most likely would correspond to
the gluon.
When $x_2\to 1$ ($x_1\to 1$) the vector-gluon distribution is singular,
but the corresponding scalar-gluon distribution is not
because at that point
$x_3 = (1-x_1) + (1-x_2) \to 0$.
The measured distribution agrees very well with the QCD predictions with
a spin-1 gluon;
a scalar gluon is clearly excluded.

\begin{figure}[bth]
\vfill\centerline{
\begin{minipage}[b]{.47\linewidth}
\centerline{\mbox{\epsfxsize=7.5cm\epsffile{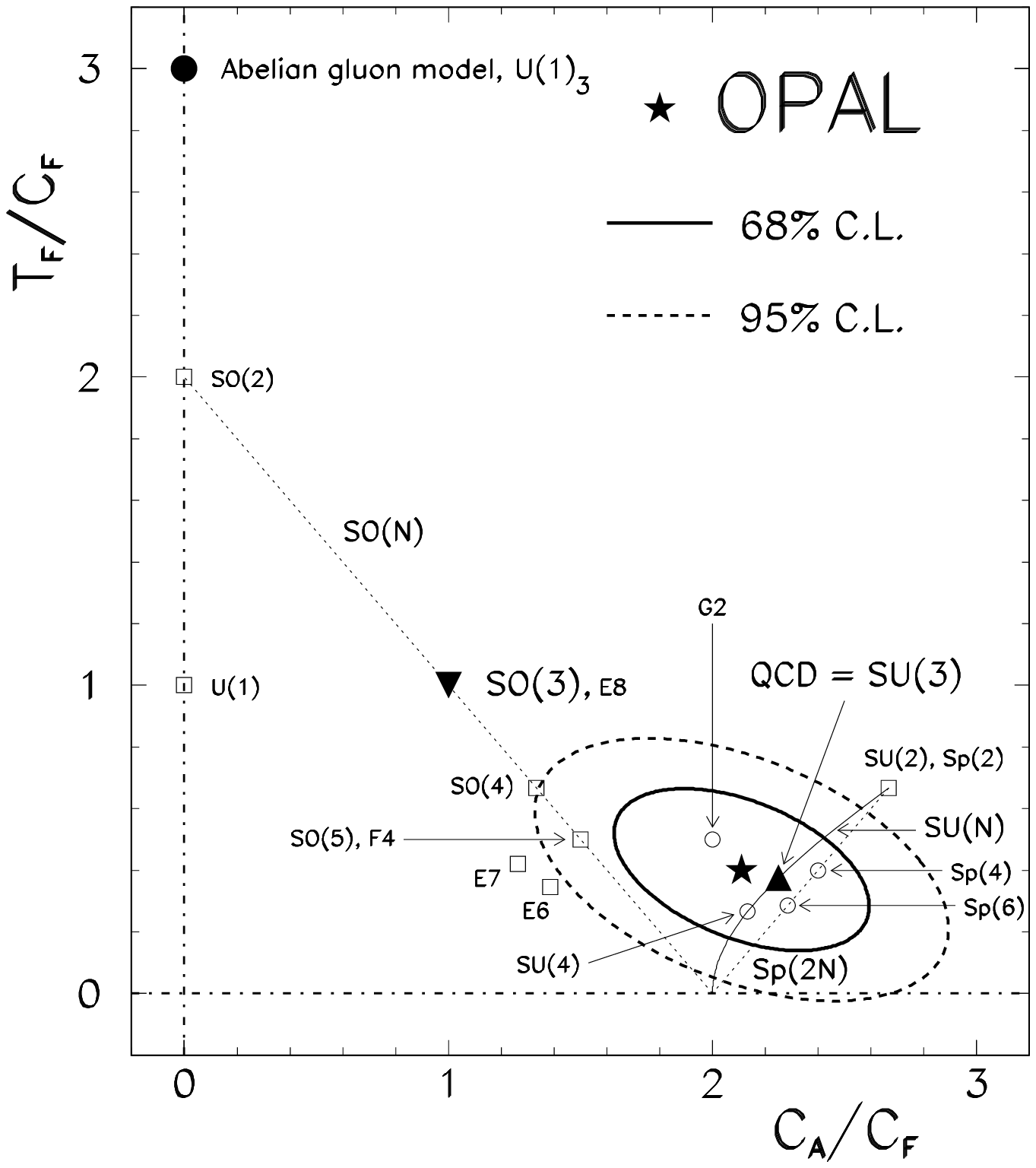}}}
\vspace{-0.2cm}
\caption{68\% and 95\% CL contours in the
$T_F/C_F$ versus $C_A/C_F$ plane, from OPAL data \protect\cite{OPAL:94}.
Expectations from various gauge models are also shown.}
\label{fig:colour_factors}
\end{minipage}
\hspace{0.7cm}
\begin{minipage}[b]{.47\linewidth}
 \centerline{\mbox{\epsfysize=9.5cm\epsffile{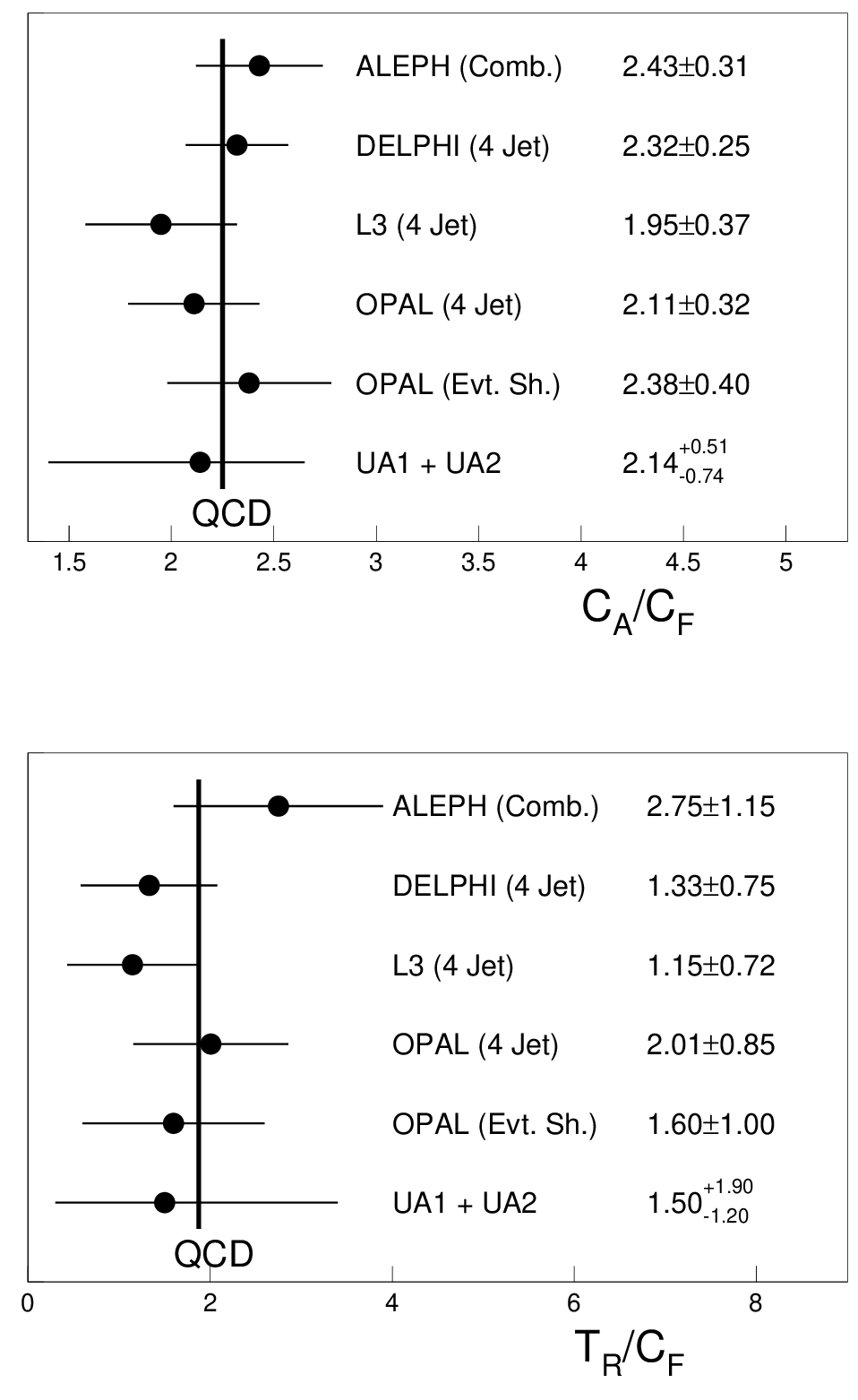}}}
 \vspace{-0.5cm}
\caption{Summary of colour-factor measurements \protect\cite{WE:94}.
The results refer to 5 active flavours with
$T_R\equiv N_f T_F =5 T_F$.}
\label{fig:colour_factors_comb}
\end{minipage}
}\vfill
\end{figure}

The predictions for jet distributions and event shapes are
functions of the colour-group factors $T_F=1/2$,
$C_F = (N_C^2-1)/(2 N_C)$ and $C_A=N_C$. These quantities,
defined in Eq.~\eqn{eq:invariants}, result from the
colour algebra associated with the different interaction vertices,
and characterize the colour-symmetry group.
If the strong interactions were based on a different gauge group,
the resulting predictions would differ in the values of these
three factors.
Since the vertices contribute in a different way to
different observables, these colour factors can be measured by
performing a combined fit to the data.
Fig.~\ref{fig:colour_factors} compares a recent OPAL
determination \cite{OPAL:94} of $C_A/C_F$ and $T_F/C_F$ with the values
of these two ratios for different colour groups.
The data is in excellent agreement with the $SU(3)$ values,
and rules out the Abelian model and many classical
Lie groups. Notice that
those groups shown by the open squares and circles are already excluded
because they do not contain three colour degrees of freedom for quarks.
Similar results have been presented by the other LEP
experiments (the hadronic production of jets at $p\bar p$ colliders
has also been analyzed in a similar way).
A summary of the colour-factor ratios obtained by the different
experiments is given in Fig.~\ref{fig:colour_factors_comb}.

\setcounter{equation}{0}
\section{DEEP INELASTIC SCATTERING}
\label{sec:dis}

We saw in Section~\ref{subsec:AF} how the deep inelastic scattering (DIS)
$e^- p \to e^- X$ can be used to learn about the proton structure.
Since this involves a bound hadronic state --the proton--, non-perturbative
phenomena such as confinement plays here a crucial role.
At the same time, the data obeys Bjorken scaling
which manifests the asymptotic freedom
property of the strong interactions. Thus, DIS appears to be an
interesting place where to investigate both perturbative and
non-perturbative aspects of QCD.

DIS can be visualized as a two-step process. First, the hard intermediate
photon, which is far off its mass-shell, scatters off a quark or
gluon with a large momentum transfer; this scattering can be
adequately described by perturbation theory.
Second, the outgoing partons recombine into hadrons
in a time of $\cO(1/\Lambda)$.
Although this recombination is not calculable in perturbation theory,
the details of the non-perturbative hadronization can be
avoided, by considering fully inclusive rates, so that perturbative
QCD can be applied.
However, the hadronic bound-structure of the initial proton state,
still introduces a non-perturbative ingredient: the proton structure
functions.

\subsection{Free parton model}

Let us ignore any QCD interactions and let us assume that the nucleon
(either proton or neutron) constituents are free spin-$\frac{1}{2}$
partons. Within the quark model, the nucleons have three point-like
constituents ($p = u_v u_v d_v$, $n = u_v d_v d_v$), which we will call
{\it valence} quarks.
Gluons are of course there; however, they do not interact directly with
the photon probe. The photon--gluon interaction only occurs through
the virtual $q$-$\bar q$ pairs coupled to the gluon constituents.
Thus, instead of gluons, the photon feels a {\it sea} of
$q$-$\bar q$ partons within the nucleon.

Let us denote $u(x)$, $\bar u(x)$, $d(x)$, $\bar d(x)$,
$s(x)$, $\bar s(x)$, $\ldots\, $ the probability distributions for
$u$, $\bar u$, $d$, $\bar d$, $s$, $\bar s$, $\ldots\, $
quarks with momentum fraction $x$ in the proton.
We have seen in Section~\ref{subsec:AF} that, within the parton model,
the proton structure functions have a simple form
in terms of parton distributions:
\bel{eq:F2_ep}
F_2^{ep}(x)/x = 2 F_1^{ep}(x) =
\frac{4}{9}\, [u(x) + \bar u(x)] + \frac{1}{9}\, [d(x) + \bar d(x)]
+ \frac{1}{9}\, [s(x) + \bar s(x)] + \cdots
\ee
The same parton distributions occur in other DIS processes such as
$\nu p\to l^- X$ or $\bar\nu p\to l^+ X$. However, since the
quark couplings of the intermediate
bosonic probe (a $W^\pm$ in that case) are not the same,
different combinations of these functions are measured:
\beqn\label{eq:F_nup}
F_2^{\nu p}(x)/x = 2 F_1^{\nu p}(x) &\!\!\! =&\!\!\!
2 \, [d(x) + s(x) + \bar u(x) + \bar c(x) + \cdots ] \ ,
\no\\
F_3^{\nu p}(x) &\!\!\! =&\!\!\!
2\,  [d(x) + s(x) - \bar u(x) - \bar c(x) + \cdots ]\ ,
\no\\  &&\\
F_2^{\bar\nu p}(x)/x = 2 F_1^{\bar\nu p}(x) &\!\!\! =&\!\!\!
2 \, [u(x) + c(x) + \bar d(x) + \bar s(x) + \cdots ] \ ,
\no\\
F_3^{\bar\nu p}(x) &\!\!\! =&\!\!\!
2 \, [u(x) + c(x) - \bar d(x) - \bar s(x) + \cdots ]\ . \no
\eeqn
Using isospin symmetry, we can further relate the up- and down-quark
distributions in a neutron to the ones in a proton:
\bel{eq:n_dis}
u^n(x) = d^p(x) \equiv d(x) \; ;\qquad d^n(x) = u^p(x) \equiv u(x) \; ;
\ee
the remaining parton distributions being obviously the same.
Thus, combining data from different DIS processes, it is possible
to obtain separate information on the individual parton distribution
functions.

The quark distributions must satisfy some constraints.
Since both the proton and the neutron have zero strangeness,
\bel{eq:s_SR}
\int_0^1 dx \, [s(x) - \bar s(x)] \, = \, 0 \ .
\ee
Similar relations follow for the heavier flavours ($c$, \ldots).
The proton and neutron electric charges imply two additional sum rules,
\bel{eq:Q_SR}
\int_0^1 dx \, [u(x) - \bar u(x)] \, = \, 2 \ , \qquad
\int_0^1 dx \, [d(x) - \bar d(x)] \, = \, 1 \ ,
\ee
which just give the excess of $u$ and $d$ quarks over antiquarks.

The quark-model concept of valence quarks gives further
insight into the nucleon structure. We can decompose the $u$ and $d$
distribution functions into the sum of valence and sea contributions,
and take the remaining parton distributions to be pure sea.
Since gluons are flavour singlet, one expects the sea to be flavour
independent. In this way, the number of independent distributions
is reduced to three:
\beqn\label{eq:V_S}
u(x) &\!\!\! =&\!\!\! u_v(x) + q_s(x) \ ,
\no\\
d(x) &\!\!\! =&\!\!\! d_v(x) + q_s(x) \ ,
\\
\bar u(x) &\!\!\! =&\!\!\! \bar d(x) = s(x) = \bar s(x) = \ldots = q_s(x) \ .
\no
\eeqn
Within this model, the strangeness sum rule \eqn{eq:s_SR}
is automatically satisfied, while \eqn{eq:Q_SR} imply constraints
on the valence-quark distributions alone.

In the analogous situation of quasi-elastic
electron--deuterium scattering, the
observed structure function shows a narrow peak around $x=\frac{1}{2}$.
This is to be expected, since the deuteron has two nucleon
constituents with $M_N\approx \frac{1}{2} M_d$ which share the total
momentum in equal terms.
A simple three-quark model for the nucleon would suggest the
existence of a similar peak
at $x=\frac{1}{3}$ in the proton and neutron structure functions.
However, the distribution shown in Fig.~\ref{fig:W2x} does not
show such behaviour.
The difference can be easily understood as originating from the
parton-sea contributions.
Taking the difference between the proton and neutron structure
functions, where the contribution from the sea cancels,
the data exhibits indeed a broad peak around
$x=\frac{1}{3}$.

Our isospin symmetric parton model implies the so-called
Gottfried sum rule \cite{GO:67}:
\bel{eq:gottfried}
\int_0^1 {dx\over x} \, \left[ F_2^{ep}(x)
- F_2^{en}(x) \right]\, =\,
\frac{1}{3} \,\int_0^1 dx \, \left[ u_v(x) - d_v(x) \right] \, = \,
\frac{1}{3} \, ,
\ee
which is well satisfied by the data.
Another interesting quantity is the ratio
\bel{eq:F2_ratio}
{F_2^{en}(x) \over F_2^{ep}(x)} \, = \,
{4 d_v(x) + u_v(x) + \Sigma_{\!\!\mbox{\rms sea}} \over
4 u_v(x) + d_v(x) + \Sigma_{\!\!\mbox{\rms sea}}} \ ,
\ee
where $\Sigma_{\!\!\mbox{\rms sea}}$ is the total sea contribution.
Since all probability distributions must be positive-definite,
this ratio should satisfy the bounds
$\frac{1}{4} \leq F_2^{en}(x) / F_2^{ep}(x) \leq 4$,
which are consistent with the data.
The measured ratio appears to tend to 1 at small $x$, indicating
that the sea contributions dominate in that region.

The conservation of the total proton momentum implies an
important sum rule:
\bel{eq:momentum_SR}
\int_0^1 dx\, x\, \left[u(x) + \bar u(x) + d(x) + \bar d(x)
+ s(x) + \bar s(x) + \cdots\right] \, = \, 1-\epsilon \ ,
\ee
where $\epsilon$ is the fraction of momentum that is not carried by quarks.
One finds experimentally that $\epsilon\approx\frac{1}{2}$
(at $Q^2\sim 10$--40 GeV$^2$),
suggesting that about half of the momentum is carried by gluons.
This shows the important role of gluons in the proton structure.
Although the naive quark model works very well in many cases, it is
a too gross simplification as a model of hadrons, at least al large $Q^2$.


\subsection{QCD-improved parton model}

At lowest order the DIS process occurs through the hard scattering
between the virtual photon ($W$, or $Z$) and one constituent parton.
The obvious first QCD corrections will be due to real gluon emission
by either the initial or final quark. To get rid of infrared divergences,
the one-loop virtual gluon contribution should also be taken into account.

\begin{figure}[htb]
\centerline{\epsfysize = 5cm \epsfbox{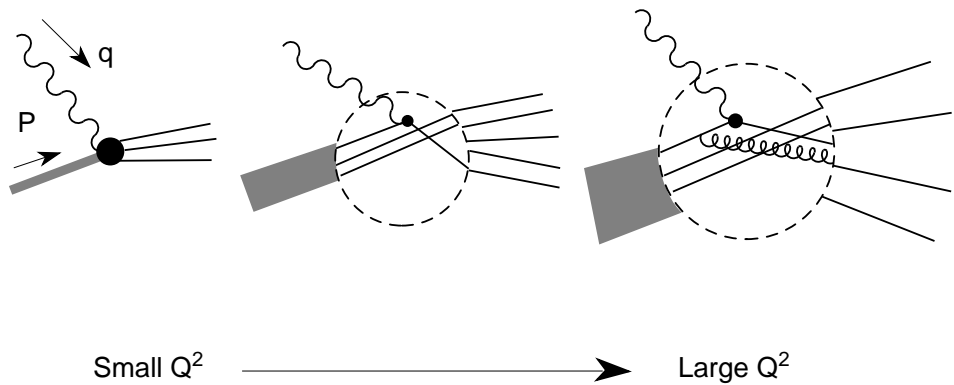}}
\caption{Resolution of the photon probe as function of $Q^2$.}
\label{fig:DISview}
\end{figure}

One can easily understand the main qualitative features of
gluon emission, with a few kinematical considerations.
At very low values of momentum transfer, the proton behaves as
a single object, either point-like (at $Q^2\approx 0$), or
with a finite size.
At higher energies, the photon is sensitive to
shorter distances and scatters with the constituent partons.
Increasing further the momentum transfer, the photon probe
has a greater sensitivity to smaller distances, and it is able to
resolve the scattered quark into a quark and a gluon.
Thus, a parton with momentum fraction $x$ can be resolved
into a parton and a gluon of smaller momentum fractions, $x'<x$ and
$x-x'$, respectively.
In a similar way, a gluon with momentum fraction $x$ can be resolved
into a quark and an antiquark.

This simple picture implies that increasing the $Q^2$,
the photon will notice some
qualitative changes in the parton distributions:
\bi
\item Gluon bremsstrahlung will shift the valence and sea distributions
to smaller $x$ values.
\item The splitting of a gluon into a quark--antiquark pair will
increase the amount of sea (mostly at small $x$).
\ei
Thus, without any detailed calculation, one can expect to find
a definite $Q^2$ dependence in the parton distributions; i.e.
violations of Bjorken scaling due to the underlying QCD interactions.

\begin{figure}[htb]
\centerline{\epsfxsize = 14cm \epsfbox{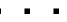}}
\caption{Leading gluonic correction to the basic DIS parton process.}
\label{fig:DIS_ge}
\end{figure}

Let us consider a quark with momentum fraction $y$. At lowest order,
its contribution to the proton structure function can be written as
\bel{eq:F_q(x)}
2 F_1^{(q)}(x) \, = \, e_q^2\,\int_0^1 dy\, q(y)\,\delta(x-y) \ .
\ee
If the quark emits a gluon before being struck by the photon, its
momentum fraction will be degraded to $yz$ ($0\leq z\leq 1$).
Assuming that the quark remains approximately on-shell,
$(q+yzP)^2\approx m_q^2 \approx 0$, implying that
$yz = Q^2/2(P\cdot q) \equiv x$. Therefore, $F_1^{(q)}(x)$
gets contributions from quarks with initial momentum fractions
$y\geq x$.

The explicit calculation of the diagrams in Fig.~\ref{fig:DIS_ge}
gives the result:
\bel{eq:DF_1_q}
2\Delta F_1^{(q)}(x) \, = \, e_q^2 \, {\alpha_s\over 2\pi}\,
\int_0^1 dy\, q(y)\,\int_0^1\, dz\, \delta(yz-x)\,
\left\{ P^+_{qq}(z) \ln(Q^2/\nu_{\!\!\mbox{\rms IR}}^2) + C(z)\right\} ,
\ee
where
\bel{eq:splitting_f}
P_{qq}(z) \,\equiv\, C_F \, \left( {1+z^2\over 1-z}\right)
\ee
is called the quark {\it splitting function}.

The important feature in Eq.~\eqn{eq:DF_1_q} is the appearance
of a {\it scaling violation} through
the logarithmic $\alpha_s$ correction.
A careful analysis of the different Feynman diagrams shows that
ultraviolet divergences are absent in the total contribution.
Therefore,
this logarithm has a completely different origin than the
ultraviolet ones found in Section~\ref{sec:loops}
The logarithmic behaviour is now generated by infrared singularities
of the type discussed in Section~\ref{subsec:jets}.
More precisely, there is a collinear singularity associated with the
gluon emission process, which has been regulated with the infrared
cut-off $\nu_{\!\!\mbox{\rms IR}}$.

The general theorems on the cancellation of infrared divergences
do not protect the structure function $F_1^{(q)}(x)$, because
this quantity is not {\it inclusive enough}. The divergence
shows up when one tries to resolve the original
quark with momentum fraction $y$ into a quark with momentum fraction
$yz$ and a gluon. $P_{qq}(z)$ is just
the coefficient of the logarithmic divergence associated with
the splitting process $q\to qG$.
Physical observables should not depend on any cut-off,
however, our definition of a parton distribution obviously depends
on the power resolution of our photon probe.
While at low $Q^2$ the photon was testing a single parton with momentum
fraction $y=x$, now it {\it feels} the splitting of a quark with $y>x$
into a quark and a gluon with separate parton distributions.

The divergence should then be reabsorbed into the {\it observable}
parton distribution function:
\bel{eq:pdf}
q(x,Q^2) = q(x,\nu_{\!\!\mbox{\rms IR}}^2) +
{\alpha_s\over 2\pi} \,
\ln(Q^2/\nu_{\!\!\mbox{\rms IR}}^2)
\int_x^1 \, {dy\over y}\, q(y) \,
 P^+_{qq}\left({x/ y}\right) \,  .
\ee
Both the {\it bare} distribution $q(x,\nu_{\!\!\mbox{\rms IR}}^2)$
and the $\alpha_s$ correction depend on the infrared cut-off, but
this dependence cancels out and does not show up in the {\it physical}
distribution function\footnote{\small
Notice, however, that the precise definition of $q(x,Q^2)$ is
{\it factorization-scheme dependent}, since we could always
include some arbitrary non-logarithmic $\alpha_s$ correction
into $q(x,Q^2)$, by simply shifting the
$C\left({x/ y}\right)$ correction factor in
\protect\eqn{eq:F_1_q_phys}.}
$q(x,Q^2)$. Instead, the parton distribution is now a $Q^2$-dependent
quantity, which fits with our intuitive picture that the photon probe
increases its resolution power with the scale.
In terms of $q(x,Q^2)$, the contribution of the quark $q$ to the
proton structure function is given by:
\bel{eq:F_1_q_phys}
2 F_1^{(q)}(x) \, = \, e_q^2 \,\left\{ q(x,Q^2) +
{\alpha_s\over 2\pi} \int_x^1 \, {dy\over y}\,
q(y)\, C\left(x/y\right)\right\} .
\ee

The individual diagrams in Fig.~\ref{fig:DIS_ge} have also a soft-gluon
singularity, which manifests in the divergent behaviour of $P_{qq}(z)$
at $z=1$. This singularity cancels exactly
in the total sum of the gluon-emission and
virtual-gluon-exchange contributions.
The net result is a slight modification in the definition of
the splitting function:
\bel{eq:spf_qq}
P_{qq}^+(z) \,\delta(yz-x)\,\equiv\, P_{qq}(z)\,
\left[\delta(yz-x) - \delta(y-x)\right] \ .
\ee

Eq.~\eqn{eq:pdf} shows an important thing: although perturbative QCD
is not able to predict the actual value of the distribution function,
it does predict how this distribution evolves in $\ln (Q^2)$. Thus,
given its value at some reference point $Q_0^2$, one can compute the
quark distribution at any other value of $Q^2$ (high-enough for perturbation
theory to be valid).
Including the leading higher-order logarithmic corrections into the
running coupling, the $Q^2$-evolution of the parton distribution is
given by \cite{LI:75,AP:77}:
\bel{eq:dQ2_pdf}
Q^2\, {d\over dQ^2} q(x,Q^2) \, = \,
{\alpha_s(Q^2)\over 2\pi} \int_x^1 {dy\over y}\,
q(y,Q^2)\, P^+_{qq}(x/y) \ .
\ee
Thus, the change in the distribution for a quark with momentum fraction $x$,
which interacts with the virtual photon, is given by the integral over
$y$ of the corresponding distribution for a quark with momentum fraction
$y\geq x$ which, having radiated a gluon, is left with
a fraction $x/y$ of its original momentum.
The splitting function has then a very intuitive physical interpretation:
$(\alpha_s/2\pi) P^+_{qq}(x/y)$ is the probability associated
with the splitting process $q(y)\to q(x) G$.
This probability is high for large momentum fractions; i.e.
high-momentum quarks lose momentum by radiating gluons.
Therefore, increasing $Q^2$, the quark distribution function will
decrease at large $x$ and will increase at small $x$.

\begin{figure}[htb]
\centerline{\epsfxsize =14cm \epsfbox{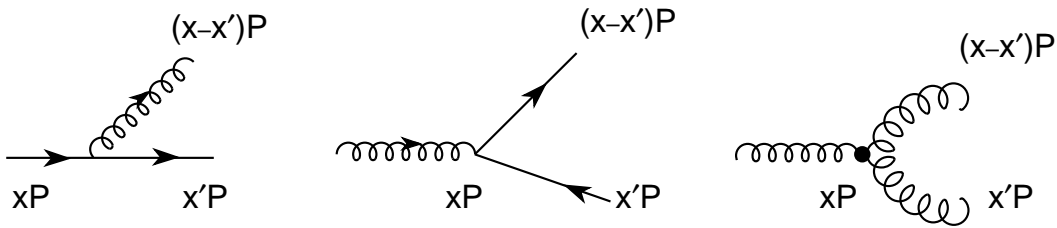}}
\caption{Basic parton-splitting processes.}
\label{fig:splitting}
\end{figure}

The evolution equation \eqn{eq:dQ2_pdf} is only correct for non-singlet
distributions such as $q_i(x)-q_j(x)$, where the (flavour-singlet) gluon
contribution cancels out. In general, one needs also to consider
the effects coming from the splitting of a gluon into a quark and an
antiquark, which interact with the photon probe.
The obvious generalization is \cite{AP:77}:
\bel{eq:evol_q_g}
Q^2 {d\over dQ^2} \left( \ba q(x,Q^2) \\ G(x,Q^2) \ea \right) =
{\alpha_s(Q^2)\over 2\pi} \int_x^1 {dy\over y}\,
\left[ \bat P^+_{qq}(x/y) & P_{qG}(x/y) \\ P_{Gq}(x/y) & P_{GG}(x/y) \ea\right]
\left( \ba q(y,Q^2) \\ G(y,Q^2) \ea \right) ,
\ee
where $P_{Gq}(z) = P_{qq}(1-z)$ determines the probability that a
quark radiates a gluon with a fraction $z$ of the
original quark momentum,
while
\beqn\label{eq:splitting_qg}
P_{qG}(z) &\!\! = &\!\! T_F \, \left[ z^2 + (1-z)^2\right] \ ,
\\ \label{eq:splitting_gg}
P_{GG}(z) &\!\! = &\!\! 2 C_A \,\left[ {z\over (1-z)_+}
+ {1-z\over z} + z (1-z) \right]
+ {1\over 6} (11 C_A - 4 N_f T_F)\,\delta(1-z) \ , \qquad
\eeqn
are the gluon splitting functions into $q\bar q$ and $GG$, respectively.
The subindex ``+'' in the $1/(1-z)_+$ factor indicates that the $z=1$
divergence disappears through
\bel{eq:ir_plus}
\int_0^1 dz\, f(z)\, [g(z)]_+ \,\equiv\,
\int_0^1 dz\, [f(z)-f(1)]\, g(z) \ .
\ee

\subsection{Moments of the structure functions}
\label{subsec:moments}

The previous discussion has been based on rather qualitative arguments.
Nevertheless, the predicted evolution equation can be derived on a more
rigorous basis using the formal framework of the operator product
expansion \cite{WI:69}, which allows to make a full QCD analysis of the
moments
\bel{eq:moments}
M^q_N(Q^2) \equiv \int_0^1 dx\, x^{N-1}\, q(x,Q^2) \ ; \qquad
M^G_N(Q^2) \equiv \int_0^1 dx\, x^{N-1}\, G(x,Q^2) \ .
\ee
Taking moments on both sides of Eq.~\eqn{eq:evol_q_g}, one finds
\bel{eq:mom_evol}
Q^2 {d\over dQ^2}
\left( \ba M^q_N(Q^2) \\ M^G_N(Q^2) \ea\right) =
{\alpha_s(Q^2)\over 2\pi} \left[
\bat \gamma_{qq}^N & \gamma_{qG}^N \\ \gamma_{Gq}^N & \gamma_{GG}^N \ea\right]
\left( \ba M^q_N(Q^2) \\ M^G_N(Q^2) \ea\right) \ ,
\ee
where
\bel{eq:anom}
\gamma_{ij}^N \,\equiv\, \int_0^1 dz \, z^{N-1}\, P_{ij}(z) \ ,
\ee
are the so-called anomalous dimensions.
Performing the trivial integrals, one gets:
\beqn\label{eq:anom_res}
\gamma_{qq}^N &\!\!\! = &\!\!\! C_F\,\left[-\frac{1}{2} +
  {1\over N(N+1)} - 2\sum_{k=2}^N {1\over k} \right] \ ,
\no\\
\gamma_{qG}^N &\!\!\! = &\!\!\! T_F\, {2 +N+N^2\over N(N+1)(N+2)}\ ,
\no\\
\gamma_{Gq}^N &\!\!\! = &\!\!\! C_F\, {2 +N+N^2\over N(N^2-1)}\ ,
\\
\gamma_{GG}^N &\!\!\! = &\!\!\! 2C_A \,\left[ -\frac{1}{12} +
 {1\over N(N-1)} + {1\over (N+1)(N+2)} - \sum_{k=2}^N {1\over k} \right] \ .
\no
\eeqn

For a non-singlet structure function, where the gluon component is absent,
the evolution differential equation leads to the solution
\bel{eq:dol_evol}
M_N^{q,\!\!\mbox{\rms ns}}(Q^2)  =  M_N^{q,\!\!\mbox{\rms ns}}(Q_0^2) \,
\left( {\alpha_s(Q^2)\over\alpha_s(Q_0^2)}\right)^{d_N} \ ;
\qquad\quad
d_N \equiv \gamma^N_{qq}/\beta_1 = {-6\gamma^N_{qq}\over 33 -2 N_f} \ .
\ee
The first moment has $d_1=0$; therefore, the Gottfried sum rule
\eqn{eq:gottfried} does not get any QCD correction at this leading order.
For $N\geq 2$, $d_N>0$ so that
$M_N^{q,\!\!\mbox{\rms ns}}(Q^2)$ decreases as $Q^2$ increases, indicating
a degradation of momentum in the non-singlet quark distribution.

Let us now consider the flavour-singlet structure function
$\Sigma(x)\equiv\sum_i\left[ q_i(x) + \bar q_i(x)\right]$.
The $N=2$ moments $M_2^\Sigma(Q^2)$ and $M_2^G(Q^2)$
give the average total fraction of
momentum carried by quarks and gluons, respectively.
The corresponding coupled evolution equations can be easily solved.
The sum of both moments does not depend on $Q^2$, since the total
momentum is conserved:
\bel{eq:mom_consv}
M_2^\Sigma(Q^2) + M_2^G(Q^2) \, = \, 1 \ .
\ee
The evolution of the $N=2$ singlet distribution then takes the simple form
\bel{eq:singlet_evol}
{M_2^\Sigma(Q^2) -{N_f\over 4C_F} M_2^G(Q^2) \over
M_2^\Sigma(Q_0^2) -{N_f\over 4C_F} M_2^G(Q_0^2)} \, =\,
{M_2^\Sigma(Q^2) -{3N_f\over 16+3N_f} \over
M_2^\Sigma(Q_0^2) -{3N_f\over 16+3N_f}} \, =\,
\left( {\alpha_s(Q^2)\over\alpha_s(Q_0^2)}\right)^{d_2^\Sigma} \ ;
\ee
with $d_2^\Sigma = 2(4C_F +N_f)/(33-2N_f)$.
If $N_f<16$, $d_2^\Sigma>0$ and the right-hand side will decrease for
increasing $Q^2$.
Thus, one gets a prediction for the asymptotic values of the
average total momentum carried by quarks and gluons:
\bel{eq:asympt_mom}
\lim_{Q^2\to\infty} M_2^\Sigma(Q^2) \, = \, {3N_f\over 16+3N_f} \ ;
\qquad
\lim_{Q^2\to\infty} M_2^G(Q^2) \, = \, {16\over 16+3N_f} \ .
\ee
For $N_f=4$, this gives $\frac{3}{7}$ and $\frac{4}{7}$, in good
agreement with the empirical observation that for $Q^2$ in the
range 10-40 GeV$^2$ each fraction is very close to $\frac{1}{2}$.

A very interesting issue is the behaviour of the parton distributions
at the end-points $x=0$ and $x=1$. The large $N$ moments probe the
$x\to 1$ region, while the low $x$ behaviour is controlled by
the $N\to 1$ limit.
As $N$ increases, $\gamma^N_{qG}$ and $\gamma^N_{Gq}$ tend to zero,
so that the evolution equations \eqn{eq:mom_evol} decouple;
i.e. the large $x$ behaviour of the quarks is independent of the
gluon evolution.
When $x\to 1$ the gluon distribution function approach zero more
rapidly than the quark ones. For large values of $x$ the quark
content of the nucleon is the relevant one.
Notice that $x=1$ means $W^2=M_p^2$, i.e. it actually corresponds
to the elastic photon--nucleon scattering.

At low $x$, $x/y\to 0$ and the splitting functions $P_{GG}(x/y)$
and $P_{Gq}(x/y)$ diverge.
The gluon distribution function becomes then dominant.
The low $x$ behaviour is controlled by the singular $N\to 1$ limit
of the gluon anomalous
dimension $\gamma^N_{GG}\sim 2 C_A/(N-1)$.
Making a saddle-point approximation, the $N\to 1$ moment can be
inverted; one finds in this way that for low $x$ the gluon distribution
function behaves as
\bel{eq:g_lx}
G(x) \, \sim\, {1\over x} \exp\sqrt{C(Q^2)\;\ln{1\over x}} \ ,
\ee
with $C(Q^2)$ a calculable function.
Obviously, this behaviour cannot be true for arbitrarily small $x$;
something must stop the growing of the gluon distribution before
running into unitarity problems.

\begin{figure}[thb]
\centerline{\epsfxsize = 10cm \epsfbox{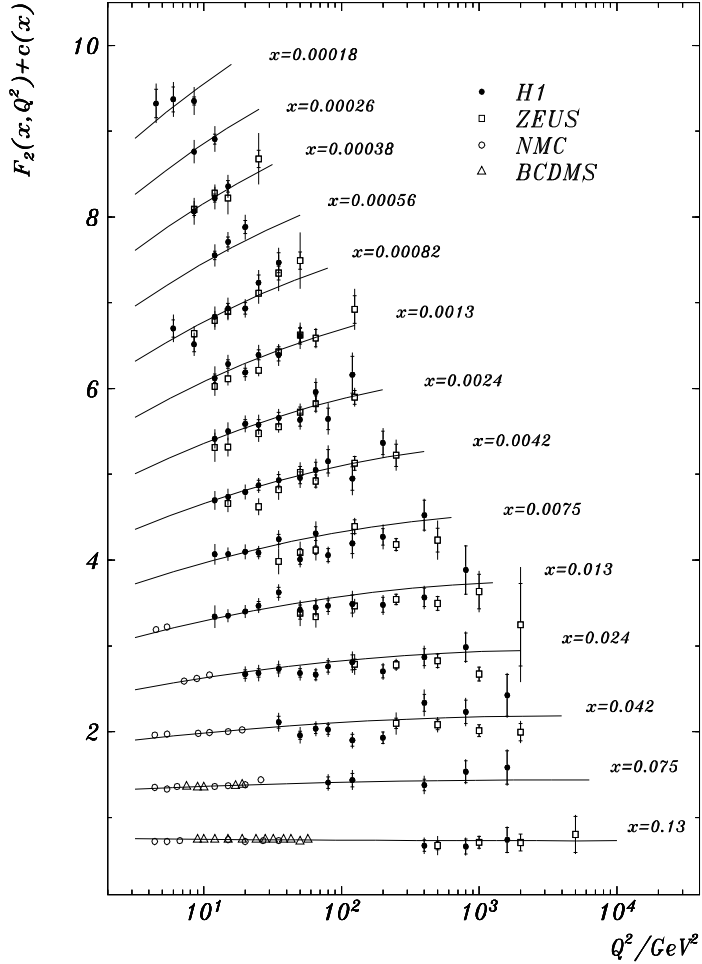}}
\caption{Recent measurements
of $F^{ep}_2(x,Q^2)$ \protect\cite{H1:95,ZEUS:95,NMC:92,BCDMS:90}.
The $F^{ep}_2(x,Q^2)$ values are plotted with all but normalization errors
in a linear scale adding a term $c(x)=0.6 (i_x-0.4)$ to $F_2$,
where $i_x$ is the bin number starting at $i_x=1$ for $x=0.13$.
The curves represent a phenomenological fit to the data.
(Taken from Ref.~\protect\cite{H1:95}).}
\label{fig:F2Q2}
\end{figure}

Kinematically, low $x$ means the high-energy (high $W^2$) limit for
the virtual photon--nucleon scattering.
The $e$-$p$ HERA collider is ideally suited for studying this region.
The HERA experiments extend the previously accessible kinematic range
up to very large squared momentum transfers, $Q^2>5\times 10^4$ GeV$^2$,
and to very small values of $x<10^{-4}$.
The measurements reported so far \cite{H1:95,ZEUS:95}
observe indeed a significant rise
of the structure function $F^{ep}_2(x,Q^2)$ with decreasing $x$,
at fixed $Q^2$.
Around $x\sim 10^{-3}$ the decrease of $x$ by an order of magnitude
amounts to a rise of $F^{ep}_2(x,Q^2)$ of about a factor of two.
The observed $Q^2$ behaviour is consistent with the expected scaling
violations, i.e. a weak rise of $F^{ep}_2(x,Q^2)$ with increasing
$Q^2$ for $x<0.1$.
The most recent data \cite{H1:95,ZEUS:95,NMC:92,BCDMS:90}
on the proton structure function
$F^{ep}_2(x,Q^2)$ are shown in Figs.~\ref{fig:F2Q2} and \ref{fig:F2x}.

\begin{figure}[htb]
\centerline{\epsfxsize = 14cm \epsfbox{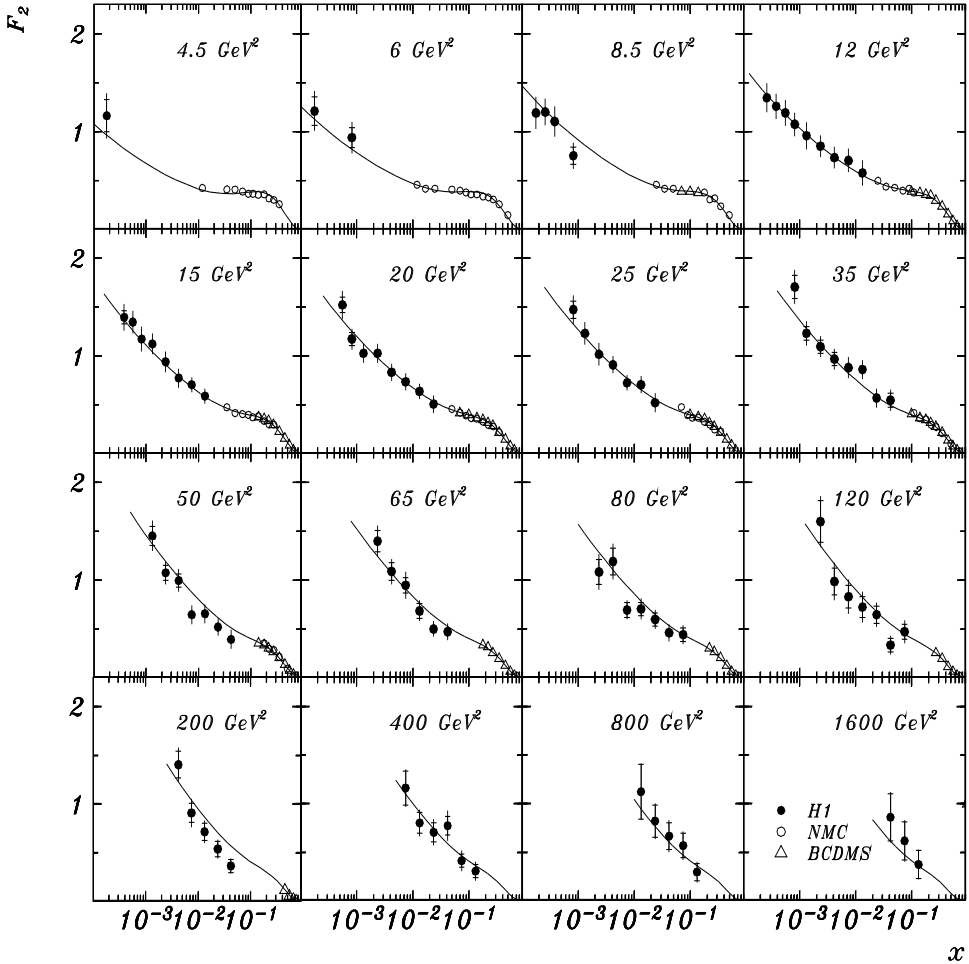}} \vspace{-1cm}
\caption{$x$ dependence of the measured structure function
$F_2^{ep}(x,Q^2)$, for different $Q^2$ values
\protect\cite{H1:95,NMC:92,BCDMS:90}.
The curves represent a phenomenological fit to the data.
(Taken from Ref.~\protect\cite{H1:95}).}
\label{fig:F2x}
\end{figure}

\subsection{QCD fits to DIS data}

There is a twofold motivation for making careful analyses of DIS data.
First, the experimental measurement of the parton distributions provides
very valuable information on the non-perturbative regime of the
strong interactions (in addition, these parton distributions are
needed for making predictions of hard-scattering processes in
hadronic collisions). Second, the measured $Q^2$ evolution (the slopes
of the distributions) can be compared with perturbative QCD predictions.

Usually, one adopts some motivated parametrization of the quark and
gluon distributions at a fixed momentum-transfer $Q_0^2$. The
evolution equations are then used to get the proton (or neutron)
structure functions at arbitrary values of $Q^2$, and
a global fit to the data is performed.

In the actual analysis one needs to worry about the unavoidable presence
of additional non-perturbative contributions. The perturbative
evolution equations can only predict the leading logarithmic dependence of the
distribution functions with $Q^2$. These distributions
have in addition uncalculable non-perturbative corrections
suppressed by inverse powers of $Q^2$,
the so-called {\it higher-twist} contributions:
\bel{eq:higher-twists}
F_i(x,Q^2) = F^{\!\!\mbox{\rms LT}}_i(x,Q^2) +
{F^{\!\!\mbox{\rms HT}}_i(x,Q^2)\over Q^2} + \cdots
\ee
The leading-twist term (LT) is the one predicted by perturbative QCD.
Since the additional $1/(Q^2)^n$ dependences have to be fitted from the
data, they increase the final uncertainties.
These corrections are numerically important for
$Q^2<\cO(10 \;\mbox{\rm GeV}^2)$ and for $x$ close to 1.
Obviously, the
perturbative QCD predictions can be better tested at large $Q^2$,
where the higher-twist effects are smaller.

Since the singlet structure functions are sensitive to the gluon distribution,
which is badly known, they suffer from rather large errors. Good data
at low values of $x$ is needed in order to perform an accurate determination.
The HERA experiments are making
an important improvement in the knowledge of
these distributions.
The latest fits \cite{MSR:95}, including the most recent HERA data,
obtain gluon and sea-quark distributions at small $x$ which are
significantly different from those in previous standard sets of
parton distributions.
The new gluon distribution is larger for $x\leq 0.01$ and smaller
for $x\sim 0.1$.
The reduction of the gluon distribution in the interval
$x\sim 0.1-0.2$ is compensated by an increase in the fitted
value of $\alpha_s$ \cite{MSR:95}, bringing the DIS determination
\cite{VM:92,MSR:95}
\bel{alpha_s_DIS}
\alpha_s(M_Z^2) \, =\, 0.114\pm 0.005
\ee
in better agreement with the
world average values, which we discuss in the next section.

\setcounter{equation}{0}
\section{DETERMINATION OF THE STRONG COUPLING}

In the massless quark limit, QCD has only one free parameter: the
strong coupling $\alpha_s$. Thus, all strong interaction
phenomena should be described in terms of this single input.
The measurements of $\alpha_s$ at different processes and at
different mass scales provide then a crucial test of QCD:
if QCD is the right theory
of the strong interactions, all measured observables should lead
to the same coupling.

\begin{figure}[hbt]
\centerline{\mbox{\epsfysize=10.0cm\epsffile{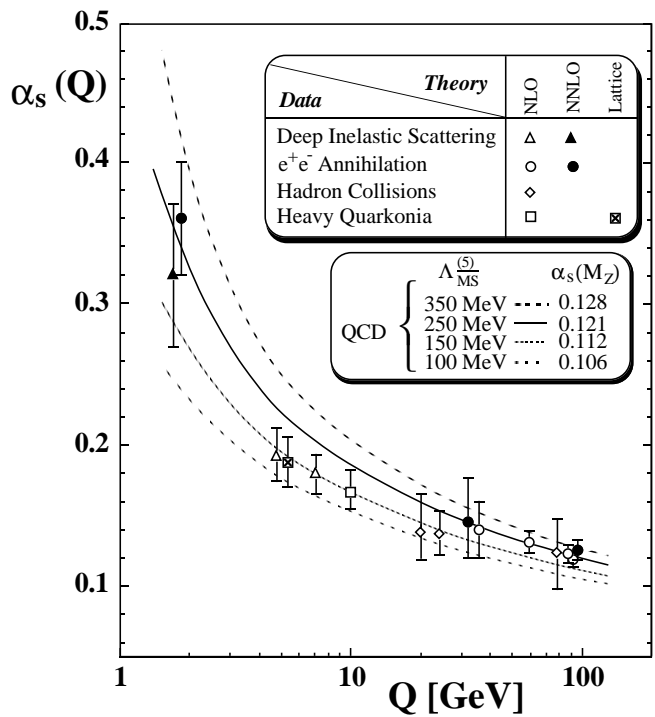}}}
\vspace{-0.3cm}
\caption{Compilation of $\alpha_s$ measurements as function of the energy scale
\protect\cite{BE:94,WE:94}.}
\label{fig:alpha_run}
\end{figure}

\begin{figure}[thb]
\centerline{\epsfysize = 13cm \epsfbox{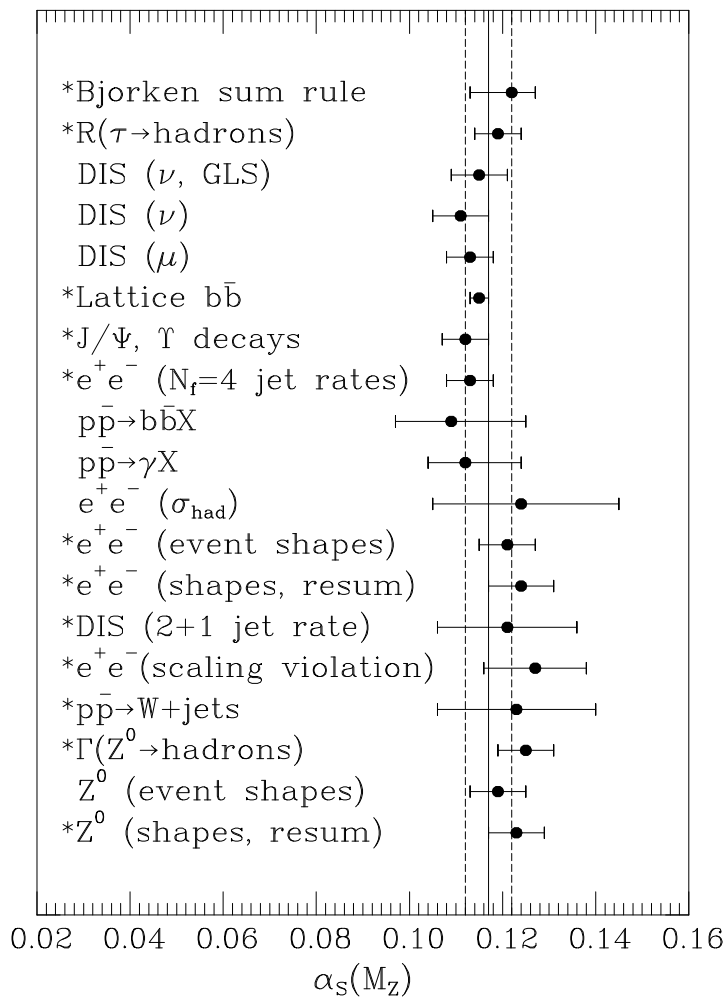}}
\caption{Summary \protect\cite{WE:94}
of $\alpha_s$ measurements, evolved to the scale $M_Z$.
Starred items include preliminary results.}
\label{fig:alpha_Z}
\end{figure}

Obviously, the test should be restricted to those processes
where perturbative techniques are reliable. Moreover, the same
definition of $\alpha_s$ 
should be taken
everywhere; the $\overline{\mbox{\rm MS}}$ scheme is usually adopted as
the standard convention.
Since the running coupling is a function of energy,
one can either compare the different determinations at the
different scales where they are measured, checking in this way
the predicted $Q^2$ dependence of the coupling, or
use this prediction to bring all measurements to a common
reference scale where they are compared. Nowadays, the $Z$-mass
scale is conventionally chosen for such a comparison.

In order to assess the significance of the test, it is very important
to have a good understanding of the uncertainties associated
with the different measurements. This is not an easy question,
because small non-perturbative effects can be present in many
observables. In addition, some quantities have been computed
to a very good perturbative accuracy (next-to-next-to-leading order),
while others are only known at the leading or next-to-leading order;
the resulting values of $\alpha_s$ refer then to different
perturbative approximations.
The estimate of theoretical uncertainties is also affected by the
plausible asymptotic (i.e. not convergent) behaviour of the
perturbative series in powers of $\alpha_s$. Although this is a
common problem of Quantum Field Theories, it is probably more
severe in QCD because the coupling is rather big (at usual energies).

Fig.~\ref{fig:alpha_run} summarizes \cite{BE:94,WE:94}
the most reliable measurements
of the strong coupling as function of the energy scale. The
agreement with the predicted running of $\alpha_s$,
indicated by the curves, is indeed very good.
The value of $\alpha_s(m_\tau^2)$, extracted from the hadronic width of the
$\tau$ lepton, provides a very important low-energy measurement;
although it has a rather
large relative error, it implies a very precise prediction at the  $M_Z$
scale, which is in excellent agreement with the direct determinations
of $\alpha_s(M_Z^2)$ performed at the $Z$ peak.
Fig.~\ref{fig:alpha_Z} \cite{WE:94}
compares the different measurements at the
common reference scale $M_Z$.
The average af all determinations gives \cite{BE:94,WE:94}:
\bel{eq:alpha_average}
\alpha_s(M_Z^2) \, = 0.117 \pm 0.005 .
\ee

\setcounter{equation}{0}
\section{CHIRAL SYMMETRY}
\label{sec:chpt}

Up to now, we have only discussed
those aspects of QCD which can be analyzed in a perturbative way.
Thus, we have restricted ourselves to the study of scattering processes at
large momentum transfers, and inclusive transitions
which avoid the hadronization problems.
The rich variety of strong-interacting phenomena governed by the confinement
regime of QCD has been completely ignored.

There are certainly many approximate tools to investigate
particular aspects of non-perturbative physics;
however, rigorous first-principle QCD calculations seem
unfortunately out of reach for present techniques.
Nevertheless, we can still investigate some general properties of QCD
using symmetry considerations.

\subsection{Flavour symmetries}

In order to build the QCD Lagrangian,
we made extensive use of the $SU(3)_C$
colour symmetry, which is the basis of
the strong interaction dynamics. The Lagrangian \eqn{eq:L_QCD}
has additional {\it global} symmetries associated with the quark
flavour numbers:
\begin{enumerate}
\item $\cL_{\!\!\mbox{\rms QCD}}$ is invariant under a global phase
redefinition of all quark flavours,
\be\label{eq:baryon_number}
q_f \, \longrightarrow\, \exp(i\theta)\, q_f \ .
\ee
This symmetry is associated with the conservation of the baryon number.
\item $\cL_{\!\!\mbox{\rms QCD}}$ is also invariant
under independent phase redefinitions
of the different quark flavours,
\bel{eq:flavour_number}
q_f \, \longrightarrow\, \exp(i\theta_f)\, q_f \ .
\ee
This symmetry implies the conservation of flavour.
\item For equal quark masses, there is a larger symmetry
under $SU(N_f)$ transformations in flavour space,
\bel{eq:isospin}
q_f \,\longrightarrow\, U_{ff'}\, q_{f'} \ , \qquad\qquad U\in SU(N_f) \ .
\ee
This is a good symmetry of the light-flavour sector ($u$, $d$, $s$), where
quark masses can be ignored in first approximation. One has then the
well-known isospin ($N_f=2$) and $SU(3)$ symmetries.
\item
In the absence of quark masses, the QCD Lagrangian splits into
two independent quark sectors,
\bel{eq:LR_sectors}
\cL_{\!\!\mbox{\rms QCD}}
\,\equiv\, -{1\over 4}\, G^{\mu\nu}_aG_{\mu\nu}^a
+ i\bar q_L \gamma^\mu D_\mu q_L  +
i\bar q_R \gamma^\mu D_\mu q_R \ .
\ee
Here, $q$ denotes the flavour (and colour) vector \
$q = \mbox{\rm column}(u,d,\ldots)$, and $L,R$ stand for the
left- and right-handed components of the quarks.
Thus, the two quark chiralities live in separate
flavour spaces which do not talk each other (gluon interactions
do not change the chirality), implying that all previous flavour
symmetries get duplicated in the two chiral sectors.
\end{enumerate}

The baryon number symmetry \eqn{eq:baryon_number} is usually called $U(1)_V$,
since both chiralities transform in the same way.
Its chiral replication is the corresponding $U(1)_A$ transformation:
\bel{eq:U1_A}
q_L \, \longrightarrow\, \exp(-i\theta)\, q_L \ ; \qquad\qquad
q_R \, \longrightarrow\, \exp(i\theta)\, q_R \ .
\ee
This symmetry of the classical (massless) QCD Lagrangian gets broken by
quantum effects (triangular loops of the type shown in Fig.~\ref{fig:triangle},
with gluons instead of photons); this is the so-called $U(1)_A$ anomaly.
Although \eqn{eq:U1_A} is not a true symmetry of QCD,
it gets broken in a very
specific way, which leads to important implications.
A discussion of the phenomenological role of anomalies is beyond the
scope of these lectures. However, let me mention that this anomaly is
deeply related to interesting low-energy phenomena such as the
understanding of the $\eta'$ mass, or
the so-called {\it proton spin crisis}\footnote{\small
This is a quite unfortunate name, because:
1) the underlying QCD dynamics has little to do
with the parton-model description  of the proton spin;
and 2) it is certainly not a crisis but rather
a success of QCD.
The failure of the naive quark-model
description of an observable where gluons
are predicted to play a crucial role (the anomaly), is indeed a clear
experimental confirmation of the QCD dynamics.}.

I want to concentrate here in the chiral extension of the old eightfold
$SU(3)_V$ symmetry, i.e. in the global
$G\equiv SU(3)_L\otimes SU(3)_R$ symmetry
of the QCD Lagrangian for massless $u$, $d$ and $s$ quarks.
This larger symmetry is not directly seen in the hadronic spectrum.
Although hadrons can be nicely classified in $SU(3)_V$ representations,
degenerate multiplets with opposite parity do not exist.
Moreover, the octet of pseudoscalar mesons
($\pi$,$K$,$\eta$) happens to be much lighter than all other
hadronic states.

There are two different ways in which a symmetry of the Lagrangian can
be realized. In the usual one (Wigner--Weyl), the ground state (the vacuum)
is also invariant. Then, all physical states can be classified in
irreducible representations of the symmetry group \cite{CO:66}.
Certainly, the hadronic spectrum does not look like that, in the case
of the chiral group.

There is a second (Nambu--Golstone),
more sophisticated, way to realize a symmetry. In
some cases, the vacuum is not symmetric. The hadronic spectrum
corresponds to energy excitations over the physical vacuum and, therefore,
will not manifest the original symmetry of the Lagrangian.
However, Goldstone's theorem \cite{GO:61} says that
in such a case there should appear a massless scalar for each
broken generator of the original symmetry group.
If the chiral symmetry is realized in this way, there should be
eight pseudoscalar massless states (Goldstone bosons)
in the hadronic spectrum; this is precisely the number of states
of the lightest hadronic multiplet: the $0^-$ octet.
Thus, we can identify the $\pi$, $K$ and $\eta$ with the Goldstone
modes of QCD; their small masses being generated by the
quark-mass matrix which explicitly breaks the global chiral symmetry
of the Lagrangian.

In the Standard electroweak model, the local
$SU(2)_L\otimes U(1)_Y$
symmetry is also realized in
the Nambu--Goldstone way. There, the symmetry-breaking phenomena is
assumed to be related to the existence of some scalar multiplet which gets
a vacuum expectation value. Since a {\it local} symmetry gets
(spontaneously) broken in that case, the Goldstone modes combine
with the gauge bosons giving massive spin-1 states plus the Higgs particle.
The QCD case is simpler, because it is a {\it global} symmetry the
one which gets broken. However, something should play the role of the
electroweak scalar field. Since quarks are the only fields carrying
flavour, they should be responsible for the symmetry breaking. The simplest
possibility is the appearance of a quark condensate
\bel{eq:quark_condensate}
v\equiv\langle 0| \bar u u |0\rangle =
\langle 0| \bar d d |0\rangle =
\langle 0| \bar s s |0\rangle < 0 \ ,
\ee
generated by the non-perturbative QCD dynamics.
This would produce a dynamical breaking
of chiral symmetry, keeping at the same time the observed
$SU(3)_V$ symmetry.

\subsection{Effective Chiral Lagrangian}

The Goldstone nature of the pseudoscalar mesons implies strong
constraints on their interactions, which can be most easily analyzed
on the basis of an effective Lagrangian.
The Goldstone bosons correspond to the zero-energy excitations
over the quark condensate; their fields can be collected in a
$3\times 3$ unitary matrix $U(\phi)$,
\be
\langle 0| \bar q^j_L q^i_R|0\rangle \, \longrightarrow \, {v\over 2}\,
U^{ij}(\phi) ,
\ee
which parametrizes those excitations.
A convenient parametrization is given by
\be
U(\phi)  \equiv \exp{\left(i \sqrt{2} \Phi / f\right)} ,
\ee
where
\be
\ba
\Phi (x) \equiv
\frac{\dis \vec{\lambda}}{\dis \sqrt 2} \, \vec{\phi}
 = \, \left( \begin{array}{ccc}
\frac{\dis \pi^0}{\dis \sqrt 2} \, + \, \frac{\dis \eta_8}{\dis \sqrt 6}
 & \pi^+ & K^+ \\
\pi^- & - \frac{\dis \pi^0}{\dis \sqrt 2} \, + \, \frac{\dis \eta_8}
{\dis \sqrt 6}    & K^0 \\
K^- & \bar K^0 & - \frac{\dis 2 \, \eta_8}{\dis \sqrt 6}
\end{array}  \right) .
\ea
\ee
The matrix
$U(\phi)$ transforms linearly under the chiral group, [$g_{L,R}\in
SU(3)_{L,R}$]
\be\label{eq:utransf}
q_L\,\stackrel{G}{\longrightarrow}\, g_L\, q_L ,\quad
q_R\,\stackrel{G}{\longrightarrow}\, g_R\, q_R \qquad
\Longrightarrow\qquad
U(\phi) \, \stackrel{G}{\longrightarrow}\, g_R \, U(\phi) \, g_L^\dagger \; ,
\ee
but the induced transformation on the Goldstone fields
$\vec{\phi}$ is highly non-linear.

Since there is a mass gap separating the pseudoscalar octet
from the rest of the hadronic spectrum,
we can build a low-energy  effective field theory
containing only the Goldstone modes.
We should write the most general Lagrangian involving the matrix
$U(\phi)$, which is consistent with chiral symmetry.
Moreover, we can
organize the Lagrangian in terms of increasing powers of
momentum or, equivalently, in terms of an increasing number of
derivatives (parity conservation requires an even number of derivatives):
\be
{\cal L}_{\!\!\hbox{\rms eff}}(U) \, = \, \sum_n \, {\cal L}_{2n} \ .
\ee
In the low-energy domain, the
terms with a minimum number of derivatives will dominate.

Due to the unitarity of the $U$ matrix, $U U^\dagger = 1$, at least
two derivatives are required to generate a non-trivial interaction.
To lowest order, the effective chiral Lagrangian is uniquely
given by the term
\be\label{eq:l2}
{\cal L}_2 = {f^2\over 4}\,
\mbox{\rm Tr}\left[\partial_\mu U^\dagger \partial^\mu U \right] .
\ee

Expanding $U(\phi)$ in a power series in $\Phi$, one obtains the
Goldstone's kinetic terms plus a tower of interactions involving
an increasing number of pseudoscalars.
The requirement that the kinetic terms are properly normalized
fixes the global coefficient $f^2/4$ in (\ref{eq:l2}).
All interactions among the Goldstones can then be predicted in terms
of the single coupling $f$:
\be
{\cal L}_2 \, = \, {1\over 2} \,\mbox{\rm Tr}\left[\partial_\mu\Phi
\partial^\mu\Phi\right]
\, + \, {1\over 12 f^2} \,\mbox{\rm Tr}\left[
(\Phi\stackrel{\leftrightarrow}{\partial}_{\!\mu}\Phi) \,
(\Phi\stackrel{\leftrightarrow}{\partial^\mu}\Phi)
\right] \, + \, \cO(\Phi^6/f^4) .
\ee

To compute the $\pi\pi$ scattering amplitude, for instance, is now
a trivial perturbative exercise. One gets the well-known \cite{WE:66}
Weinberg result  [$t\equiv (p_+' - p_+)^2$]
\be\label{eq:WE1}
T(\pi^+\pi^0\to\pi^+\pi^0) = {t/ f^2}.
\ee
Similar results can be obtained for $\pi\pi\to 4\pi, 6\pi, 8\pi, \ldots$\,
The non-linearity of the effective Lagrangian relates
amplitudes with different numbers of Goldstone bosons, allowing
for absolute predictions in terms of $f$.
Notice that the Goldstone interactions are proportional to their
momenta (derivative couplings).
Thus, in the zero-momentum limit, pions become free.
In spite of confinement, QCD has a weakly-interacting regime at low
energies, where a perturbative expansion in powers of momenta can be applied.

It is straightforward to generalize the effective Lagrangian
(\ref{eq:l2}) to incorporate electromagnetic and semileptonic
weak interactions. One learns then that $f$ is in fact
the pion-decay constant $f\approx f_\pi=92.4$ MeV, measured in
$\pi\to\mu\nu_\mu$ decay \cite{PI:95}.
The corrections induced by the non-zero quark
masses are taken into account through the term
\bel{eq:L_m}
\cL_m \, = \, {|v|\over 2}\, \mbox{\rm Tr}\left[\cM (U + U^\dagger)\right]
\ , \qquad \cM\equiv\mbox{\rm diag}(m_u,m_d,m_s) \ ,
\ee
which breaks chiral symmetry in exactly the same way as
the quark-mass term in the underlying QCD Lagrangian does.
Eq.~\eqn{eq:L_m}
gives rise to a quadratic pseudoscalar-mass term plus
additional interactions proportional to the quark masses.
Expanding in powers of $\Phi$
(and dropping an irrelevant constant), one has
\be\label{eq:massterm}
\cL_m \, = \,
 |v|\,\left\{ -{1\over f^2} \mbox{\rm Tr}\left[{\cal M}\Phi^2\right]
+ {1\over 6 f^4} \mbox{\rm Tr}\left[ {\cal M} \Phi^4\right]
+ \cO(\Phi^6/f^6) \right\} .
\ee

The explicit evaluation of the trace in the quadratic mass term provides
the relation between the physical meson masses and the quark masses:
\beqn\label{eq:masses}
M_{\pi^\pm}^2 & = & (m_u+m_d) {|v|\over f^2}\ , \nonumber\\
M_{\pi^0}^2 & = & (m_u+m_d) {|v|\over f^2} - \varepsilon +
\cO(\varepsilon^2)\ , \nonumber\\
M_{K^\pm}^2 & = & (m_u + m_s) {|v|\over f^2}\ , \\
M_{K^0}^2 & = & (m_d + m_s) {|v|\over f^2}\ , \nonumber\\
M_{\eta_8}^2 & = & {1\over 3} (m_u+m_d + 4 m_s)  {|v|\over f^2} + \varepsilon +
\cO(\varepsilon^2)\ , \no
\eeqn
where
\be
\varepsilon = {|v|\over 2f^2}\ {(m_u - m_d)^2\over  (2m_s - m_u-m_d)} \ .
\ee
Chiral symmetry relates the magnitude of the meson and quark masses
to the size of the quark condensate.
Taking out the common $|v|/f^2$ factor, Eqs.~(\ref{eq:masses}) imply
the old Current Algebra mass ratios,
\be\label{eq:mratios}
{M^2_{\pi^\pm}\over m_u+m_d} = {M^2_{K^+}\over (m_u+m_s)} =
{M_{K^0}\over (m_d+m_s)}
\approx {3 M^2_{\eta_8}\over (m_u+m_d + 4 m_s)} \ ,
\ee
and
[up to $\cO(m_u-m_d)$ corrections]
the Gell-Mann--Okubo mass relation
\be
3 M^2_{\eta_8} = 4 M_K^2 - M_\pi^2 \ .
\ee
%

Although chiral symmetry alone cannot fix the absolute values
of the quark masses, it gives information about quark-mass
ratios. Neglecting the tiny $\cO(\varepsilon)$ effects,
one gets the relations
\bel{eq:ratio1}
{m_d - m_u \over m_d + m_u} \, =\,
{(M_{K^0}^2 - M_{K^+}^2) - (M_{\pi^0}^2 - M_{\pi^+}^2)\over M_{\pi^0}^2}
\, = \, 0.29 \ ,
\ee
\bel{eq:ratio2}
{2 m_s -m_u-m_d\over 2 (m_u+m_d)} \, = \,
{M_{K^0}^2 - M_{\pi^0}^2\over M_{\pi^0}^2}
\, = \, 12.6 \ .
\ee
In (\ref{eq:ratio1}) we have subtracted the pion square-mass
difference, to take into account the electromagnetic contribution
to the pseudoscalar-meson self-energies;
in the chiral limit ($m_u=m_d=m_s=0$), this contribution is proportional
to the square of the meson charge and it is the same for $K^+$ and $\pi^+$.
The mass formulae (\ref{eq:ratio1}) and (\ref{eq:ratio2})
imply the quark-mass ratios advocated by Weinberg:
\be\label{eq:Weinbergratios}
m_u : m_d : m_s = 0.55 : 1 : 20.3 \ .
\ee
Quark-mass corrections are therefore dominated by $m_s$, which is
large compared with $m_u$ and $m_d$.
Notice that the difference $m_d-m_u$ is not small compared with
the individual up- and down-quark masses; in spite of that,
isospin turns out
to be an extremely good symmetry, because
isospin-breaking effects are governed by the small ratio
$(m_d-m_u)/m_s$.

The $\Phi^4$ interactions in (\ref{eq:massterm})
introduce mass corrections to the $\pi\pi$ scattering amplitude
(\ref{eq:WE1}),
\be\label{eq:WE2}
T(\pi^+\pi^0\to\pi^+\pi^0) = {t - M_\pi^2\over f_\pi^2} \ .
\ee
%
Since $f\approx f_\pi$ is fixed from pion decay, this result
is now an absolute prediction of chiral symmetry.

The lowest-order chiral Lagrangian encodes
in a very compact way all the Current Algebra results obtained in
the sixties \cite{currentalgebra}.
The nice feature of the chiral approach is its elegant
simplicity, which
allows to estimate higher-order corrections in a systematic way.
A detailed summary of the chiral techniques and their
phenomenological applications can be found in
Ref.~\cite{PI:95}.

\section{SUMMARY}

Strong interactions are characterized by three basic properties:
asymptotic freedom, confinement and dynamical chiral symmetry breaking.

Owing to the gluonic self-interactions, the QCD coupling becomes
smaller at short distances, leading indeed to an asymptotically-free
quantum field theory.
Perturbation theory can then be applied at large momentum transfers.
The resulting predictions have achieved a remarkable success,
explaining a wide range of phenomena in terms of a single coupling.
The running of $\alpha_s$ has been experimentally tested at
different energy scales, confirming the predicted QCD behaviour.

The growing of the running coupling at low-energies
makes very plausible
that the QCD dynamics generates the required confinement
of quarks and gluons into colour-singlet hadronic states.
A rigorous proof of this property is, however, still lacking.
At present, the dynamical details of hadronization are completely unknown.

Non-perturbative tools, such as QCD sum rules and lattice calculations,
provide indirect evidence that QCD also implies the proper pattern
of chiral symmetry breaking.
The results obtained so far support the existence of a non-zero
$q$-$\bar q$ condensate in the QCD vacuum, which dynamically breaks
the chiral symmetry of the Lagrangian.
However, a formal understanding of this
phenomena has only been achieved in some approximate limits.

Thus, we have at present an overwhelming experimental and theoretical
evidence that the $SU(3)_C$ gauge theory correctly describes the hadronic
world.
This makes QCD
the established theory of the strong interactions.
Nevertheless,
the non-perturbative nature of its low-energy limit
is still challenging our theoretical capabilities.

\section*{ACKNOWLEDGEMENTS}

I would like to thank the organizers for the charming atmosphere of this
school, and the students for their many interesting questions and
comments. I am also grateful to
B. Gavela, J. Fuster and A. Santamar\'{\i}a
for their help with the postscript figures.
This work has been supported in part by CICYT (Spain) under grant
No. AEN-93-0234.

\newpage

\renewcommand{\theequation}{A.\arabic{equation}}
\setcounter{equation}{0}
\section*{\boldmath APPENDIX A: $SU(N)$ ALGEBRA}

$SU(N)$ is the group of $N\times N$ unitary matrices,
$U U^\dagger = U^\dagger U =1$, with $\det U=1$.
The generators of the $SU(N)$ algebra,
$T^a$ ($a=1,2,\ldots,N^2-1$), are hermitian,
traceless matrices satisfying the commutation relations
\bel{eq:T_com}
[T^a, T^b] \, = \, i f^{abc}\, T^c \, ,
\ee
the structure constants $f^{abc}$ being real and totally antisymmetric.

The fundamental representation $T^a = \lambda^a/2$ is $N$-dimensional.
For $N=2$, $\lambda^a$ are the usual Pauli matrices, while
for $N=3$, they correspond to the eight Gell-Mann matrices:
\beqn\label{eq:GM_matrices}
\lambda^1 \! = \!\left( \bath 0 & 1 & 0 \\ 1 & 0 & 0 \\ 0 & 0 & 0 \ea \right) ,
\quad\,
\lambda^2 &\!\!\! \! = &\!\!\!\!
\left( \bath 0 & -i & 0 \\ i & 0 & 0 \\ 0 & 0 & 0 \ea \right) ,
\quad
\lambda^3 \! = \!\left( \bath 1 & 0 & 0 \\ 0 & -1 & 0 \\ 0 & 0 & 0 \ea \right)
,
\quad
\lambda^4 \! = \!\left( \bath 0 & 0 & 1 \\ 0 & 0 & 0 \\ 1 & 0 & 0 \ea \right) ,
 \no\\   && \\
\lambda^5 \! = \!
\left( \bath 0 & 0 & -i \\ 0 & 0 & 0 \\ i & 0 & 0 \ea \right) \! ,
\;\;
\lambda^6 &\!\!\!\! = &\!\!\!   
\left( \bath 0 & 0 & 0 \\ 0 & 0 & 1 \\ 0 & 1 & 0 \ea \right)\! , \;\;
\lambda^7 \! = \!\left( \bath 0 & 0 & 0 \\ 0 & 0 & -i \\ 0 & i & 0 \ea \right)
\! , \;\;
\lambda^8 \! = \!     
{1\over\sqrt{3}}\!
\left( \bath 1 & 0 & 0 \\ 0 & 1 & 0 \\ 0 & 0 & -2 \ea \right) \! . \no
\eeqn
They satisfy the anticommutation relation
\bel{eq:anticom}
\left\{\lambda^a,\lambda^b\right\} \, = \,
{4\over N} \,\delta^{ab} \, I_N \,
+ \, 2 d^{abc} \, \lambda^c \, ,
\ee
where $I_N$ denotes the $N$-dimensional unit matrix and the constants
$d^{abc}$ are totally symmetric in the three indices.

For $SU(3)$, the only non-zero (up to permutations)
$f^{abc}$ and $d^{abc}$ constants are
\beqn\label{eq:constants}
&&{1\over 2} f^{123} = f^{147} = - f^{156} = f^{246} = f^{257} = f^{345}
= - f^{367} = {1\over\sqrt{3}} f^{458} = {1\over\sqrt{3}} f^{678} =
{1\over 2}\, , \hfil
\no\\
&&d^{146} = d^{157} = -d^{247} = d^{256} = d^{344} =
d^{355} = -d^{366} = - d^{377} = {1\over 2}\, ,
\\
&&d^{118} = d^{228} = d^{338} = -2 d^{448} = -2 d^{558} = -2 d^{688} =
-2 d^{788} = -d^{888} = {1\over \sqrt{3}}\, . \qquad\qquad
\no
\eeqn

The adjoint representation of the $SU(N)$ group is given by
the $(N^2-1)\!\times\! (N^2-1)$ matrices
$(T^a_A)_{bc} \equiv - i f^{abc}$.
The relations
\beqn\label{eq:invariants}
{\rm Tr}\left(\lambda^a\lambda^b\right) =  4 T_F
\, \delta_{ab}
\, , \qquad\qquad\quad && T_F = {1\over 2} \, ,
\no\\
\left(\lambda^a\lambda^a\right)_{\alpha\beta} = 4
C_F \, \delta_{\alpha\beta} \, , \qquad\qquad\quad
&& C_F = {N^2-1\over 2N} \, ,
\\ \;
{\rm Tr}(T^a_A T^b_A) = f^{acd} f^{bcd} = C_A \,\delta_{ab} \, , \quad
&& C_A = N \, , \qquad\no
\eeqn
define the $SU(N)$ invariants $T_F$, $C_F$ and $C_A$.
Other useful properties are:
\beqn
\left(\lambda^a\right)_{\alpha\beta}
\left(\lambda^a\right)_{\gamma\delta}
= 2 \delta_{\alpha\delta}\delta_{\beta\gamma}
 -{2\over N} \delta_{\alpha\beta}\delta_{\gamma\delta} \, ;
\qquad &&
{\rm Tr}\left(\lambda^a\lambda^b\lambda^c\right)
 =  2 (d^{abc} + i f^{abc})\, ;
\qquad\no\\ \;\;
{\rm Tr}(T^a_A T^b_A T^c_A)  =  i \, {N\over 2} f^{abc} \, ;
\qquad \sum_b d^{abb} = 0\, ; \quad &&
d^{abc} d^{ebc}  =  \left( N - {4\over N}\right) \delta_{ae} \, ;
\\
f^{abe} f^{cde} + f^{ace} f^{dbe} + f^{ade} f^{bce} = 0 \, ; \qquad\qquad &&
f^{abe} d^{cde} + f^{ace} d^{dbe} + f^{ade} d^{bce} = 0 \, . \qquad\quad\no
\eeqn

\newpage

\renewcommand{\theequation}{B.\arabic{equation}}
\setcounter{equation}{0}
\section*{APPENDIX B: GAUGE-FIXING AND GHOST FIELDS}


The fields $G^\mu_a$ have 4 Lorentz degrees of freedom, 
while a massless spin-1 gluon has 2 physical polarizations only.
Although gauge invariance makes the additional degrees of freedom irrelevant,
they give rise to some technical complications when quantizing the gauge
fields.

The canonical momentum associated with $G^\mu_a$,
$\Pi^a_\mu(x)\equiv
\delta\cL_{\!\!\mbox{\rms QCD}}/\delta (\partial_0G^\mu_a)
= G^a_{\mu 0}$,
vanishes identically for $\mu=0$.
The standard commutation relation
\bel{eq:quantization}
\left[ G^\mu_a(x), \Pi^\nu_b(y) \right] \delta(x^0-y^0) \, = \,
i g^{\mu\nu} \delta^{(4)}(x-y) \, ,
\ee
is then meaningless for $\mu=\nu=0$. In fact, the field $G^0_a$ is just a
classical quantity, since it commutes with all the other fields.
This is not surprising, since we know that there are 2 unphysical components
of the gluon field, which should not be quantized. Thus, we could just impose
two gauge conditions, such as $G^0_a = 0$ and $\vec\nabla \vec G_a = 0$,
to eliminate the 2 redundant degrees of freedom,
and proceed working with the physical gluon polarizations only. However,
this is a (Lorentz) non-covariant procedure, which leads to a very
awkward formalism. Instead, one can impose a Lorentz-invariant gauge
condition, such as $\partial_\mu G^\mu_a = 0$. The simplest way to
implement this is to add to the Lagrangian the gauge-fixing term
\bel{eq:L_GF}
\cL_{\!\!\mbox{\rms GF}} \, = \, -{1\over 2\xi} \, (\partial^\mu G_\mu^a)\,
 (\partial_\nu G^\nu_a)
\ee
where $\xi$ is the so-called gauge parameter.
The 4 Lorentz components of the canonical momentum
\bel{eq:canonical}
\Pi^a_\mu(x) \equiv {\delta\cL_{\!\!\mbox{\rms QCD}}\over\delta
(\partial_0 G^\mu_a)}
= G^a_{\mu 0} - {1\over \xi}\, g_{\mu 0}\, (\partial^\nu G_\nu^a)
\ee
are then non-zero, and one can develop a covariant quantization formalism.
Since \eqn{eq:L_GF} is a quadratic $G_a^\mu$ term, it modifies the gluon
propagator:
\bel{eq:propagator}
\langle 0 | T[G^\mu_a(x) G^\nu_b(y)]|0\rangle =
i \delta_{ab} \int {d^4k\over (2\pi)^4} {\e^{-ik(x-y)}\over k^2 + i\varepsilon}
\left\{ -g^{\mu\nu} + (1-\xi) {k^\mu k^\nu\over k^2 + i\varepsilon}\right\}
\, .
\ee
Notice, that the propagator is not defined for $\xi=\infty$, i.e. in
the absence of the gauge-fixing term \eqn{eq:L_GF}.

In QED, this gauge-fixing procedure is enough for making a consistent
quantization of the theory. The initial gauge symmetry of the Lagrangian
guarantees that the redundant photon polarizations do not generate
non-physical contributions to the scattering amplitudes, and the final
results are independent of the arbitrary gauge parameter $\xi$.
In non-abelian gauge theories, like QCD, a second problem still remains.

\begin{figure}[htb]
\centerline{\epsfysize =3.5cm \epsfbox{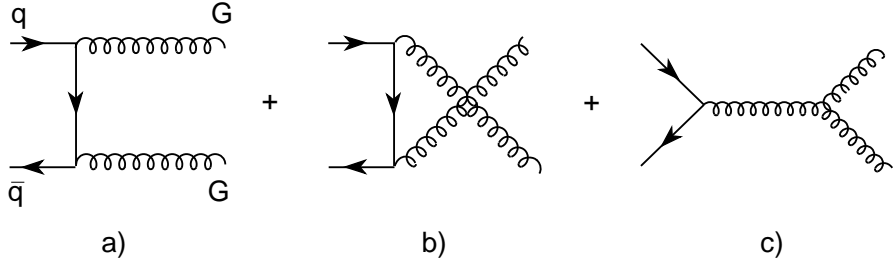}}
\vspace{-0.2cm}
\caption{Tree-level Feynman diagrams contributing to $q\bar q\to GG$.}
\label{fig:qqGG}
\end{figure}

Let us consider the scattering process $q\bar q\to G G$, which proceeds
through the three Feynman graphs shown in Fig.~\ref{fig:qqGG}. The scattering
amplitude has the general form
$T = J^{\mu\mu'} \varepsilon_{\mu\phantom{'}}^{(\lambda)}
\varepsilon_{\mu'}^{(\lambda')}$.
The probability associated with the scattering process
\bel{eq:prob}
\cP \sim {1\over 2} J^{\mu\mu'} (J^{\nu\nu'})^\dagger \,
\sum_\lambda \varepsilon_\mu^{(\lambda)}\varepsilon_\nu^{(\lambda)*}\,
\sum_{\lambda'} \varepsilon_{\mu'}^{(\lambda')}\varepsilon_{\nu'}^{(\lambda')*}
\ee
involves a sum over the final gluon polarizations.
One can easily check that the physical probability $\cP_T$, where only the
two transverse gluon polarizations are considered in the sum, is different
from the covariant quantity $\cP_C$, which includes a sum over all
polarization components:  $\cP_C > \cP_T$.
In principle, this is not a problem because only $\cP_T$ has physical meaning;
we should just sum over the physical transverse polarizations to get the
right answer. However, the problem comes back at higher orders.

\begin{figure}[hbt]
\centerline{\epsfysize =3.5cm \epsfbox{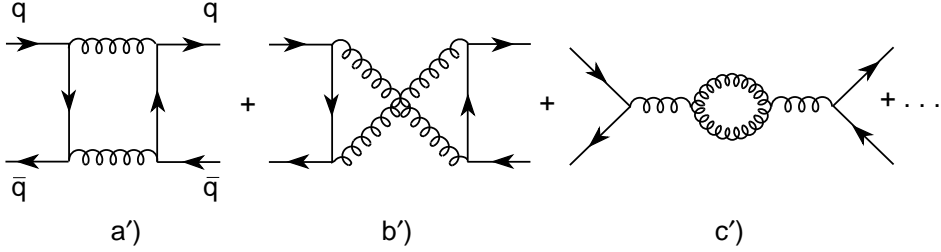}}
\vspace{-0.2cm}
\caption{1-loop diagrams contributing to $q\bar q\to q\bar q$.}
\label{fig:loop}
\end{figure}

The covariant gluon propagator \eqn{eq:propagator} contains the 4 polarization
components; therefore higher-order graphs such as the ones in
Fig.~\ref{fig:loop} get unphysical contributions from the longitudinal and
scalar gluon polarizations propagating along the internal gluon lines.
The absorptive part of these 1-loop graphs (i.e. the imaginary part obtained
putting on-shell the two gluon lines within the loop)
is equal  to $|T(q\bar q\to GG)|^2$.
Thus, these loops suffer the same probability problem than the tree-level
$q\bar q\to GG$ amplitude.
The propagation of unphysical gluon components implies then a violation
of unitarity (the two fake polarizations contribute a positive probability).

In QED this problem does not appear because the gauge-fixing condition
$\partial^\mu A_\mu=0$ still leaves a residual gauge invariance
under transformations satisfying $\Box\theta = 0$. This guarantees that
(even after adding the gauge-fixing term) the electromagnetic current
is conserved, i.e.
$\partial_\mu J^\mu_{\mbox{\rms em}}=\partial_\mu (eQ\bar\Psi\gamma^\mu\Psi) =
0$.
If one considers the $e^+e^-\to\gamma\gamma$ process, which proceeds through
diagrams identical to a) and b) in Fig.~\ref{fig:qqGG},
current conservation implies
$k_\mu J^{\mu\mu'} = k'_{\mu'} J^{\mu\mu'} = 0$, where $k_\mu$ and $k'_{\mu'}$
are the momenta of the photons with polarizations $\lambda$ and $\lambda'$,
respectively (remember that the interacting vertices contained in
$J^{\mu\mu'}$ are in fact the corresponding electromagnetic currents).
As a consequence, the contributions from the
scalar and longitudinal photon  polarizations vanish
and, therefore, $\cP_C = \cP_T$.

The reason why $\cP_C\not=\cP_T$ in QCD stems from the third diagram in
Fig.~\ref{fig:qqGG}, involving a gluon self-interaction.
Owing to the non-abelian character of the $SU(3)$ group, the gauge-fixing
condition $\partial_\mu G^\mu_a=0$ does not leave any residual
invariance\footnote{\small
To maintain $\partial_\mu (G^\mu_a)'=0$ after the gauge transformation
\eqn{eq:inf_transf}, one would need
$\Box \delta\theta_a = g_s f^{abc}\partial_\mu(\delta\theta_b) G^\mu_c$,
which is not possible because $G^\mu_c$ is a quantum field.}.
Thus, $k_\mu J^{\mu\mu'}\not= 0$.

\begin{figure}[hbt]
\centerline{\epsfysize =3cm \epsfbox{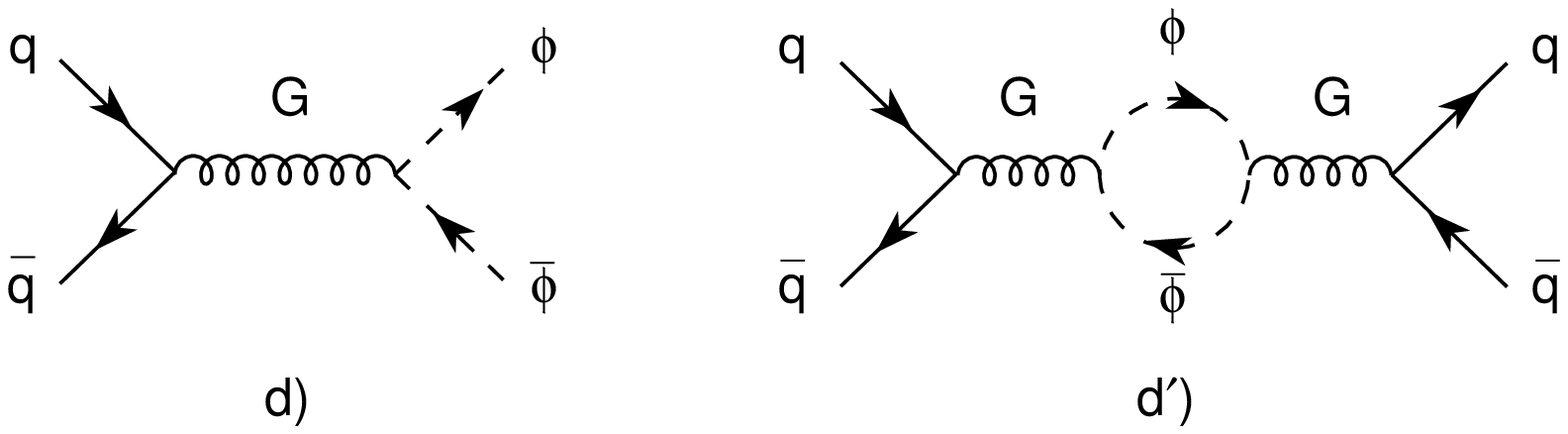}}
\caption{Feynman diagrams involving the ghosts.}
\label{fig:ghosts}
\end{figure}

Again, the problem could be solved adopting a non-covariant quantization where
only the physical transverse polarizations propagate; but the resulting
formalism would be awful and very inconvenient for performing
practical calculations.
A more clever solution    
consist \cite{FE:63} in adding additional unphysical fields,
the so-called {\it ghosts}, with a coupling to the gluons
such that exactly cancels {\it all} unphysical contributions from
the scalar and longitudinal gluon polarizations.
Since a positive fake probability has to be cancelled, one needs fields
obeying the wrong statistics (i.e. of negative norm)
and thus giving negative probabilities.
The magic cancellation is achieved by adding to the Lagrangian the
Faddeev--Popov term \cite{FP:67},
\bel{eq:L_FP}
\cL_{\!\!\mbox{\rms FP}} \, = \, -\partial_\mu\bar\phi_a D^\mu\phi^a \, ,
\qquad
\quad D^\mu\phi^a\equiv \partial^\mu\phi^a - g_s f^{abc} \phi^b G^\mu_c \, ,
\ee
where $\bar\phi^a$, $\phi^a$ ($a=1,\ldots,N_C^2-1$) is a set of
anticommuting (i.e. obeying the Fermi-Dirac statistics), massless,
hermitian, scalar fields.
The covariant derivative $D^\mu\phi^a$ contains the needed coupling to the
gluon field. One can  easily check that
diagrams d) and d') in Fig.~\ref{fig:ghosts} exactly cancel the unphysical
contributions from diagrams c) and c') of Figs.~\ref{fig:qqGG} and
\ref{fig:loop}, respectively; so that finally $\cP_C=\cP_T$.
Notice, that the Lagrangian \eqn{eq:L_FP} is necessarily not Hermitian,
because one needs to introduce an explicit violation of unitarity
to cancel the unphysical probabilities and restore the unitarity of the
final scattering amplitudes.

The exact mechanism giving rise to the $\cL_{\!\!\mbox{\rms FP}}$ term
can only be understood (in a simple way) using the more powerful
path-integral formalism, which is beyond the scope of these lectures.
The only point I would like to emphasize here, is that the addition of
the gauge-fixing and Faddeev--Popov Lagrangians is just a mathematical
trick, which allows to develop a simple covariant formalism, and
therefore a set of simple Feynman rules, making easier to perform
explicit calculations.


\end{document}